\documentclass[a4paper,12pt]{article}

\usepackage[a4paper,
            left=2cm,
            right=2cm,
            top=2cm,
            bottom=3cm]{geometry}
            
\usepackage{epsfig,latexsym,amsfonts,amsmath,amsthm,amssymb,amsbsy,array,makecell,multirow,slashed,wasysym,textcomp,subfigure,wrapfig,datetime,comment,mathtools,cancel,cite,mathrsfs,arydshln,rotating,float,multirow,booktabs,tabularx,arydshln,bm,dsfont}
\usepackage[font={footnotesize}]{caption}
\usepackage{braket}
\usepackage[normalem]{ulem}

\def\be{\begin{equation}}
\def\ee{\end{equation}}
\def\bea{\begin{eqnarray}}
\def\eea{\end{eqnarray}}
 \newcommand{\badat}{\begin{alignedat}}
 \newcommand{\eadat}{\end{alignedat}}

\long\def\new#1\endnew{{\bf #1}}		
\long\def\del#1\enddel{}

\def\del{\partial}

\RequirePackage[dvipsnames]{xcolor}
\definecolor{oldmauve}{rgb}{0.4, 0.19, 0.28}
\definecolor{pansypurple}{rgb}{0.47, 0.09, 0.29}
\definecolor{burgundy}{rgb}{0.5, 0.0, 0.13}
\definecolor{carminepink}{rgb}{0.92, 0.3, 0.26}
\definecolor{blue(pigment)}{rgb}{0.2, 0.2, 0.6}
\definecolor{darkseagreen}{rgb}{0.56, 0.74, 0.56}
\definecolor{darkspringgreen}{rgb}{0.09, 0.45, 0.27}
\definecolor{ceruleanblue}{rgb}{0.16, 0.32, 0.75}

\RequirePackage[dvipsnames]{xcolor}
\definecolor{arXiv}{named}{OliveGreen}
\definecolor{ColorCite}{named}{BrickRed}
\definecolor{ColorLink}{named}{NavyBlue}
\definecolor{ColorURL}{named}{RoyalBlue}

\usepackage[colorlinks=false,citecolor=ColorCite,linkcolor=ColorLink,urlcolor=ColorURL]{hyperref}

 \newcommand{\virg}{\hspace{1 mm}, \hspace{8 mm}}
\newcommand{\p}{\partial}

\def\bz{{\bar z}}
\def\bw{{\bar w}}

\newcommand{\C}{\mathscr{C}}
\newcommand{\N}{\mathscr{N}}

\renewcommand{\S}{\mathcal{S}}
\newcommand{\T}{\mathcal{T}}
\renewcommand{\P}{\mathcal{P}}
\newcommand{\E}{\mathcal{E}}
\newcommand{\Q}{\mathcal{Q}}
\newcommand{\q}{q_e}

\newcommand{\D}{d}

\newtheorem{theorem}{Theorem}[section]

\theoremstyle{remark}

 \makeatletter
\def\@part[#1]#2{\par
  \addcontentsline{toc}{part}{#1}%
    {\parindent0pt\raggedright   
  \LARGE \bfseries #2\par}%
  \vskip 2ex}
\makeatletter
\makeatletter
\renewcommand*{\toclevel@part}{0}   
\makeatother

\begin{document}
\author{Laura Donnay}
\numberwithin{equation}{section} 

\begin{titlepage}
  \thispagestyle{empty}
  
  \begin{center} 
  \vspace*{5cm}
{\LARGE\textbf{Infrared physics of QED and gravity}}\\  \vspace*{0.5cm}
{\LARGE\textbf{from representation theory}}

\vskip1cm

   \centerline{Laura Donnay$^{a,b}$\footnote{ldonnay@sissa.it}, Yannick Herfray$^{c}$\footnote{yannick.herfray@univ-tours.fr}}
   
\vskip1cm

\it{$^a$SISSA, Via Bonomea 265, 34136 Trieste, Italy}\\
\it{$^b$INFN, Sezione di Trieste, Via Valerio 2, 34127 Trieste, Italy}\\
\it{$^c$Institut Denis Poisson, Université de Tours, Université d’Orléans,\\ CNRS, IDP, UMR 7013, Parc de Grandmont,
37200 Tours, France}\\

\end{center}

\vskip1cm

\begin{abstract}
The infrared structure of QED and gravity is known to be governed by an infinite-dimensional symmetry group which extends the Poincaré group to include, respectively, large $U(1)$ transformations and BMS supertranslations.
We describe how the unitary irreducible representations (UIRs) of these asymptotic symmetry groups encode universal infrared features of a scattering process. Motivated by the goal of defining an infrared-finite $S$-matrix based on these UIRs, we also study supermomentum eigenstates and contrast our construction with the dressed-state approach for infrared-safe amplitudes.
\end{abstract}

\end{titlepage}

\tableofcontents

\newpage
\phantomsection
\setcounter{footnote}{0}

\part{Introduction}

Despite its status as one of the most precise physical theories constructed so far, QED features scattering amplitudes that suffer from inherent infrared (IR) divergences, rendering the $S$-matrix between conventional Fock states mathematically ill-defined\footnote{After resummation, the amplitude goes to zero as the IR cutoff is removed.}~\cite{Bloch:1937pw}. Infrared divergences occur because standard perturbation theory assumes that asymptotic incoming and outgoing states can be identified with states that evolve freely at asymptotic times, which is clearly incorrect for theories with long-range forces mediated by massless bosons.  
The traditional way to deal with these IR divergences, which traces back to the classic work of Bloch and Nordsieck~\cite{Bloch:1937pw}, is to rather consider inclusive cross-sections, where photons of energy smaller than the detector resolution are traced over. 
Remarkably, the cross-section method provides infrared-safe observables~\cite{Yennie:1961ad,Kinoshita:1962ur,Lee:1964is,Weinberg:1965nx,Weinberg:1995mt,Grammer:1973db}; however, it also gives up on a mathematically well-defined notion of an $S$-matrix.

Foundational contributions toward resolving this issue were made in~\cite{Chung:1965zza,Dollard:1964cok,Greco:1967zza,Blanchard1969,Kibble:1968sfb,KibbleII,KibbleIII,KibbleIV},  ultimately resulting in the Faddeev-Kulish (FK) construction \cite{Kulish:1970ut}, which introduced a framework for constructing asymptotic states that diagonalize the interacting asymptotic Hamiltonian. FK states describe charged particles dressed by ``photon clouds'', and their associated $S$-matrix elements were shown to be free of IR divergences\footnote{This construction does not apply to QED with massless charged particles.}.
Despite this progress, the FK construction still has major shortcomings. First of all, although IR divergences are removed from the scattering elements, the associated singular behavior now resides in the states themselves~\cite{Hannesdottir:2019opa}. As a result, FK dressed states are not normalizable elements of a Fock space. Rather, they belong to an enlarged Hilbert space $\mathcal H_{\text{FK}}$, which is only separable and gauge-invariant \cite{Kulish:1970ut} for zero net electric charge\footnote{This precludes the existence of an analogous statement for gravity, as pointed out in \cite{Prabhu:2022zcr}.}. Another issue with FK states is that they feature inherent ambiguities in the dressing factor and, consequently in the amplitude. It has been shown that different choices of dressing can lead to different analytic properties of the associated $S$-matrix~\cite{Hannesdottir:2019opa,Lippstreu:2025jit}. Finally, and related to the previous points, to date there exist almost no explicit examples -- aside from the works \cite{Forde:2003jt,Hannesdottir:2019opa,Hannesdottir:2019rqq} -- of computed $S$-matrix elements for FK states, and basic questions, such as the identification of the quantum numbers carried by dressed states, remain unresolved.

A novel viewpoint on IR divergences in QED was put forward in \cite{Kapec:2017tkm} (see also \cite{Gabai:2016kuf}), building on the realization that the infrared sector of abelian gauge theories with charged particles is governed by an infinite-dimensional symmetry group~\cite{He:2014cra,Campiglia:2015qka}. This group extends the Poincaré group to include an infinite set of $U(1)$ transformations that asymptote to an arbitrary function on the celestial sphere (see \cite{Strominger:2017zoo} for a review). It was then understood that IR divergences arise, for conventional states, as a necessity to respect the infinite set of conservation laws associated with these asymptotic symmetries. Once reconsidered in this new light, the dressed states of the original FK construction \cite{Kulish:1970ut} can be understood as implementing the conservation laws in a rather drastic manner, namely by setting the total asymptotic charges to zero. In this sense, the FK states were considered to be overly restrictive, as IR-finiteness of the amplitudes only requires the \emph{total conservation} of asymptotic charges~\cite{Kapec:2017tkm,Arkani-Hamed:2020gyp}. 

The analogue of this story in the context of gravity underwent a very different historical development: the fact that the asymptotic symmetry group of gravity in flat spacetimes spans an infinite-dimensional extension of the Poincaré group was discovered as early as the 1960s \cite{bondi_gravitational_1962,sachs_asymptotic_1962}. This extension, whose additional symmetry generators correspond to supertranslations, is known as the BMS group. However, it took several decades before the relevance of this group to the gravitational $S$-matrix was fully appreciated \cite{Strominger:2013jfa,He:2014laa}. Despite this different history, the status of infrared divergences for gravitational theories is now largely on par with that in QED: the analogue of the FK construction for perturbative quantum gravity was developed in \cite{Ware:2013zja}, and IR-finite dressed states were shown to be respect the infinite set of BMS conservation laws~\cite{Choi:2017bna,Choi:2017ylo}. \\

These developments lead to the following perspective, which underlies the present work.
The facts that (i) the standard Fock space is insufficient to accommodate asymptotic states that would yield infrared-safe scattering elements, and (ii) IR-safe dressed states satisfy the conservation laws associated with the infinite-dimensional asymptotic symmetries, strongly suggest that, if a unitary and IR-finite $S$-matrix exists, its states should be associated with unitary irreducible representations (UIRs) of the asymptotic symmetry group of QED/gravity.
The program of developing a systematic framework, based on representation theory, for scattering amplitudes involving massless interactions was initiated in \cite{Bekaert:2024uuy,Bekaert:2025kjb} for the gravitational setting. It builds on earlier pioneering works \cite{Mccarthy:1972ry,McCarthy_72-I,McCarthy_73-II,McCarthy_73-III,McCarthy:1974aw,Girardello:1974sq,McCarthy_75,McCarthy_76-IV,McCarthy_78,McCarthy_78errata,PiardBMS,Barnich:2014kra,Barnich:2015uva,Oblak:2016eij} that classified UIRs of the BMS group, i.e. ``BMS particles'', in terms of the allowed BMS little groups and associated supermomenta. These results were reconsidered in \cite{Bekaert:2024uuy,Bekaert:2025kjb} in light of the more recent developments relating asymptotic symmetries to the $S$-matrix. One of the key outcomes of these works is the realization of a unique, Lorentz-invariant but nonlinear decomposition of the BMS supermomentum into a hard and a soft part. This decomposition allows one to understand in which ways generic BMS particles carry additional degrees of freedom compared to hard UIRs, the latter being in one-to-one correspondence with familiar Poincaré representations.

A particularly encouraging aspect of these investigations is the discovery of a rich landscape of admissible asymptotic representations, which offers substantial room for the construction of a mathematically well-defined and infrared-finite gravitational $S$-matrix. However, among the set of all generic BMS representations, not all of them are expected to be physically relevant. A similar situation arises for Poincaré particles: although Wigner’s classification of Poincaré UIRs is mathematically complete \cite{Wigner:1939cj}, additional physical input is required to discard exotic representations, such as tachyonic or continuous-spin states.
This suggests that further physical criteria are needed to navigate the landscape of BMS UIRs and identify the 
subset of physically relevant representations. In this context, a first step is to revisit and reformulate the standard understanding of infrared divergences in conventional scattering -- such as soft theorems and virtual divergences from soft exchanges -- entirely within the framework of representation theory for both the BMS and QED asymptotic symmetry groups. Another important question is to understand whether and how the Faddeev-Kulish construction (and its generalizations) fits into the framework of representation theory, and to compare dressed states with supermomentum eigenstates, which are the natural objects arising from representation theory. These are the main goals of this work. In the case of QED, this first requires extending the most salient aspects of the corresponding representation theory, along the lines of what was carried out in \cite{Bekaert:2024uuy,Bekaert:2025kjb} for the BMS group. This is developed in the first part of the present work, while a detailed classification of the UIRs of the QED asymptotic symmetry group will be presented in \cite{Bekaert2026}.\\

This paper is organized into two main parts: one devoted to QED and the other to gravity, anticipating that readers may wish to focus on only one. At the same time, the two parts intentionally mirror each other, thereby emphasizing the parallels between them.

In Part I, we focus on QED (Part II applies to gravity and follows the same structure). We start in Section \ref{sec:hard_reps} (resp. \ref{sec:hardBMS_reps}) with a presentation of the so-called hard representations of the asymptotic symmetry group of QED (gravity) and introduce the corresponding hard supermomentum. 
Among generic representations, hard UIRs are very special as they are in one-to-one correspondence with usual Poincaré UIRs. We also explain why hard representations alone are insufficient to address infrared-safe scattering, as they fail to conserve supermomentum. In Section \ref{sec:generic_QED} (resp. Section \ref{sec:generic_BMS}), we turn to generic representations of the asymptotic symmetry groups. After introducing the notion of a generic QED (gravitational) supermomentum and the associated asymptotic little groups, we describe hard and soft representations and demonstrate the hard/soft decomposition of supermomenta. We also present the corresponding wavefunctions and supermomentum eigenstates.
Section \ref{sec: IR div QED} (resp. Section \ref{sec: IR div gravity}) revisits infrared divergences in conventional hard scattering processes. We recall how photons (gravitons) beyond the hard Fock space naturally arise, reinterpret the soft photon (graviton) theorem as a statement of supermomentum conservation, and reproduce the exponentiation formula for virtual soft divergences within the representation-theoretic framework. In the last Section \ref{sec: dressed to kill QED} (resp. Section \ref{sec: dressed to kill gravity}), we turn to dressed states and their relationship with supermomentum conservation laws. After a review and discussion on the (revisited) Faddeev-Kulish construction, we compare and contrast dressed states with supermomentum eigenstates.

We conclude with a discussion of the implications of our results and directions for future work. The paper also contains several appendices collecting conventions and technical material used throughout. Appendix \ref{app:conventions} summarizes our conventions, including conformal densities and spin weights on the celestial sphere, and properties of the $\eth$ operator and delta functions. It also includes useful identities involving hard supermomenta and reviews our parametrizations of massless and massive momenta. Appendix \ref{Appendix: section representations} provides a brief review of induced-representation methods for unitary irreducible representations of semi-direct product groups of the form $SO(3,1) \ltimes A$. Finally, Appendix \ref{App:distrib_id} collects distributional identities on the celestial sphere, including the ones relevant for hard supermomenta as well as inner products in the soft sector. 

\phantomsection
\part{Part I. QED}

The asymptotic symmetry group of abelian gauge theories with charged particles was shown to include an infinite amount of $U(1)$ gauge transformations that asymptotically approach an arbitrary function $\varepsilon(z,\bz)$ of the celestial sphere~\cite{Strominger:2013lka,Barnich:2013sxa,He:2014cra,Campiglia:2015qka}. Since the asymptotic gauge parameter $\varepsilon(z,\bz)$ transforms as a conformal density of weights $(h,\bar{h})=(0,0)$ under the action of the Lorentz group, the QED asymptotic symmetry group is\footnote{We denote by $SO(3,1)$ the proper, orthochronous Lorentz group. Throughout this paper, $(h,\bar{h})$ stand for the (anti-)holomorphic conformal weights as defined in the CFT literature and $\E[w]$ is space of smooth conformal densities of weight $w$ on the sphere, as is standard in conformal geometry. Thus, elements of $\E[w]$ correspond to fields of conformal weights $(h,\bar{h}) = (-\frac{w}{2},-\frac{w}{2})$; see Appendix \ref{app:conformal_density} for our conventions. Here, for QED, the notation $\E[0]$ is favored (over $\C^{\infty}(S^2)$) to maintain uniformity with the gravity case.}
\begin{equation}\label{eq:AGS}
\tag{1}
    SO(3,1) \ltimes \Big( \mathbb{R}^{3,1} \times \mathcal E[0] \Big)\,,
\end{equation}
where $\mathcal E[0]=\C^{\infty}(S^2)$ denotes the vector space of smooth conformal densities of weight zero on the celestial sphere. $\mathbb{R}^{3,1} \times \mathcal E[0]$ is an abelian group which can be thought of as an infinite-dimensional extension of translations; elements of this group are given by pairs $\left(T^{\mu}, \varepsilon(z,\bz) \right)$ of translations $T^{\mu} \in \mathbb{R}^{3,1}$ together with asymptotic gauge parameters $\varepsilon(z,\bz) \in \mathcal E[0]$. 

\section{Hard representations of the QED asymptotic symmetry group}
\label{sec:hard_reps}

The general representation theory for UIRs of groups of the form $SO(3,1) \ltimes A$, with $A$ an (infinite-dimensional) abelian group, is briefly reviewed in Appendix \ref{Appendix: section representations}. For the representations that we will consider in this section, it will be enough to know that the abelian group $\mathbb{R}^{3,1} \times \mathcal E[0]$ acts on the usual creation/annihilation operator of QED, of momentum $p_{\mu}$ and electric charge $\q$, as
\begin{equation}\label{eq:action of hard reps}
    \hat b(p) \;\mapsto\; e^{i\,\left( p_{\mu}T^{\mu} \,+\, \langle Q, \varepsilon \rangle \right)}\, \hat b(p)\,.
\end{equation}
In particular, for a constant symmetry parameter $\varepsilon$, one will recover the usual global $U(1)$ representation\footnote{For hard representations, this will follow from the identities discussed in Appendix \ref{app:hard_id}.}: 
\begin{equation}
   \hat b(p) \to \,e^{i\,\left( p_{\mu}T^{\mu} \,+\, \q \varepsilon \right)}\,\hat b(p).
\end{equation}
In \eqref{eq:action of hard reps}, the phase is the contraction of $\left(T^{\mu}, \varepsilon(z,\bz) \right)$ with a dual element
\begin{equation}\label{Supermomentum}
   \Big(p_{\mu}, Q(z,\bz) \Big) \in \left(\mathbb{R}^{3,1}\right)^* \times \mathcal E[-2] .
\end{equation}
In particular, $Q(z,\bz)$ is dual to the asymptotic gauge parameter $\varepsilon(z,\bz)$: it is a conformal density of weights $(1,1)$ with pairing defined as\footnote{Here and everywhere in this article, $d^2z: = i dz \wedge d\bar{z}$; see Appendix \ref{app: delta function}.}
\be
\langle Q, \varepsilon\rangle= \int d^2 z \,Q(z,\bz) \varepsilon(z,\bz) \quad \in \mathbb R\,.
\ee
Dual elements such as \eqref{Supermomentum} will be referred to as ``supermomenta'' in the rest of the article; as we will see, this terminology has the advantage of treating the representation theory of the group \eqref{eq:AGS} and of the BMS group on the same footing (see Appendix \ref{Appendix: section representations}).\\ 

In this section, we show that the familiar Poincar\'e UIRs lift to representations of the infinite-dimensional asymptotic symmetry group \eqref{eq:AGS}.
They give what we will refer to as the ``hard UIRs'' of the asymptotic symmetry group of QED. In practice, this means that, for such representations, the corresponding hard supermomentum \eqref{Supermomentum} is completely determined by both the usual momentum $p_{\mu}$ and electric charge $\q$. The explicit expressions of hard UIRs for a massless and massive scalar, which we will explain in the rest of this section, are summarized in Table \ref{tbl:Dic}.  Among \emph{generic} representations (which will be presented in Section \ref{sec:generic_QED}), hard UIRs play a very special role as they are the ones which are in one-to-one correspondence with the usual Poincaré UIRs. 
\vspace{10pt}
\begin{table}[H]
\centering
\begin{minipage}{0.8\textwidth} 
\centering
	\begin{tabular}{c|l}
		$\,\,$  Hard rep. $\,\,$ & \hspace{1cm} $\hat b(p) \to e^{i\left( p_{\mu}T^{\mu}+ \langle Q, \varepsilon \rangle\right)}\, \hat b(p)$ $\,\,$ \\[0.4cm]
         \hline \\[-0.2cm]    
massless & $\Big(p^{\mu}, Q(z,\bz) \Big) =   \Big( \omega q^{\mu}(\zeta, \bar{\zeta}),  \q \delta^{(2)}(z-\zeta) \Big)$\\[0.4cm]
massive& $\Big(p^{\mu}, Q(z,\bz) \Big) =   \Big( p^{\mu}, \dfrac{\,\q\, m^2}{4\pi\, (q(z,\bz)\cdot p)^2}\Big)$\\		[0.3cm]
	\end{tabular}
	\caption{Hard UIRs of the asymptotic symmetry group of QED \eqref{eq:AGS} for a scalar of massless ($p^\mu=\omega q^\mu(\zeta,\bar \zeta)$, $q^2=0$) and massive ($p^2=-m^2$) momentum and charge $q_e$. Group elements consist of translations $T^\mu \in \mathbb R^{3,1}$ and asymptotic gauge parameters $\varepsilon(z,\bz) \in\E[0]$. The hard supermomentum is $\left( p^{\mu}, Q(z,\bz)\right) \in \mathbb{R}^{3,1} \times \E[-2]$.}
	\label{tbl:Dic}
    \end{minipage}
\end{table}

Although QED hard representations can be understood through their realization in terms of the scattering data of massless \cite{He:2014cra,Strominger:2017zoo} and massive~\cite{Campiglia:2015qka,Campiglia:2015lxa} fields at, respectively, null and timelike infinity, our presentation avoids relying on explicit asymptotic expansions. We hope that this will make the discussion natural for readers who are not already familiar with the technicalities of asymptotic analysis. We start the presentation of hard UIRs in Section \ref{subsec:hard_QED_scalar} for a scalar field and then turn to spinning fields in Section \ref{subsec:hard_QED_spinning}. 

\subsection{Scalar fields}
\label{subsec:hard_QED_scalar}
Let $\phi(X)$ a complex scalar field of electric charge $q_e$ and momentum $p^\mu=(p^0,\vec p)$,
\begin{equation}
    \phi(X) =\int \frac{\D^3 p}{(2\pi)^3\, 2p^0} \Big[ \hat b(\vec p) \, e^{i p \cdot X} +  \hat d(\vec p)^\dagger \, e^{-i p \cdot X}\Big] \label{complex_scalar}\,,
\end{equation}
where $\hat b(\vec p)$ and $\hat d(\vec p)$ are the annihilation operators for the particles and antiparticles, respectively. They obey the usual commutation relations
\begin{equation}
   \left[ \hat b(\vec p),\hat b(\vec p\,')^\dagger \right]  =\left[ \hat d(\vec p),\hat d(\vec p\,')^\dagger \right] =(2\pi)^3\,2p^0\,\delta^{(3)}(\vec p-\vec p\,')\,.
\end{equation}

\subsubsection{Poincaré UIRs}
Let us first review some elementary facts about the familiar UIRs of the Poincaré group.
In momentum space, the momentum and Lorentz operators given by
\begin{equation}
\badat{2}
\label{eq:Poin_gen1}
&\hat P^\mu=\int \frac{\D^3 p}{(2\pi)^3\, 2p^0} \,p^\mu \left(\hat b^\dagger(\vec p) \hat b(\vec p)+\hat d^\dagger(\vec p) \hat d(\vec p)\right)\\
&\hat J^{\mu\nu}=i\int \frac{\D^3 p}{(2\pi)^3\, 2p^0} \left[\hat b^\dagger(\vec p)\Big(p^\mu \frac{\partial}{\p p_\nu}-p^\nu \frac{\partial}{\p p_\mu}\Big) \hat b(\vec p)+\hat d^\dagger(\vec p) \Big(p^\mu \frac{\partial}{\p p_\nu}-p^\nu \frac{\partial}{\p p_\mu}\Big)
\hat d(\vec p)\right]\,,
\eadat
\end{equation}
form an explicit representation of the Poincaré algebra\,,
\begin{equation}
\badat{2}
&\big[\hat{P}^{\mu} , \hat{P}^{\nu}  \big]=0\\
&\big[\hat{J}^{\mu\nu} , \hat{P}^{\rho}  \big]=i \eta^{\nu \rho}\hat P^\mu -i \eta^{\mu \rho}\hat P^\nu\\
&\big[\hat{J}^{\mu\nu} , \hat{J}^{\rho\sigma}  \big] = i\big(\eta^{\sigma\mu}\hat J^{\nu\rho}  -\eta^{\rho\mu} \hat J^{\nu \sigma} + \eta^{\rho\nu} \hat J^{\mu\sigma} -  \eta^{\sigma\nu} \hat  J^{\mu\rho}   \big)\,,
\eadat
\end{equation}
and a corresponding representation of the Poincaré group via $U(\omega,a)=\exp(\frac i 2 \omega_{\mu \nu}\hat{J}^{\mu \nu}+ia_\mu \hat{P}^\mu)$.
It defines an (infinite-dimensional) unitary irreducible representation (UIR) acting on the Hilbert space of one-particle states. 

The conserved current associated with the global $U(1)$ symmetry of QED is given by 
\begin{equation}\label{eq:current_U1}
   j_\mu=\,:\!iq_e(\phi \p_\mu \phi^*-\phi^* \p_\mu \phi)\!: 
\end{equation}
and the associated charge operator $\hat{Q}_e=\int d^3 x \,j^0$ is
\begin{equation}
\badat{2}\label{eq:Q_e}
&\hat{Q}_e =q_e \int \frac{\D^3 p}{(2\pi)^3\, 2p^0} \,\left(\hat b^\dagger(\vec p) \hat b(\vec p)-\hat d^\dagger(\vec p) \hat d(\vec p)\right)\,.
\eadat
\end{equation}
Since $\hat{Q}_e$ commutes with all Poincaré generators \eqref{eq:Poin_gen1}, it trivially extends the algebra of symmetries, resulting in a direct sum of spacetime and gauge symmetries,
\begin{equation}\label{eq:AGS_0}
   \Big( \mathfrak{so}(3,1) \ltimes  \mathbb{R}^{3,1} \Big) \oplus \mathfrak u(1) \,.
\end{equation}
The creation/annihilation operators are eigenvalue of both momentum and the $U(1)$ charge,
\begin{equation}
\badat{2}
&[\hat{Q}_e, \hat b^{\dagger}(p)] = q_e \hat b^{\dagger}(p) \virg &[\hat{P}^{\mu}, \hat b^{\dagger}(p)] = p^{\mu}\hat b^{\dagger}(p)\,,\\
&[\hat{Q}_e, \hat d^{\dagger}(p)] = -q_e \hat d^{\dagger}(p) \virg &[\hat{P}^{\mu}, \hat d^{\dagger}(p)] = p^{\mu} \hat d^{\dagger}(p).   
\eadat
\end{equation}

\subsubsection{Hard representations of the asymptotic symmetry group}
We now show that the familiar Poincaré representations can be lifted, at no cost, to what we will refer to as ``hard'' UIRs of the enhanced symmetry group \eqref{eq:AGS}.

To each particle of momentum $p^{\mu}$ and electric charge $q_e$, one can associate a weight--$(1,1)$ density on the celestial sphere, denoted $Q(z,\bz)$, via\footnote{Note that there also exists a hard supermomentum for the tachyonic case $p^2>0$, which we do not consider here.}

\vspace*{0.1cm}
\begin{equation} \label{eq:Qhard}
Q(z,\bar z) =
\left\{\begin{aligned}
  &\hspace{0.5cm} \quad q_e\,\delta^{(2)}(z-\zeta) & \quad \text{if } &p^2=0,  \;p_{\mu} = \omega q_{\mu}(\zeta,\bar{\zeta})\\[0.6em]
  &\hspace{0.5cm} \dfrac{q_e m^2}{4\pi\,(q(z,\bar z)\cdot p)^2} & \quad\, \text{\,if } &p^2=-m^2\,.\\[0.4
  em]
\end{aligned}\right.
\end{equation}
In the above, $q^\mu(z,\bz)$ is the canonical null vector which realizes the identification of the celestial sphere as the projective null cone; see Appendix \ref{app: q}.
For both massless and massive particles, one has
\be \label{eq:surjection_qe}
q_e=\int d^2z \,Q(z,\bz)\,.
\ee
The proof of identity \eqref{eq:surjection_qe} is trivial for the massless case and easily showed for the massive case in Appendix \ref{app:hard_id}.
Importantly, let us emphasize that $Q(z,\bar z)$ depends on the momentum, as the second line of \eqref{eq:Qhard} shows explicitly and (less explicitly) in the expression for the massless case through the delta function $\delta^{(2)}(z-\zeta)$. 

 The key property of \eqref{eq:Qhard} is that both expressions satisfy the following identities (see Appendix \ref{app:hard_id})
\begin{equation}
    2p_{[\mu} \partial_{p^\nu]} \,Q(z,\bz)  = \Big(  \mathcal{Y}_{\mu\nu}^{z} \partial_z +  \mathcal{Y}_{\mu\nu}^{\bz} \partial_{\bz}  +  \partial_z\mathcal{Y}_{\mu\nu}^{z} + \partial_{\bz}\mathcal{Y}_{\mu\nu}^{\bz}\Big)\,Q(z,\bz)\,,
\end{equation}
where $\partial_{p^\mu} := \frac{\partial}{\partial p^{\mu}}$ and
\begin{equation}
    \mathcal{Y}_{\mu\nu}^{z}(z,\bar{z}) \partial_z:=  2\partial_{\bz}q_{[\mu} q_{\nu]} \partial_z \virg  \mathcal{Y}_{\mu\nu}^{\bz}(z,\bar{z}) \partial_\bz:=  2\partial_{z}q_{[\mu} q_{\nu]} \partial_\bz 
\end{equation}
are the infinitesimal generators of Lorentz transformations on the celestial sphere.

We can now use the above facts to construct an explicit representation of the asymptotic gauge symmetry algebra
\begin{equation}\label{eq:AGS_1}
 \mathfrak{so}(3,1) \ltimes     \Big( \mathbb{R}^{3,1}  \oplus \E[0] \Big)\,.
\end{equation}
We define the hard supercharge operator
\begin{equation}\label{eq:hatQ}
\badat{2}
&\hat Q(z,\bz)=\int \frac{\D^3 p}{(2\pi)^3\, 2p^0} \,Q(z,\bz) \left(\hat b^\dagger(\vec p) \hat b(\vec p)-\hat d^\dagger(\vec p) \hat d(\vec p)\right)\,,
\eadat
\end{equation}
which extends the usual expression \eqref{eq:Q_e}.
A direct computation shows that, together with the Poincaré generators $\hat{P}^{\mu}$, $\hat J^{\mu \nu}$ in \eqref{eq:Poin_gen1}, it realizes the algebra \eqref{eq:AGS_1}:
{
\setlength{\jot}{8pt}
\begin{equation}
\badat{2}\label{eq:AGS_hard}
&\big[\hat{P}^{\mu} , \hat{P}^{\nu}  \big]=0 \virg \big[\hat{J}^{\mu\nu} , \hat{P}^{\rho}  \big]=i \eta^{\nu \rho}\hat P^\mu -i \eta^{\mu \rho}\hat P^\nu\,,\\
&\big[\hat{J}^{\mu\nu} , \hat{J}^{\rho\sigma}  \big] = i\big(\eta^{\sigma\mu}\hat J^{\nu\rho}  -\eta^{\rho\mu} \hat J^{\nu \sigma} + \eta^{\rho\nu} \hat J^{\mu\sigma} -  \eta^{\sigma\nu} \hat  J^{\mu\rho}   \big)\,,\\
& \big[\hat{P}^{\mu} , \hat{Q}(z,\bz)  \big] =0\virg \big[\hat{Q}(z,\bz) , \hat{Q}(w,\bw) \big]=0\,,\\
 &\big[\hat{J}_{\mu\nu} , \hat{Q}(z,\bz)  \big] = i \big( \mathcal{Y}_{\mu\nu}^{z} \partial_z  + \mathcal{Y}_{\mu\nu}^{{\bz}} \partial_{\bz} +    \partial_z \mathcal{Y}_{\mu\nu}^{z} + \partial_{\bz}\mathcal{Y}_{\mu\nu}^{\bz} \big)\, \hat{Q}(z,\bz)\,.
\eadat
\end{equation}}
In particular, the last line of \eqref{eq:AGS_hard} shows that  $\hat{Q}(z,\bz) $ indeed transforms as a conformal primary of weights $(1,1)$ under Lorentz transformations\footnote{Which implies that $\mathcal F[\epsilon]=\int d^2z\, \hat Q(z,\bz) \varepsilon(z,\bz)$ satisfies $\big[\hat{J}_{\mu\nu} , \mathcal F[\varepsilon] \big] =i \,\mathcal F\big[  \mathcal{Y}_{\mu\nu}^{z} \partial_z  \varepsilon + \mathcal{Y}_{\mu\nu}^{{\bz}} \partial_{\bz}  \varepsilon\,\big]$.}. We thus see that, thanks to the existence of the canonical projection on the usual electric charge operator \eqref{eq:Q_e}
\begin{equation}
    \hat{Q}(z,\bz)  \mapsto \hat{Q}_e = \int d^2z \,\hat{Q}(z,\bz),
\end{equation}
each familiar Poincaré UIR can in fact be thought of as a UIR of the enhanced asymptotic gauge symmetry group \eqref{eq:AGS}. In this representation, the usual creation/annihilation operators are eigenvectors of both the supercharge and momentum 
\begin{equation}
\badat{2}
[\hat{Q}(z,\bz), \hat b^{\dagger}(p)] &= Q(z,\bz)\,\hat b^{\dagger}(p)\virg
& [\hat{P}^{\mu}, \hat b^{\dagger}(p)] &= p^{\mu}\,\hat b^{\dagger}(p)\,, \\
[\hat{Q}(z,\bz), \hat d^{\dagger}(p)] &= -Q(z,\bz)\,\hat d^{\dagger}(p) \virg 
& [\hat{P}^{\mu}, \hat d^{\dagger}(p)] &= p^{\mu}\,\hat d^{\dagger}(p)\,,
\eadat
\end{equation}
the eigenvalue $Q(z,\bz)$ being given by \eqref{eq:Qhard}. Therefore, and as announced in the introduction of this section, the finite action of the group, generated by 
\begin{equation}
    U(\omega, T, \varepsilon) = \exp\Big(\frac i 2 \omega_{\mu \nu}\hat{J}^{\mu \nu}+ia_\mu \hat{P}^\mu +i\int d^2 z \,\varepsilon(z,\bz) \hat{Q}(z,\bz)\Big)\,,
\end{equation}
induces an action of the form \eqref{eq:action of hard reps}.
 
\subsubsection{Hard representations as scattering data}
As we saw, the above discussion makes no reference to the asymptotic expansion of the field.  Nevertheless, it is also possible to understand the hard representations as the scattering data of a scalar field at null and timelike infinity, for massless and massive hard UIRs, respectively. 
For the massless case, the hard operator \eqref{eq:hatQ} can be seen to coincide with the expression in \cite{He:2014cra}
\be
\hat Q(z,\bz)=-\int_{-\infty}^{+\infty} du \,j_u^{(2)}(u,z,\bz) \virg 
\ee
where $j_u^{(2)}(u,z,\bz)=\lim_{r \to \infty}r^2 j_u(u,r,z,\bz)$ is the charged matter current at null infinity (see Appendix~\ref{app:momenta_massless} for our coordinate conventions).
For the massive case~\cite{Campiglia:2015qka}, the hard operator \eqref{eq:hatQ} coincides with
\be
\hat Q(z,\bz)=-\int_{\mathbb H^3} \frac{d^3 y}{\sqrt{1+|y|^2}} \,\frac{j^{(3)}_\tau(y^{\alpha})}{4\pi\,\big(q(z,\bar z)\cdot \hat{p}(y^{\alpha})\big)^2}\, \virg 
\ee
where $j^{(3)}_\tau(y^\alpha)=\lim_{\tau \to \infty}\tau^3 j_\tau(\tau,y^\alpha)$ is the charged matter current at timelike infinity, $y^\alpha$ denoting the coordinates on the hyperboloid $\mathbb H^3$ (see Appendix \ref{app:momenta_massive} for conventions).

\subsection{Spinning fields}
\label{subsec:hard_QED_spinning}
For completeness, we now turn to the discussion of hard representations for spinning fields, which follow in a very similar way from the scalar case presented in the previous section.
\subsubsection*{Dirac fields}
Let us start with a Dirac field,
\begin{equation}
\badat{2}
    &\psi(X) =\sum_{\gamma=\pm}\int \frac{\D^3 p}{(2\pi)^3\, 2p^0} \Big[ \hat b_\gamma(\vec p) u^\gamma(\vec p) \, e^{i p \cdot X} +  \hat d_\gamma(\vec p)^\dagger v^\gamma(\vec p) \, e^{-i p \cdot X}\Big] \,,
\eadat
\end{equation}
with creation and annihilation operators obeying the anticommutation rules
\begin{equation}
   \left\{ \hat b_\gamma(\vec p),\hat b_\delta(\vec p\,')^\dagger \right\}  =\left\{ \hat d_\gamma(\vec p),\hat d_\delta(\vec p\,')^\dagger \right\} =\delta_{\gamma \delta}(2\pi)^3\,2p^0\,\delta^{(3)}(\vec p-\vec p\,')\,.
\end{equation}
The momentum and Lorentz generators are given by
\begin{equation}
\badat{2}
\label{eq:Poin_gen1_s1/2}
&\hat P^\mu=\int \frac{\D^3 p}{(2\pi)^3\, 2p^0} \,p^\mu \sum_{\gamma=\pm}\left(\hat b^\dagger_\gamma(\vec p) \hat b_\gamma(\vec p)+\hat d^\dagger_\gamma(\vec p) \hat d_\gamma(\vec p)\right)\\
&\hat J^{\mu\nu}=\int \frac{\D^3 p}{(2\pi)^3\, 2p^0} \,\sum_{\gamma,\gamma'=\pm}\left[\hat b^\dagger_\gamma(\vec p)\Big(i\big(p^\mu \frac{\partial}{\p p_\nu}-p^\nu \frac{\partial}{\p p_\mu}\big)\delta_{\gamma \gamma'}+\bar u_\gamma(\vec p)\Sigma^{\mu \nu} u_{\gamma'}(\vec p)\Big)  \hat b_{\gamma'}(\vec p)+(\hat b \to \hat d, u \to v)\right]\,,
\eadat
\end{equation}
with $\Sigma^{\mu \nu}$ the spin generator in the spinor representation, and the charge operator is
\begin{equation} \label{eq:charge_dirac}
\badat{2}
&\hat Q_e=\int \frac{\D^3 p}{(2\pi)^3\, 2p^0} \,\sum_{\gamma=\pm}\left(\hat b^\dagger_\gamma(\vec p) \hat b_\gamma(\vec p)-\hat d^\dagger_\gamma(\vec p) \hat d_\gamma(\vec p)\right)\,.
\eadat
\end{equation}

The lift from usual Poincaré UIRs of momentum $p^\mu$ and spin $\frac 12$ to UIRs of the asymptotic group \eqref{eq:AGS} proceeds analogously to the scalar case. We promote the charge operator \eqref{eq:charge_dirac} to the celestial operator
\begin{equation} \label{eq:charge_dirac_lifted}
\badat{2}
&\hat Q(z,\bz)=\int \frac{\D^3 p}{(2\pi)^3\, 2p^0} \,Q(z,\bz) \,\sum_{\gamma=\pm}\left(\hat b^\dagger_\gamma(\vec p) \hat b_\gamma(\vec p)-\hat d^\dagger_\gamma(\vec p) \hat d_\gamma(\vec p)\right)\,,
\eadat
\end{equation}
with $Q(z,\bz)$ again given by \eqref{eq:Qhard}. Since the spin part of the Lorentz generator  commutes with $\hat Q(z,\bz)$, we can automatically see that the generators \eqref{eq:Poin_gen1_s1/2}, \eqref{eq:charge_dirac_lifted} close the algebra \eqref{eq:AGS_hard}.

\subsubsection*{Spin-one field}

Consider now a real massless spin-one field
\begin{equation}
    A_\mu(X) = e \sum_{\alpha=\pm} \int \frac{\D^3 p}{(2\pi)^3\, 2p^0} \Big[\varepsilon_\mu^{*\alpha}(\vec p)\, \hat a_\alpha(\vec p) \, e^{i p \cdot X} + \varepsilon_\mu^{\alpha}(\vec p)\, \hat a_\alpha(\vec p)^\dagger \, e^{-i p \cdot X}\Big] \,,\label{eq:spin-one}
\end{equation}
where $\alpha=\pm$ are the two helicities, $e$ denotes the absolute value of the electron charge and the ladder operators obey  $\left[ \hat a_\alpha(\vec p),\hat a_\beta(\vec p\,')^\dagger \right]  =(2\pi)^3\,2p^0\,\delta_{\alpha \beta}\delta^{(3)}(\vec p-\vec p\,')$.
The Poincaré generators are given by
\begin{equation}
\badat{2}
\label{eq:Poin_gen1_s1}
&\hat P^\mu=\int \frac{\D^3 p}{(2\pi)^3\, 2p^0} \,p^\mu \sum_{\alpha=\pm}\hat a^\dagger_\alpha(\vec p) \hat a_\alpha(\vec p)\,,\\
&\hat J^{\mu\nu}=\int \frac{\D^3 p}{(2\pi)^3\, 2p^0} \,\sum_{\alpha,\beta}\hat a^\dagger_\alpha(\vec p)\bigg(i\big(p^\mu \frac{\partial}{\p p_\nu}-p^\nu \frac{\partial}{\p p_\mu}\big)\delta_{\alpha \beta}+ S^{\mu\nu}_{\alpha \beta}(\vec p)\bigg) \hat  a_\beta(\vec p)\,,
\eadat
\end{equation}
with $S^{\mu\nu}_{\alpha\beta}$ the spin generator.

Since the field has no global $U(1)$ symmetry, it follows that it carries a hard UIR of the QED asymptotic symmetry group \eqref{eq:AGS} with trivial hard supermomentum $\big(p_\mu, Q(z,\bar z)=0\big)$.
As we shall see later, however, physically relevant, IR finite, states shall receive contributions coming from photons which do not belong to the hard representations.

\subsection{Why hard representations are not enough}
\label{subsec:hard_not_enough}
In the above, we saw that usual Poincaré particles are in 1:1 correspondence with the hard UIRs of the asymptotic symmetry group of QED.
However, the issue with hard representations is that they \emph{cannot} preserve supermomentum.

To see this, let us consider a scattering involving a family $(p^{\mu}_i, q_i)$ of momenta and electric charges satisfying momentum and charge conservation
\begin{align}
    \sum_i \eta_i p^{\mu}_i&=0, & \sum_i \eta_i q_i&=0\,,
\end{align}
where $\eta=+1$ ($\eta=-1$) if the particle is incoming (outgoing). Denoting by $\Big( p^{\mu}_i, Q_i(z,\bz)\Big)$ their corresponding hard supermomenta (as discussed in Table \ref{tbl:Dic}), then, as we show below,
\begin{equation}\label{Hard reps dont conserve supermomentum}
    \sum_i \eta_i \Big( p^{\mu}_i, Q_i(z,\bz)\Big) = \left( 0^{\mu} , \partial_z \partial_{\bz} \S \right)\,.
\end{equation}
The quantity $\S(z,\bar z) \in \E[0]$ is always non-zero and therefore constitutes the \emph{obstruction to supermomentum conservation}. It is given  explicitly by
\begin{equation}\label{QED: soft factor}
    \S(z,\bz) = \frac{1}{2\pi}\sum_i \eta_i\, q_i\ln |p_i \cdot q(z,\bz) |.
\end{equation}

The proof follows from charge conservation, together with the use of the following distributional identity on the celestial sphere, proved in Appendix \ref{app:distrib_id} and valid for any hard supermomentum $(p^{\mu}, Q(z,\bz))$ of electric charge $\q$:
\begin{align}\label{QED: non linearity and distributional identity}
Q(z,\bz)&=\frac{q_e}{2\pi}\p_\bz\p_z \ln|p\cdot q(z,\bz)|+q_e\,\delta^{(2)}(z-\infty,\bz-\infty)\,.
\end{align}
The above identity is meant as a \emph{global} identity, $Q= \eth\bar{\eth}A + \delta$ relating conformally weighted distributions on the celestial sphere $Q\in \E[-2]$, $A\in \E[0]$. In other terms, a more precise -- though perhaps more cumbersome -- formulation of the identity \eqref{QED: non linearity and distributional identity}  would be\footnote{Note that this expression is coherent with $\hat{Q}(\hat{z},\hat{\bz})=\frac{q_e}{2\pi}\hat{\p}_{\hat{\bz}}\hat{\p}_{\hat{z}}\ln|p\cdot \hat{q}(\hat{z},\hat{\bz})|$. Also note that the absence of `hat' on $q$ in the second expression comes as result of the definition of $A\in \E[0]$ (which is implicitly making use of the chart $(z,\bz)$). Both, directly related, facts that 1) $z=\infty$ plays a special  role in \eqref{QED: non linearity and distributional identity} and that  2) the definition of $A$ here depends on the chart $(z,\bz)$, point to the fact that this way of writing the hard supermomentum breaks Lorentz invariance. However this is \emph{not} the case of \eqref{QED: soft factor} which is a perfectly well-defined, Lorentz-invariant, global weight-zero density on the sphere; in particular $\hat{\S}(\hat{z},\hat{\bz}) = \frac{1}{2\pi}\sum_i \eta_i\, q_i\ln |p_i \cdot \hat{q}(\hat{z},\hat{\bz})|$. This being said, one needs to keeps in mind that \eqref{QED: soft factor} is not necessarily smooth: if the $i$-th momentum is massless, $p_i^\mu = \omega_i q^\mu(\zeta_i,\bar{\zeta}_i)$, $\S(z,\bz)$ develops a singularity at $z=\zeta_i$.}
\begin{equation}
    \begin{cases}
        Q(z,\bz)=\frac{q_e}{2\pi}\p_\bz\p_z  \ln|p\cdot q(z,\bz)| &\text{in the chart} \quad (z,\bz), \\
        \hat{Q}(\hat{z},\hat \bz)=\frac{q_e}{2\pi}\hat{\p}_{\hat \bz}\hat{\p}_{\hat{z}} \ln
        |p\cdot q(\tfrac{1}{\hat{z}},\tfrac{1}{\hat \bz})|+q_e\,\delta^{(2)}(\hat{z},\hat \bz) &\text{in the chart} \quad (\hat{z},\hat \bz)= (z^{-1},\bz^{-1})\,. 
    \end{cases}
\end{equation}
Equivalently, the distributional identity \eqref{QED: non linearity and distributional identity} means that for any (smooth, globally-defined) weight-zero conformal densities $\varepsilon(z,\bar z)$,
\begin{align}\label{eq:distrib1}
\int d^2z \,\varepsilon(z,\bar z)  Q(z,\bz)&=\frac{q_e}{2\pi} \int d^2 z \; \p_\bz\p_z \varepsilon(z,\bar z)\, \ln|p\cdot q(z,\bz)|+q_e\,\varepsilon(\infty,\infty)\,.
\end{align}

Therefore, what the above shows is that, for a scattering involving only hard UIRs, momentum and electric charge conservation does \emph{not} imply supermomentum conservation $ \sum_i \eta_i \big( p^{\mu}_i, Q_i(z,\bz)\big)=0$. As we will see in Section \ref{section: QED soft theorems}, this statement, when applied to conventional Fock states, directly relates to Weinberg's soft photon theorem, with the soft factor being given by the Lorentz-invariant quantity $\S\in \E[0]$. This justifies the need to go beyond hard UIRs and instead consider more general representations that can accommodate supermomentum conservation.

\section{Generic representations}
\label{sec:generic_QED}
We now turn to generic representations of the QED asymptotic symmetry group $SO(3,1) \ltimes \left( \mathbb{R}^{3,1} \times \E[0] \right)$. We highlight below those aspects of the representation theory for QED that will be relevant for our subsequent analysis. We refer the reader to \cite{Bekaert2026} for the detailed classification of UIRs and to Appendix \ref{Appendix: section representations} for general considerations. 

\subsection{QED supermomentum}
On general ground (see Appendix \ref{Appendix: section representations}), the starting point for constructing induced representations of $SO(3,1) \ltimes \left( \mathbb{R}^{3,1} \times \E[0] \right)$ is to consider elements $\P$ of the dual space
\begin{equation}
   \P \in \left( \mathbb{R}^{3,1} \oplus \E[0] \right)^*\,,
\end{equation}
which will be referred to as \emph{supermomenta}. Supermomenta will be written as pairs\footnote{The second element should really be thought of as a distribution. Notice also the notation $\mathcal Q$, as opposed to the previous one, $Q$, meant to denote only hard representations.} 
\begin{equation}
\P = \big( p_{\mu}, \Q(z,\bz)\big)  \in (\mathbb{R}^{3,1})^* \times \E[-2]\,.
\end{equation} By definition, they are dual to the symmetry parameters $\T := \big(T^{\mu}, \varepsilon(z,\bz) \big) \in  \mathbb{R}^{3,1} \times \E[0]$ with duality pairing given by
\begin{equation}
    \langle \P , \T \rangle := p_{\mu} T^{\mu} + \int d^2z\, \Q(z,\bz) \varepsilon(z,\bz)\,.
\end{equation}
An important property of supermomenta is that they can always be projected, in an invariant manner, on pairs $(p_{\mu},\q)$ of a usual momentum and electric charge via
\begin{equation}\label{QED: projection from supermomentum to charge}
\pi: \quad\left|
    \begin{array}{ccc}
    \mathbb{R}^{3,1} \times \E[-2] &  \to &\mathbb{R}^{3,1} \times \mathbb{R}\\[0.5em]
        \big( p_{\mu}, \Q(z,\bz)\big) &  \mapsto &\Big( p_{\mu}, \q= \int d^2z \,\Q(z,\bz)\Big)
    \end{array}\right..
\end{equation}

\subsection{Asymptotic little group}
The asymptotic little group $\ell_{\P} \subseteq SL(2,\mathbb{C})$ of a representation of supermomentum $\P$ is the subgroup of Lorentz whose elements $B\in \ell_{\P}$ stabilize the supermomentum
\begin{equation}
    \Big( B \cdot p_{\mu}, B\cdot \Q(z,\bz)\Big)= \Big( p_{\mu}, \Q(z,\bz)\Big).
\end{equation}
By construction, since $B$ must in particular always stabilize the momentum $p^{\mu}$, the asymptotic little group is always a subgroup of the Poincaré little group $\ell_p$,
\begin{equation}
    \ell_{\P} \subset \ell_p= \begin{cases}
        SU(2) \\ ISO(2) \\
        SL(2,\mathbb{R})\\
        SL(2,\mathbb{C})
    \end{cases}
\end{equation}
where $SU(2)$, $ISO(2)$, $SL(2,\mathbb{R})$ are the Poincaré little groups stabilizing, respectively, timelike (massive), null (massless), and spacelike (tachyonic) four-momenta, while $SL(2,\mathbb{C})$ corresponds to the special case where $p^{\mu}=0$.

\subsection{Hard representations}
We already encountered hard representations in Section \ref{sec:hard_reps}, which are summarized in Table \ref{tbl:Dic}. From the perspective of the general theory, they are very special: their asymptotic little group $\ell^{hard}_{\P}$ has maximal dimension\footnote{Leaving aside the degenerate case $\ell_{\P}^{hard} = SL(2,\mathbb{C})$ corresponding to $\P=(0,0)$.}
\begin{equation}\label{eq:samelittle}
    \ell_{\P}^{hard} = \ell_p= \begin{cases}
        SU(2) \\ ISO(2) \\
        SL(2,\mathbb{R})
    \end{cases}.
\end{equation}
This special property singles out the hard representations from other typical representations and places them in one-to-one correspondence with Poincaré UIRs. In fact, the property \eqref{eq:samelittle} can almost be taken as a definition for hard representations; see \cite{Bekaert2026} for more details.

\subsection{Soft representations}
Another type of representations which are quite special, but less so than the hard ones, are the ``soft'' representations\footnote{This terminology of ``hard'' and ``soft'' is chosen to align as closely as possible with the existing literature (e.g., \cite{Strominger:2017zoo} and references therein). In QFT, the designation ``soft'' is determined by the relevant energy scale; in the present context, it is meant in the sense specified in \eqref{eq:QED_soft}.}. A representation is called soft if both its momentum and electric charge are zero, i.e. 
\begin{equation}\label{eq:QED_soft}
    \pi(\P) = \left(p_{\mu} , \q = \int d^2z \,\Q(z,\bz) \right) = \left( 0_{\mu},0 \right).
\end{equation}
One can show (see \cite{Bekaert2026}) that soft supermomenta must be of the form $\P = \left( 0^{\mu} , \partial_z \partial_{\bz} \N \right)$
for some function $\mathcal N(z,\bz) \in \E[0]$ on the sphere. In other terms,
\begin{equation}\label{QED: Equivalence between soft of image of eth}
    \P \;\;\text{is soft} \quad  \Leftrightarrow\qquad p_{\mu}=0 \quad \text{and} \quad\exists \mathcal N \in \E[0] \quad \text{s.t.}\quad \Q =  \eth\bar{\eth} \mathcal N.
\end{equation}
Here, the conformal Laplacian $\eth\bar{\eth}$ is the primary operator which, in a chart and for the flat metric, simply reads
\begin{equation}
    \eth \bar{\eth} \left| \begin{array}{ccc}
    \E[0] &  \to & \E[-2]\\
        f &  \mapsto & \partial_z \partial_{\bz}f
    \end{array}\right. .
\end{equation}
To prevent any later confusion\footnote{\label{footnote:distrib}For example, there is inherently nothing wrong with writing expressions such as $\delta^{(2)}(z-w) = \frac{1}{2\pi}\partial_z \partial_{\bz} \ln|z-w|^2$, as long as one is aware that, in this expression, $\ln |z-w|^2$ does not stand for a globally well-defined distribution on $S^2$ and that, as a closely related fact, integration by part will produce a boundary term. See the discussion below equation \eqref{QED: non linearity and distributional identity} and Appendix \ref{app: delta function} for more details.  As a result, this notation does not imply that $\delta^{(2)}(z-w)$ is soft: $1=\int \delta^{(2)}(z-w) \neq0$, as it should.}, it is essential to emphasize that, in the equivalence \eqref{QED: Equivalence between soft of image of eth}, we really mean $\Q = \eth\bar{\eth} \mathcal N$, with $\Q \in \E[-2]$, $\mathcal N\in \E[0]$ as a \emph{global} statement on the sphere: i.e. not only $\Q(z,\bz) = \partial_z\partial_{\bz}\mathcal N(z,\bz)$ in the north patch coordinates, but also $\hat{\Q}(\hat{z},\hat \bz) = \partial_{\hat{z}}\partial_{\hat \bz}\hat{\mathcal N}(\hat{z},\bar{\hat{z}})$ in the south patch coordinates $\hat{z}=z^{-1}$, see Appendix \ref{app:global conformal densities}   (and Appendix \ref{app: delta function} for a discussion on why the delta function is not soft). In particular, under these assumptions and as a result of the fact that the equality holds globally on the sphere, integration by parts always holds without any further concerns about boundary terms\footnote{Thus, if $\mathcal N_1 \in \E[0]$ and $\mathcal N_2\in\E[0]$, it is always true that $\int \mathcal N_1 \eth \bar{\eth} \mathcal N_2 = -\int \eth \mathcal N_{1} \bar{\eth} \mathcal N_2  = \int \bar{\eth} \eth\mathcal N_1  \mathcal N_2$.}.

An important property of the space of soft supermomenta is that it forms a Hilbert space, with the norm $\|\mathcal N\|^2$ induced by the scalar product
\begin{align}\label{eq: Lorentz invariant norm QED}
    \langle \partial \mathcal N_1, \bar \partial \mathcal N_2  \rangle &:= \int d^2z \,\partial_z \mathcal N_1 \partial_{\bz} \mathcal N_2\,.
\end{align}
An alternative, useful, form for this Lorentz-invariant norm is in term of the following two-point function (see section 6.4. in \cite{Gelfand2})
\begin{equation}
    \langle \partial \mathcal N_1, \bar \partial \mathcal N_2  \rangle = \int d^2z_1 \int d^2z_2 \, \partial_{z_1} \partial_{\bz_1}\mathcal N_1\, \partial_{z_2} \partial_{\bz_2} \mathcal N_2 \;\ln|q(z_1,\bz_1)\cdot q(z_2,\bz_2)|\,.
\end{equation}
Although this expression may appear singular and non-invariant at first sight, it is in fact well defined and invariant; see Appendix \ref{Ssection: apdx 2-point function E[0]} for a discussion. 

\subsection{Hard/soft decomposition of supermomenta}\label{section: Hard/soft decomposition of supermomenta QED}

\subsubsection{Definition}

Let $\P = (p_{\mu}, \Q(z,\bz))$ be a generic supermomentum of momentum $p_{\mu}$ and electric charge $\q$. In general, it will neither be hard (i.e it will be not contained in Table \ref{tbl:Dic}) nor soft (i.e $(p_{\mu}, \q)\neq 0 $). However, as we shall see, one can always uniquely decompose it in the form
\begin{equation}\label{QED: hard/soft decomposition}
    \P = \Big( p_{\mu}, Q(z,\bz)\Big) +\Big( 0_{\mu}, \partial_z\partial_{\bz}\mathcal N(z,\bz)\Big)\,,
\end{equation}
with $\mathcal N\in \E[0]$ and  $\big(p_{\mu}, Q(z,\bz)\big)$ the hard supermomentum of momentum $p_{\mu}$ and electric charge $\q$.  This decomposition is unique and Lorentz-invariant. With the tools developed above, the proof should appear relatively straightforward:

A supermomentum uniquely defines, by the projection \eqref{QED: projection from supermomentum to charge}, a momentum $p_{\mu}$ and an electric charge $\q$. In turn, these uniquely define, via Table \ref{tbl:Dic}, a hard supermomentum $\big( p_{\mu}, Q(z,\bz)\big)$. Now, since the projection \eqref{QED: projection from supermomentum to charge} is linear, this means that
\begin{equation}
    \pi \Big(  \big(p_{\mu}, \Q(z,\bz)\big) - \big( p_{\mu}, Q(z,\bz)\big) \Big) = \left( p_{\mu} , \q\right)- \left( p_{\mu} , \q\right) =\left( 0_{\mu} , 0\right)\,,
\end{equation}
and thus $\P - \big( p_{\mu}, Q(z,\bz)\big)$ is soft. Finally, due to the equivalence \eqref{QED: Equivalence between soft of image of eth}, this means that there exists $\mathcal N \in \E[0]$ such that $\P - \big( p_{\mu}, Q(z,\bz)\big) = \big( 0_{\mu}, \partial_z\partial_{\bz}\mathcal N(z,\bz)\big)$.

It is perhaps worth stressing that the decomposition \eqref{QED: hard/soft decomposition} presented here has essentially nothing to do with a spherical harmonics decomposition, $\P = \Big( p_{\mu}, \Q(z,\bz)\big|_{\ell =0}\Big) +\Big( 0_{\mu}, \Q(z,\bz)\big|_{1\leq\ell}\Big)$. The decomposition \eqref{QED: hard/soft decomposition} is Lorentz-invariant and nonlinear, whereas the spherical harmonics decomposition is linear but not Lorentz-invariant. In this sense, the two decompositions are as far apart as possible.

\subsubsection{Non-linearity of the decomposition}
As already mentioned, an essential feature of the decomposition \eqref{QED: hard/soft decomposition} is that it is nonlinear. If $\P_1$ and $\P_2$ are two supermomenta such that
\begin{align}
\nonumber
        \P_1 = \Big( p^{\mu}_1, Q_1(z,\bz)\Big) +\Big( 0^{\mu}, \partial_z\partial_{\bz}\mathcal N_1(z,\bz)\Big) \virg   \P_2 = \Big( p^{\mu}_2, Q_2(z,\bz)\Big) +\Big( 0_{\mu}, \partial_z\partial_{\bz}\mathcal N_2(z,\bz)\Big)\,,
\end{align}
then
\begin{equation}
    \P_1 +\P_2 = \Big( p_3^{\mu}, Q_3(z,\bz)\Big) +\Big( 0^{\mu}, \partial_z\partial_{\bz} \left( \mathcal N_1(z,\bz)  + \mathcal N_{2}(z,\bz) + \S(z,\bz) \right) \Big)\,,
\end{equation}
with $\big( p_3^{\mu}, Q_3(z,\bz)\big)$ the hard supermomentum of momentum $p^{\mu}_3 = p^{\mu}_1 + p^{\mu}_2$ and electric charge $q_3= q_1+ q_2$, and
\begin{equation}
 \S(z,\bz) =   \frac{q_1}{2\pi}\ln|p_1\cdot q(z,\bz)| + \frac{q_2}{2\pi}\ln|p_2\cdot q(z,\bz)| - \frac{q_3}{2\pi}\ln|p_3\cdot q(z,\bz)|\,.
\end{equation}
Note that this phenomenon directly arises as a consequence of \eqref{Hard reps dont conserve supermomentum} and is directly responsible for the fact that hard UIRs cannot conserve supermomentum.

\subsection{Wavefunctions and supermomentum eigenstates}
\label{sec:supermomentum eigenstates QED}
Let $|\P\rangle$ be a supermomentum eigenstate, $\hat{P}^{\mu}|\P\rangle = p^{\mu}|\P\rangle$, $\hat{\Q}(z,\bz)|\P\rangle = \Q(z,\bz)|\P\rangle$. Since the decomposition \eqref{QED: hard/soft decomposition} is unique, we are not loosing any information by rewriting this state as $|\P\rangle = |p , \partial_z\mathcal N\rangle$. By definition, it satisfies
\begin{align}
    \hat{P}^{\mu}|p , \partial_z\mathcal N\rangle &= p^{\mu}|p , \partial_z\mathcal N\rangle,&  \hat{\Q}(z,\bz)|p , \partial_z\mathcal N\rangle = \left(Q(z,\bz) +\partial_{\bz}\partial_z\mathcal N \right)|p , \partial_z\mathcal N\rangle\,.
\end{align}
Just like momentum eigenstates, supermomentum eigenstates are singular states which do not have a finite norm. The one-particle Hilbert space realizing the UIR is given be the space of normalizable states of the form\footnote{For simplicity we restrict ourselves to scalar states, see \cite{Bekaert2026} for more details.}
\begin{equation}\label{New state for QED}
    |\Psi\rangle = \int_{SL(2,\mathbb{C})/\ell_{\P}} \frac{d M}{\text{Vol}(\ell_{\P})} \Psi(M)|M\cdot p , M\cdot \partial_z\mathcal N\rangle\,,
\end{equation}
where $M\in SL(2,\mathbb{C})$, $dM$ is the Haar measure and the wavefunction $\Psi(M)$ must satisfy $\Psi(MB) = \Psi(M)$ for any $B\in \ell_{\P}$. The norm of the state is simply
\begin{equation}
    \langle \Psi|\Psi\rangle = \int_{SL(2,\mathbb{C})/\ell_{\P}} \frac{d M}{\text{Vol}(\ell_{\P})} |\Psi(M)|^2\,.
\end{equation}

For hard representations, this reproduces the usual scalar wavefunction $\Psi(M)=\Psi(k)$, where we introduced $k^{\mu} = M^{\mu}{}_{\nu}p^{\nu}$. These are functions on
\begin{align}
   \frac{SL(2,\mathbb{C})}{\ell_{p}}=\begin{cases}
        H_3 &= \frac{SL(2,\mathbb{C})}{SU(2)}, \qquad\text{if}\; p^2<0\\  \mathbb{R}\times S^2 &= \frac{SL(2,\mathbb{C})}{ISO(2)}, \qquad\text{if}\; p^2=0
   \end{cases}.
\end{align}
However, for generic representations, the asymptotic little group $\ell_{\P}$ is strictly smaller than the Poincaré little group $\ell_{p}$ and the wavefunctions $\Psi(M)$ are functions on the total space of a fibre bundle
\begin{equation}
    \frac{SL(2,\mathbb{C})}{\ell_{\P}} \to \frac{SL(2,\mathbb{C})}{\ell_{p}}
\end{equation}
whose typical fibre $\ell_{\P}/\ell_{p}$ encodes the extra degrees of freedom of the representation (as compare to the corresponding hard one). Note that, while they have more degrees of freedom than the usual hard wavefunctions, they belong to a separable Hilbert space (because they are $L^2$ functions on a finite-dimensional manifold). In that sense, the Hilbert space to which these new ``particles'' belong is mathematically just as good as the usual one.

\paragraph{Asymptotic particles}
As we previously argued, if the asymptotic symmetry group \eqref{eq:AGS} is a symmetry of the $S$-matrix of QED, then hard representations cannot be the end of the story since they cannot conserve supermomentum (see the discussion in section  \ref{subsec:hard_not_enough}). Since, as we shall review in the next section, Weinberg's soft theorem for QED is in fact equivalent to the conservation of supermomentum, it strongly suggests\footnote{The alternative would be that the symmetry is broken is a rather non-standard way, likely implying along the way that the Lorentz group itself is broken; see the discussion at the end.} that the Hilbert space of asymptotic states must be extended to include new representations, including states of the form \eqref{New state for QED}. In fact, the idea of extending the Hilbert space of QED is not new: it is well known that it \emph{must} be extended in order to have IR-finite $S$-matrix elements~\cite{Chung:1965zza,Kibble:1968sfb,KibbleII,KibbleIII,KibbleIV,Kulish:1970ut}. Rather, the point that we want to make here is that the representation theory of the asymptotic symmetry group provides an appropriate and natural framework to realize this extension as a Fock space built from a new notion of particles (which, as already emphasized, extends beyond hard UIRs). Since there are infinitely many representations of the asymptotic symmetry group -- as many as $(\mathbb{R}^{3,1}\times \E[-2])/SL(2,\mathbb{C})$, with the hard ones forming only a tiny special corner -- additional physical input is required to avoid getting lost in this vast landscape. We now turn to IR physics, which should provide precisely such input.

\section{Infrared divergences in hard scattering processes}
\label{sec: IR div QED}
The upshot of Section \ref{sec:hard_reps} was the following: while every usual QFT particle can be thought of as a hard UIR of the asymptotic symmetry group of QED, hard representations alone \emph{cannot} by themselves conserve supermomentum. We will in fact now see that the quantity $\mathcal S(z,\bar z)$ given in \eqref{QED: soft factor} that characterizes the obstruction for supermomentum conservation, directly relates both to real and virtual infrared divergences for the scattering of conventional particles: $\mathcal S(z,\bar z)$ gives the soft factor of Weinberg's theorem, while its Lorentz-invariant norm controls the exponentiated virtual divergences. 

\subsection{Photons beyond the hard Fock space}
As we already saw in Section \ref{subsec:hard_QED_spinning}, since they are electrically neutral, the usual Fock space of photons is such that they have a rather trivial hard supermomentum $(p^\mu,Q(z,\bz))=(\omega q^{\mu}(\zeta,\bar{\zeta}),0)$. However, photons can also carry a non-zero soft supermomentum $(0^\mu,\partial_z\partial_{\bz} \mathcal N(z,\bz))$. Indeed, the Noether analysis for abelian gauge transformations that approach an arbitrary function $\varepsilon(z,\bz)$ on the celestial sphere leads to a non-vanishing Noether charge~\cite{He:2014cra,Strominger:2017zoo}. For a gauge field $A_\mu$ expanded in modes as in \eqref{eq:spin-one}, the charge is given by $\mathcal F_\varepsilon=\int d^2 z \,\varepsilon(z,\bz)\, \partial_z\partial_{\bz} \hat{\mathcal N}^{\text{ph}}(z,\bz)$, with\footnote{Note that $\lim_{\omega \to 0} \omega\,\left( \hat a_-(\omega,z,\bz)^{\dagger} + \hat a_+(\omega,z,\bz)  \right)$ has conformal weight $w=-1$ and spin weight $s=-1$. Since $\eth : (w=0,s=0) \to (w=-1,s=-1)$ is surjective (see \cite{Penrose:1985bww} and our conventions in Appendix \ref{app:conformal_density}, \ref{app: eth operator}), this can always be written as $\eth\hat{\mathcal N}$ with $\hat{\mathcal N} \in \mathcal E[0]$.
Since the Noether charge is real by construction, $\hat{\mathcal N}$ is Hermitian and thus 
\begin{equation}\label{eq:Qph alt}
    \partial_z\partial_{\bz} \hat{\mathcal N}^{\text{ph}}(z,\bz)=  \frac{1}{4\pi e} \p_z \left( \,\lim_{\omega \to 0} \omega\,\left( \hat a_+(\omega,z,\bz)^{\dagger} + \hat a_-(\omega,z,\bz)  \right)\,\right).
\end{equation}
Imposing the corresponding reality condition is equivalent to assuming that the soft magnetic charge
\begin{equation}\label{eq:Qph magnetic}
8\pi i e \hat{\mathcal Q}^M(z,\bz) := \p_\bz \left( \,\lim_{\omega \to 0} \omega\,\left( \hat a_-(\omega,z,\bz)^{\dagger} + \hat a_+(\omega,z,\bz)  \right)\,\right) - \p_z \left( \,\lim_{\omega \to 0} \omega\,\left( \hat a_+(\omega,z,\bz)^{\dagger} + \hat a_-(\omega,z,\bz)  \right)\right)
\end{equation} vanishes. In practice, it seems enough for us to suppose that $\hat{\mathcal Q}^M(z,\bz) |\psi \rangle=0$ on all our states.}
\begin{equation}\label{eq:Qph}
    \partial_z\partial_{\bz} \hat{\mathcal N}^{\text{ph}}(z,\bz)=  \frac{1}{4\pi e} \p_\bz \left( \,\lim_{\omega \to 0} \omega\,\left( \hat a_-(\omega,z,\bz)^{\dagger} + \hat a_+(\omega,z,\bz)  \right)\,\right)\,.
\end{equation}
And indeed, together with the operators \eqref{eq:Poin_gen1_s1}, this forms a basis
\begin{equation}
\left( \hat{P}^{\mu}, \hat{J}^{\mu\nu}, \hat{\Q}^{\text{ph}}(z,\bz):=\partial_z\partial_{\bz} \hat{\mathcal N}^{\text{ph}}(z,\bz)\right)
\end{equation}
of generators of the asymptotic symmetry algebra \eqref{eq:AGS_hard}. 

However, in order for the asymptotic symmetry to be non-trivially represented, one needs to make a radical move. To see this, let be $|\psi\rangle := \hat{\psi} |0 \rangle$ be a photon state, with
\begin{equation}\label{Photons are different: normalizable state}
\badat{2}
  \hat{\psi} &= 
   e  \int \frac{d^3k}{(2\pi)^3 2k^0} \,\left( \psi_-(k) \hat{a}^{\dagger}_{+}(k) + \psi_+(k) \hat{a}^{\dagger}_{-}(k)  \right)\\
   &=  \frac{e}{16 \pi^3} \int \omega d\omega d^2z \,\left( \psi_-(\omega,z,\bz) \hat{a}^{\dagger}_{+}(\omega,z,\bz) + \psi_+(\omega,z,\bz) \hat{a}^{\dagger}_{-}(\omega,z,\bz)  \right).
  \eadat
\end{equation}
 Then one has,
 \begin{equation}\label{Photons are different: action of soft charge}
     \partial_z\partial_{\bz} \hat{\mathcal N}^{\text{ph}}(z,\bz)|\psi\rangle = \left[ \partial_z\partial_{\bz} \hat{\mathcal N}^{\text{ph}}(z,\bz) , \hat{\psi} \right] | 0\rangle + \hat{\psi} \, \partial_z\partial_{\bz} \hat{\mathcal N}^{\text{ph}}(z,\bz)|0\rangle\,,
 \end{equation}
 where the first contribution only comes from the pole of the wavefunction\footnote{This can be derived by making use of \eqref{eq:Qph}, \eqref{Photons are different: normalizable state} and \eqref{Canonical commutation relation in omega,z}. Note that the requirement that the state has no magnetic charge:   \begin{equation}
  0= \hat{\Q}^{M}(z,\bz) |\psi\rangle=  \left[ \hat{\Q}^{M}(z,\bz) , \hat{\psi} \right]  | 0 \rangle  = \frac{1}{8\pi ie} \lim\limits_{\omega\to 0}\omega\Big( \partial_{\bz} \psi_-(\omega,z,\bz) - \partial_{z} \psi_+(\omega,z,\bz) \Big)  \,  | 0 \rangle
\end{equation} means that $\lim\limits_{\omega\to 0}\omega \partial_{z} \psi_+(\omega,z,\bz) = \lim\limits_{\omega\to 0}\omega \partial_{\bz} \psi_-(\omega,z,\bz) $.}
 \begin{equation}
   \left[ \partial_z\partial_{\bz} \hat{\mathcal N}^{\text{ph}}(z,\bz) , \hat{\psi} \right]  | 0 \rangle  = \frac{1}{4\pi} \lim\limits_{\omega\to 0}\omega \partial_{\bz} \psi_-(\omega,z,\bz)  \;  | 0 \rangle.
\end{equation}
However, for any state of finite norm,
\begin{align}\label{Photons are different: finite norm of hard state}
\langle 0 | \hat{\psi}^{\dagger} \hat{\psi} | 0 \rangle 
&= \frac{e^2}{16 \pi^3}  \int \omega d\omega d^2z\,\left(|\psi_{+}(\omega,z,\bz)|^2+ |\psi_{-}(\omega,z,\bz)|^2 \right) < \infty,
\end{align}
the first term in \eqref{Photons are different: action of soft charge} must in fact be zero, while the second term can be non-zero only if some sort of ``spontaneous symmetry breaking''\footnote{This wording should be taken with caution in this context, as the degeneracy of the vacuum does not necessarily imply that the $S$-matrix factorizes into ``superselection sectors''; see \cite{Strominger:2017zoo}  and the discussion at the end.} occurs (since in that case the symmetry generator acts non-trivially on the vacuum). Therefore, the asymptotic symmetry algebra can be represented
non-trivially only in one of two ways: i) by enlarging the Hilbert space (as compared to the
Fock space of hard particle representations), or ii) by invoking spontaneous symmetry breaking.
Both perspectives have appeared throughout the literature, often in alternation and not always
clearly distinguished. The latter viewpoint underpins the discussion of asymptotic symmetries
in the context of soft theorems and relates to infrared divergences of the $S$-matrix \cite{He:2014cra,Strominger:2017zoo}, while the first scenario is essentially the core of the infrared-finite Faddeev-Kulish construction~\cite{Chung:1965zza,Kibble:1968sfb,KibbleII,KibbleIII,KibbleIV,Kulish:1970ut,Gabai:2016kuf,Kapec:2017tkm}.

\paragraph{Relationship with the memory effect in electromagnetism}  

If a classical electromagnetic field $A_{\mu}(\omega,z,\bz)$ passes on a test charge at rest, it will in general induce a memory effect \cite{Bieri:2013hqa}: the test field will experience a velocity kick. This effect can be directly related to the pole in $\omega$ in $A_{\mu}(\omega,z,\bz)$ \cite{Pasterski:2015zua}, i.e. precisely to the contribution that would be given by the first term in \eqref{Photons are different: action of soft charge}. However, while this is a perfectly reasonable classical effect, it cannot be accommodated by the usual hard state, due to \eqref{Photons are different: finite norm of hard state}. This problem, namely the impossibility for the quantum Hilbert space to realize a perfectly legitimate classical observable\footnote{Which was much later understood to be a memory effect.}, was pointed out very early on by Ashtekar \cite{Ashtekar:1987tt} as the source of IR divergences. In this sense, the possibility of implementing the asymptotic symmetries \eqref{eq:AGS} on the Hilbert space directly relates to the possibility of encoding memory at quantum level and to IR divergences. 

\subsection{Soft photon theorem as supermomentum conservation}\label{section: QED soft theorems}
Let us consider a scattering process  $\langle \text{out} | \hat{S} | \text{in} \rangle $ in momentum basis involving $n$ incoming and $m$ outgoing particles.
Supermomentum-invariance of the $S$-matrix is equivalent to the usual conservation of momentum, together with
\begin{equation}\label{eqref:WI_qed}
    \langle \text{out} |\hat{\Q}^{\text{out}}(z,\bz)\hat{S}-\hat{S} \hat{\Q}^{\text{in}}(z,\bz) | \text{in} \rangle = 0\,,
\end{equation}
with $\hat{\Q}(z,\bz)$ the total (incoming or outgoing) supercharge generator.

Let us focus on the case of charged scalar particles \eqref{complex_scalar} of momentum $p_i$ and electric charge $q_i$ (the case of spinning particles follows in a similar way). Conventional asymptotic states belong to the Fock space associated with the free field operators $b(\vec p)$ and $d(\vec p)$.
As we have seen in Section~\ref{sec:hard_reps}, charged scalars have a non-zero hard supercharge operator given by
\begin{equation}\label{eq:Qin0}
    \hat{Q}(z,\bar z) = \int  \frac{d^3p}{(2\pi)^32p^0}\, Q(z,\bz)\left(\hat b^\dagger(\vec p) \hat b(\vec p)-\hat d^\dagger(\vec p) \hat d(\vec p)\right) \,,
\end{equation}
with $Q(z,\bz)$ given by \eqref{eq:Qhard}.
The incoming hard operator acts on incoming states $| \text{in} \rangle =|p_1^\text{in},q_1^\text{in}; \dots; p_n^\text{in},q_n^\text{in} \rangle$ as
\begin{equation}\label{eq:Qin1}
    \hat{Q}^\text{in}(z,\bar z) | \text{in} \rangle=\sum_{i=1}^n Q^\text{in}_i(z,\bz)| \text{in} \rangle\,,
\end{equation}
where $Q_i(z,\bar z)$ is the hard charge of the $i$-th incoming particle, namely
\begin{equation} 
Q_i(z,\bar z) =
\left\{\begin{aligned}
  &\hspace{0.5cm} \quad q_i\,\delta^{(2)}(z-\zeta_i) & \quad \text{if } &p_i^2=0,  \;p^{\mu}_i = \omega_i q^{\mu}_i(\zeta_i,\bar{\zeta}_i)\\[0.6em]
  &\hspace{0.5cm} \dfrac{q_i m_i^2}{4\pi\,(q(z,\bar z)\cdot p_i)^2} & \quad\, \text{\,if } &p_i^2=-m_i^2\,.\\[0.4
  em]
\end{aligned}\right.
\end{equation}

For a scattering event involving charged scalars interacting through an electromagnetic process, the total incoming supercharge operator is therefore the sum of \eqref{eq:Qph} and \eqref{eq:Qin0},
\begin{equation}\label{eq:Qin2}
\badat{2}
    \hat{\Q}^\text{in}(z,\bz) &= \hat{Q}^\text{in}(z,\bar z) + \partial_z \partial_{\bz} \hat{\mathcal N}^{ph}(z,\bz)\\ &= \hat{Q}^\text{in}(z,\bar z)+ \frac{1}{4\pi e} \lim_{\omega \to 0} \omega\,\p_\bz\left( \hat a_-^\text{in}
    (\omega,z,\bz)^{\dagger}+  \hat a_+^\text{in}(\omega,z,\bz) \right) \,,
    \eadat
\end{equation}
where only the electromagnetic field contributes to the second term. A similar expression holds for the total outgoing operator.\\
We now want to use the fact that, for both massive and massless cases, $Q_i(z,\bz)$ can be identically rewritten as
{\setlength{\belowdisplayshortskip}{12pt} 
\begin{equation}\label{eq:Qfact}
Q_i(z,\bz)=\frac{q_i}{2\pi}\p_\bz \left( \frac{\varepsilon^+(z,\bz)\cdot p_i}{q(z,\bz)\cdot p_i}\right)+q_i\delta^{(2)}(z-\infty,\bz-\infty)\,.
\end{equation}}
\par\noindent
The above equality holds in the sense of distributions and is proven in Appendix \ref{app:distrib_id}.
Using this together with \eqref{eq:Qin1}, \eqref{eq:Qin2} and $\displaystyle \lim_{\omega \to 0} \omega\, \langle \text{out} | \hat a_+^{\text{out}}\hat{S}   | \text{in} \rangle = - \displaystyle\lim_{\omega \to 0} \omega\, \langle \text{out} | \hat{S} \hat a_-^{\text{in} \dagger}   | \text{in} \rangle$, the Ward identity \eqref{eqref:WI_qed} becomes 
\begin{equation}\label{eq:eu}
\p_\bz\left(\; \frac{1}{e}\lim_{\omega \to 0} \omega\, \langle \text{out} | \hat a_+^{\text{out}}(\omega,z,
\bz)\hat{S}   | \text{in} \rangle-   \left(\sum_{i\in \text{in}} q_i  \, \frac{\varepsilon^+\cdot p_i}{q\cdot p_i} - \sum_{i\in \text{out}} q_i \, \frac{\varepsilon^+\cdot p_i}{q\cdot p_i} \right)\langle \text{out} | \hat{S} | \text{in} \rangle  \right) =0\,.
\end{equation}
Notice that the distributional terms $q_i\delta^{(2)}(z-\infty,\bz-\infty)$ have canceled out in the total sum by virtue of charge conservation.
The equality \eqref{eq:eu} is \emph{equivalent} to Weinberg's soft photon theorem~\cite{Weinberg:1965nx}
\begin{equation}\label{eq:soft_photon_theorem}
\lim_{\omega \to 0} \omega\, \langle \text{out} | \hat a_+^{\text{out}}(\omega,z,
\bz) \hat{S} | \text{in} \rangle= e\left(\sum_{i\in \text{in}} q_i  \, \frac{\varepsilon^+\cdot p_i}{q\cdot p_i} - \sum_{i\in \text{out}} q_i \, \frac{\varepsilon^+\cdot p_i}{q\cdot p_i} \right)\langle \text{out} | \hat{S} | \text{in} \rangle  \,.
\end{equation}
The equivalence follows from the fact that the $\bar \eth$ operator is \emph{injective} when acting on a density $f$ of spin-weight $s=-1$, hence  $\bar \eth f=0$ implies $f=0$ (see Proposition (4.15.59) in \cite{Penrose:1985bww} and our conventions \ref{app:conformal_density}, \ref{app: eth operator}).

The above offers a revisiting of the (by now classical) argument for the equivalence between the soft photon theorem and the Ward identity associated with the asymptotic symmetries of QED, with the advantage of treating, at the same time, both the massless \cite{He:2014cra} and massive cases \cite{Campiglia:2015qka}. The derivation is also substantially shorter as it directly shows the equivalence (thanks to the property of the edth operator), as opposed to the original two-steps derivations that make use of a convenient choice of gauge parameter. 

The derivation presented here also makes very clear that, from the point of view of representation theory, the soft photon theorem is simply the manifestation of the fact that a scattering process involving $N$ Poincaré particles which satisfy momentum and charge conservation fails to conserve supercharge and hence supermomentum conservation. The soft factor
\begin{equation}\label{eq:soft_factor_ph}
\sum_{i=1}^N  \eta_i\, q_i  \, \frac{\varepsilon^+\cdot p_i}{q\cdot p_i} = 2\pi \p_z \mathcal S(z,\bz) \virg \mathcal S(z,\bz) =\frac{1}{2\pi}\sum_{i=1}^N \eta_i \,q_i \ln |q(z,\bz)\cdot p_i|\,,
\end{equation}
where $\eta=+1$ ($\eta=-1$) if the particle is incoming (outgoing), exactly provides the missing contribution required for the supermomentum conservation law to hold; see equations \eqref{Hard reps dont conserve supermomentum} and \eqref{QED: soft factor}.

\subsection{Exponentiation formula for virtual divergences}
On top of the divergences arising when attaching an external soft photon, conventional scattering amplitudes suffer from IR divergences due to the exchange of virtual soft photons between external legs. In the seminal works~\cite{Yennie:1961ad,Weinberg:1965nx,Weinberg:1995mt}, those IR divergences were shown to  factorize in the amplitude $\mathcal A$ according to the abelian exponentiation theorem\footnote{These divergences are one-loop exact.}
\begin{equation}\label{eq:fact1}
    \mathcal A= e^{\mathcal W} \mathcal A^{I.F.}_\Lambda\,,
\end{equation}
where $\mathcal A^{I.F.}_\Lambda$ is infrared-finite and the exponent is
\begin{equation}
\label{weinfact}
\mathcal W=\log\left(\frac{\Lambda}{\lambda}\right) \frac{1}{8\pi^2}\sum_{i,j} \frac{\eta_i \eta_j\, q_i q_j}{\beta_{ij}} \left(\frac{1}{2}\ln \frac{1+\beta_{ij}}{1-\beta_{ij}} -i\pi \delta_{\eta_i,\eta_j} \right) \,,
\end{equation}
with $\beta_{ij}$ the relative velocity of particle $i$ and $j$ 
\begin{equation}
\label{betaij}
\beta_{ij}\equiv \sqrt{1-\frac{m_i^2 m_j^2}{ (p_i \cdot p_j)^{2}}}\,.
\end{equation}
The scale $\Lambda$ defines what is meant by a ``soft'' particle  (namely one whose momentum is less than $\Lambda$), while the lower scale $\lambda \ll \Lambda$ is introduced as a regulator cutoff.
The imaginary part of $\mathcal W$ is often referred to as the Coulomb phase divergence (it does not affect the decay rate), while the real part is the term that cancels the divergences associated with the emission of soft photons in inclusive cross-sections\footnote{It was recently shown that only the combination of the real and Coulomb phase terms satisfies crossing symmetry \cite{Lippstreu:2025jit}.}.
It was shown in \cite{Nande:2017dba} that the soft factorization theorem for abelian gauge theory \eqref{eq:fact1} is the same as a $U(1)$ Kac-Moody current algebra factorization on the celestial sphere. There, the exponentiated soft divergence was reproduced from expectation values of Wilson line operators, with the current algebra level identified with the cusp anomalous dimension. 

As we will now prove, the real part of the infrared factor is, in fact, for both massive and massless particles, 
\begin{equation}\label{eq:W is the norm QED}
    \Re(\mathcal W)  = - \frac{\log(\Lambda/\lambda) }{8\pi} \,\| \S \|^2\,,
\end{equation}
where $\S(z,\bz)$ is the obstruction to supermomentum conservation, defined in \eqref{eq:soft_factor_ph},
and $\|\S\|^2 = \int d^2z \,\partial_z \S \partial_{\bz} \S$ its Lorentz-invariant norm squared \eqref{eq: Lorentz invariant norm QED}\footnote{For massless particles, it was previously noted in \cite{Agrawal:2025bsy} that the Goldstone two-point function of \cite{Nande:2017dba} relates to the inner product of unitary discrete series representations of the Lorentz group.}.

To see this, we first recall that the real part of \eqref{weinfact} can be written as~\cite{Weinberg:1965nx}
\begin{equation}
\label{we1}
\Re(\mathcal W)= -\frac{1}{4(2\pi)^3} \log\Big(\frac{\Lambda}{\lambda}\Big) \sum_{i,j} \eta_i\eta_j\, q_i q_j  \int d^2 z \,\frac{ (p_i\cdot p_j)}{(p_i\cdot q)(p_j\cdot q)}\,.
\end{equation}
This formula is valid for both massive and massless cases, but the massless case is plagued with further divergences, as we recall below.
To show that the norm of $\mathcal S$ coincides with \eqref{we1}, we write 
\begin{equation}
    \badat{2}
\|\S\|^2 &=
     \sum_{i,j} \eta_i\eta_j \frac{q_i q_j}{(2\pi)^2} \int d^2 z \,\partial_z\Big(\ln|p_i \cdot q(z,\bz)|\Big)\, \partial_{\bz}\Big(\ln|p_j \cdot q(z,\bz)|\Big)\\
    &=\sum_{i,j} \eta_i\eta_j\frac{q_i q_j}{(2\pi)^2}\int d^2 z \, \frac{(\varepsilon^+ \cdot p_i)}{q\cdot p_i}\, \frac{(\varepsilon^- \cdot p_j)}{q\cdot p_j} \;\;= \sum_{i,j} \eta_i\eta_j\frac{q_i q_j}{(2\pi)^2}\int d^2 z \, \frac{(p_i \cdot p_j) }{(q\cdot p_i) \, (q\cdot p_j)}\,,
    \label{eq:norm compu}
\eadat
\end{equation}
where to obtain the last equality we made use of the completeness relation \eqref{eq:completeness}, which yields $p_i\cdot p_j= 2 p_{i}^{\mu}p_{j}^{\nu} \left( \varepsilon^-_{(\mu}\varepsilon^+_{\nu)} +  q_{(\mu} \,\partial_z\partial_{\bz}q_{\nu)}\right)$, as well as charge conservation $\sum_i \eta_i q_i=0$ (see also \cite{Nande:2017dba}). It thus shows \eqref{eq:W is the norm QED}.

 The above expression formally applies for both massive and massless momenta. Now, let us suppose that the $k$-th particle is massless, $p_k^{\mu} = \omega_k q^{\mu}(z_k,\bz_k)$ and see what it implies. Integrating by part the first line of \eqref{eq:norm compu}, which is allowed since $\S \in \E[0]$ is a global function on the celestial sphere (see the discussion in Section \ref{subsec:hard_not_enough}), and making use of \eqref{QED: non linearity and distributional identity}, we find\footnote{Notice that the distributional terms supported at $z=\infty$ drop in the second line by virtue of charge conservation.}
\begin{equation}
    \badat{2}
   \|\S\|^2 & =- \sum_{i,j} \eta_i\eta_j \frac{q_i q_j}{(2\pi)^2}\int d^2 z\,  \partial_{\bz}\partial_z\Big(  \ln|p_i \cdot q(z,\bz)|\Big)\,\ln|p_j \cdot q(z,\bz)|\\
    &=-\sum_{i,j} \eta_i\eta_j \frac{q_j}{2\pi}\int d^2 z \, Q_i(z,\bz)\,\ln|p_j \cdot q(z,\bz)|\,.
\eadat
\end{equation}
Now if, the $k$-th particle is massless, $Q_k(z,\bz) = q_k \delta^{(2)}(z-z_k)$, then the corresponding contribution is
\begin{equation}
\badat{3}
    -\, \eta_k \;\sum_{j} \eta_j \frac{q_j}{2\pi}\int d^2 z \, Q_k(z,\bz)\,\ln|p_j \cdot q(z,\bz)| &= -\frac{ \eta_k q_k}{2\pi}\;\sum_{j} \eta_j q_j\,\ln|p_j \cdot q(z_k,\bz_k)|\\
    &= -\frac{ \eta_k q_k}{2\pi}\;\sum_{j} \eta_j q_j\,\ln|p_j \cdot p_k |\,.
\eadat
\end{equation}
Since $p_k^2 =0$, this last expression clearly gives a divergent contribution for the $j=k$ term; this reflects the collinear divergence associated with massless charged particles, which in turn leads to a logarithmic divergent emission rate~\cite{Weinberg:1965nx,Weinberg:1995mt}\footnote{This issue does not arise in gravity~\cite{Weinberg:1965nx}, see Section \ref{sec:virtual}.}. 

\section{Dressed to kill IR divergences}
\label{sec: dressed to kill QED}
The equivalence between Weinberg's soft theorem for QED and supermomentum conservation lies at the core of the argument supporting the view that the asymptotic symmetry group \eqref{eq:AGS} is a genuine symmetry of the QED $S$-matrix~\cite{He:2014cra,Campiglia:2015qka,Strominger:2017zoo}. In which case, as we argued in Section \ref{subsec:hard_not_enough}, this must imply that the Hilbert space must be extended to include representations of the asymptotic symmetry group that are not hard. In fact, it is well known that IR divergences arise precisely from the inadequacy of the usual Fock space of QED for dealing with long-range interactions and that, on the other hand, dressed states -- which allow for IR-finite $S$-matrix elements -- precisely provide examples of states which lie outside of the conventional free Fock space~\cite{Chung:1965zza,Kibble:1968sfb,KibbleII,KibbleIII,KibbleIV,Kulish:1970ut}. 

The main purpose of this section is to compare and contrast the Faddeev–Kulish (FK) construction, where charged states are dressed with clouds of photons, with supermomentum eigenstates, which we introduced in Section \ref{sec:supermomentum eigenstates QED}.
In Section \ref{sec: dressed state approach QED}, we recall that the dressed-state construction leads to asymptotic states which are eigenstates of the soft charge \cite{Gabai:2016kuf,Kapec:2017tkm}. However, general FK states are not supermomentum eigenstates, for the simple reason that they are not momentum eigenstates. Nevertheless, we show in Section \ref{sec: dressed states vs} that dressed states can be turned into supermomentum eigenstates through a specific limiting procedure. As we will demonstrate, the net effect of the dressing is effectively to \emph{linearize} the supermomentum of the asymptotic particle. Since the nonlinearity of supermomentum constitutes the obstruction to supermomentum conservation, this automatically ensures that, for scattering amplitudes involving such dressed states, charge and momentum conservation become equivalent to supermomentum conservation, and hence to IR-finiteness.  

\subsection{The dressed-state approach}
\label{sec: dressed state approach QED}

In this section, we review the dressed-state approach to obtain IR-finite scattering amplitudes: we first recall the original construction, which culminated in the work of Faddeev-Kulish \cite{Kulish:1970ut}, and then discuss more recent generalizations \cite{Gabai:2016kuf,Kapec:2017tkm}.

\subsubsection{Faddeev-Kulish (FK) dressing}
The FK dressing for a single particle state $|\vec p\,\rangle=\hat b^{\dagger}(\vec 
p)  |0\rangle$ of momentum $p^\mu(\omega, \zeta,\bar \zeta) $ and charge $q_e$ is given by~\cite{Kulish:1970ut}
\begin{align}\label{eq:dressed}
     |\vec p\,\rangle_{\text{FK}}=\, e^{\hat{R}}\,|\vec p\,\rangle,
\end{align}
with the dressing operator
\begin{equation}\label{R_factorQED}
\badat{2}
  \hat{R}&=-e  \int \frac{d^3k}{(2\pi)^3 2k^0} \,\left[f^{\mu}(k,p) \hat a^{\dagger}_{\mu}(k) - f^{*\mu}(k,p)\hat a_{\mu}(k)\right]\,,
  \eadat
\end{equation}
where $k^{\mu}= \varpi q^{\mu}(z,\bz)$ denotes the photon momentum.
The dressing factor $f^{\mu}(k,p)$ takes the form
\begin{equation}\label{FK: f term QED}
    f^\mu(k,p)=q_e\Bigg(\frac{p^\mu}{p\cdot k}-c^\mu\Bigg)\psi(k,p)\,,
\end{equation}
where $\psi(k,p)$ is a smooth function such that $\psi(k,p)=1$ in a neighborhood of $\varpi=0$. 
The vector $c^\mu(k,p)$  is null
\begin{equation}\label{IR dressing QED: constraint on C}
    c^{\mu}c_{\mu}=0\,,
\end{equation}
and must satisfy $c\cdot k=1$, which ensures the transverse condition $f\cdot k=0$ and, as a result, the invariance of the expression \eqref{R_factorQED} under gauge transformations.
The anti-Hermitian operator $\hat R$  is said to dress each charged particle state with a ``cloud'' of photons. 

Let us now comment on the different choices made for the $c$-vector.
In \cite{Kulish:1970ut}, FK showed the equivalence of dressed states \eqref{eq:dressed} with the states considered by Chung~\cite{Chung:1965zza}, which are recovered by making the gauge choice $c\cdot \varepsilon=0$ for each asymptotic state. This amounts to the dressing \eqref{R_factorQED} where $f^\mu$ is replaced by
\begin{equation}
    F^\mu=q_e\,\frac{p^\mu}{p\cdot k}\psi(k,p)\,.
\end{equation}
Since Chung states had already been shown to be infrared finite to all orders of perturbation theory \cite{Chung:1965zza}, this implied that $S$-matrix elements in the dressed-state basis, $_{\text{FK}}\langle \vec p\, | S | \vec p\,\rangle_{\text{FK}}$, are also free of IR divergences\footnote{Notice however that the collinear divergences arising for massless charged particles are not addressed in this construction.}.
This dressing can be generalized by dressing each particle separately.

\subsubsection{Generalized dressing}

As we said above, introducing a vector $c^{\mu}$ in the dressing \eqref{FK: f term QED} is necessary to ensure that the dressing is gauge invariant, i.e. $f \cdot k=0$. There is no canonical, gauge-invariant, way to fix this $c^{\mu}$, and there is thus an inherent ambiguity when dressing the particles.  In this respect, it is more natural to allow each particle to have its own dressing, each parametrized by a different $c_i^{\mu}$. In that case, a sufficient condition for the cancellation of IR divergences was given by Gabai and Sever in~\cite{Gabai:2016kuf}:
\begin{equation}\label{QED Dressing: q conservation1}
\Delta c^{\mu} := \sum_{i \in \text{in}} q_i \hat{c}^\mu_i - \sum_{i \in \text{out}} q_i \hat{c}^\mu_i =0\,,
\end{equation}
where $c_i$ denotes the dressing of the $i$-th particle and $\hat{c}_i = \lim_{\varpi\to 0} \varpi c_i$ its pole. Clearly, when all $c_i$'s are equal (as FK chose), this condition is ensured by charge conservation. 
Now, we believe that the condition \eqref{QED Dressing: q conservation1} is in fact too strong a requirement, as it imposes a constraint on terms proportional to $k^{\mu}$ which however drop out of \eqref{R_factorQED} (since $k^{\mu}a_{\mu}(k)=0$), and are therefore pure gauge in this sense. We conjecture
that the sufficient and necessary condition for the cancellation of IR divergences should be given by\footnote{This condition is the direct analogue of the condition \eqref{BMS Dressing: IR finite condition} for the cancellation of IR divergences in gravity; a detailed derivation of \eqref{QED Dressing: q conservation2} as done for the gravity case in \cite{Choi:2017bna} would be particularly valuable.}
\begin{equation}\label{QED Dressing: q conservation2}
    \Delta c^{\mu} \Delta c_{\mu} =0\,.
\end{equation}
As we show below, \eqref{QED Dressing: q conservation2} is in fact equivalent to \eqref{QED Dressing: q conservation1} when the latter is understood modulo terms proportional to $k^{\mu}$. 

One might take the viewpoint that it is not necessary for each dressed state $|\vec p_i\,\rangle_{\text{FK}}=\, e^{\hat{R}_i}\,|\vec p_i\,\rangle$ to be separately gauge-invariant, and only require the invariance of the dressed `in' state, $|\text{in}\,\rangle_{\text{FK}}=\, e^{\hat{R}_{\text{in}}}\,| \text{in}\,\rangle$, where now
\begin{equation}\label{FK: f term QED in}
    f^\mu_{\text{in}}(k)= \sum_{i\in \text{in}} q_i\Bigg(\frac{p^\mu_i}{p_i\cdot k}-c^\mu_i\Bigg)\psi_i(k,p_i)\,,
\end{equation}
and the constraint following from gauge invariance is $f_{\text{in}}\cdot k=0$. There is then a very peculiar phenomenon arising when the total incoming electric charge is zero $\sum_{i\in \text{in}} q_i =0$: the simplest dressing of the incoming state, $c^{\mu}_i =0$, $\psi_i(k,p_i) = \psi(k)$ satisfies the constraint $f^\mu_{\text{in}}\cdot k_\mu=0$ and is thus gauge invariant. 
However, it should be clear that, when $\sum_{i\in \text{in}} q_i \neq0$, there is no clear way of choosing the $c$'s, and that this is an inherent ambiguity in the choice of dressing of the incoming state.

\subsubsection{Dressing and asymptotic symmetries}
Writing
$a^{\dagger}_{\mu}(k) = \sum_{\alpha} a^{\dagger}_{\alpha} \varepsilon^{\alpha}_{\mu}$ where the sum runs over polarizations ($\alpha=\pm$) and $f_{\alpha} = f^{\mu} \varepsilon^{(-\alpha)}_{\mu}$, the dressing \eqref{R_factorQED} can be written as
\begin{equation}\label{R_factorQED2}
\badat{2}
  \hat{R}
  &= -e \int \frac{\varpi d\varpi d^2z}{16\pi^3} \left( f_+(\hat a^{\dagger}_- - \hat a_+) + f_-(\hat a^{\dagger}_+ - \hat a_-)  \right)\,.
  \eadat
\end{equation}

The dressed state \eqref{eq:dressed} turns out to be an eigenstate of the \emph{soft part} of the supermomentum operator, $\hat{\P}^{soft} =(0^\mu, \partial_z \partial_{\bz}\hat{\mathcal N}^{\text{ph}}$)~\cite{Gabai:2016kuf,Kapec:2017tkm}. Indeed, making use of the definition of the soft charge \eqref{eq:Qph} and the canonical commutation relation one finds\footnote{This uses the canonical commutation relations \eqref{Canonical commutation relation in omega,z} and the fact that $[A,B]= c\mathds{1}$ implies $[A,e^B] = c e^B$.}
\begin{equation}
 \partial_z \partial_{\bz}\hat{\mathcal N}^{\text{ph}}(z,\bz) \,|\vec p\,\rangle_{\text{FK}} =  \partial_{z} \partial_{\bz}\mathcal N(z,\bz) \,|\vec p\,\rangle_{\text{FK}}\,,
\end{equation}
where the eigenvalue is given by\footnote{\label{Footnote: reality of FK}We would like to comment on how the reality conditions are imposed here: the eigenvalue appearing in this expression has no reason a priori to be real. In fact, had we instead used the alternative definition \eqref{eq:Qph alt} for the soft charge, we would have found the complex conjugate eigenvalue
\begin{equation}\label{eq:FK_ei 2}
   \partial_z\partial_{\bz} \mathcal N = -\frac{1}{2\pi} \lim\limits_{\varpi\to 0}\,\varpi\partial_{z} f_+=  -\frac{1}{2\pi}\big( \lim\limits_{\varpi\to 0}\,\varpi\partial_{\bz} f_-\big)^* \,.
\end{equation}
What this means is that a generic choice of dressed state \eqref{eq:dressed} has non-zero soft magnetic charge 
\begin{equation}
\Q^M(z,\bz) |\vec p\,\rangle_{\text{FK}} =  -\frac{1}{2\pi} \left(\lim\limits_{\varpi\to 0}\,\varpi \,\frac{\partial_{z} f_+ - \partial_{\bz} f_-}{2i}\right)  |\vec p\,\rangle_{\text{FK}}\,.
\end{equation}
Considering dressed states with zero magnetic charge thus yields an unambiguous, real, eigenvalue for the soft charge. This seems to have gone unnoticed in the previous literature and is likely related to the factor of two issue in \cite{Gabai:2016kuf}.}
\begin{equation}\label{eq:FK_ei}
   \partial_z\partial_{\bz} \mathcal N = -\frac{1}{2\pi} \lim\limits_{\varpi\to 0}\,\varpi\partial_{\bz} f_- \,.
\end{equation}
As a result, the dressed state with \eqref{FK: f term QED} has eigenvalue
\begin{align}
   \partial_z\partial_{\bz} \mathcal N 
   &= -\frac{q_e}{2\pi}   \partial_z\partial_{\bz}\ln|p\cdot q| + \frac{q_e}{2\pi}\;\partial_{\bz}\left(  \hat{c}\cdot \partial_{z} q\right) \,.
\end{align}

However, because the operator \eqref{R_factorQED} inserts infinitely many hard particles, \eqref{eq:dressed}
is \emph{not} an eigenstate of the full supermomentum operator $\hat{\P}=\big(\hat{P}^\mu,\hat{Q}+\partial_z \partial_{\bz}\hat{\mathcal N}^{\text{ph}}\big)$,
simply because it is not even a momentum (nor a hard electric charge) eigenstate .

\subsection{Dressed states vs supermomentum eigenstates}
\label{sec: dressed states vs}
The point of view developed in this paper is that the asymptotic one-particle states for an IR-finite unitary $S$-matrix should be given by UIRs of the asymptotic symmetry group. Supermomentum eigenstates $|\P\rangle = |p , \partial_z\mathcal N\rangle$ were presented in Section \ref{sec:supermomentum eigenstates QED}; they satisfy
\begin{equation}
\hat{P}^{\mu}|\P\rangle = p^{\mu}|\P\rangle \virg \hat{\Q}(z,\bz)|\P\rangle = \big( Q(z,\bz) + \partial_z\partial_{\bz}\mathcal N(z,\bz) \big)|\P\rangle\,,
\end{equation}
where $Q(z,\bz)$ is the hard contribution coming from $|p\rangle$ (see Table \ref{tbl:Dic}) and $\partial_z\partial_{\bz}\mathcal N$ the eigenvalue of the soft part of the supermomentum.
In this section, we discuss the relationship between supermomentum eigenstates and dressed states.
As we saw, generalized FK dressed states are rather intricate objects, with infinitely many hard particles superimposed on the initial one, and as such are not momentum nor supermomentum eigenstates. As we shall now discuss, however, it is possible via a subtle limiting procedure to produce genuine supermomentum eigenstates from dressing. We show that these dressing à la Faddeev-Kulish have the effect of making the supermomentum of the particle \emph{linear} in its momentum, thereby ensuring supermomentum conservation.\\

In order to obtain supermomentum eigenstates, we introduce the function 
\begin{equation}\label{definition: psi function}
   \psi_{\epsilon}(\varpi)= \begin{cases}
       1 & \text{if} \qquad 0 \leq \varpi <\epsilon \\
       0 & \text{otherwise}\,,
   \end{cases} 
\end{equation} 
and consider the dressing $\hat{R}$ obtained from \eqref{R_factorQED} by taking \begin{equation}\label{QED: dressing of eigenstate}
    f^\mu= q_e\Bigg(\frac{p^\mu}{p\cdot k}-\frac{\hat{c}^\mu}{\varpi}\Bigg)\,\psi_{\epsilon}(\varpi)\,,
\end{equation}
where $\hat{c}^\mu$ does not depend on $\varpi$. Taking the limit $\epsilon\to 0$, one can see that the dressing becomes
\begin{equation}\label{R_factorQED3}
\badat{2}
  \hat{R}  &= i \int d^2z \; \partial_{\bz}\partial_z \mathcal N \; \hat{\Phi}\,,
  \eadat
\end{equation}
where \begin{equation}\label{QED: FK eigenvalue}
    \partial_{\bz}\partial_z \mathcal N  = -\frac{q_e}{2\pi}   \partial_z\partial_{\bz}\ln|p\cdot q| + \frac{q_e}{2\pi}\;\partial_{\bz}\left(\hat{c}\cdot \partial_{z} q\right)\,,
\end{equation}
and $\hat{\Phi}(z,\bz) := \frac{1}{2}\left(\hat{\phi}(z,\bz) +\hat{\phi}^{\dagger}(z,\bz) \right)$ is the real part of a ``Goldstone operator'' defined
as\footnote{Note that writing the left hand side as $\partial_{\bz} \hat{\phi}(z,\bz)$ is not an assumption since the $\bar{\eth}$ operator is surjective on spin coefficient of spin-weight $s=1$, see \cite{Penrose:1985bww} taking into account our conventions in Appendix \ref{app: eth operator}. Also note that the imaginary part of $\hat{\phi}$ decouples in \eqref{R_factorQED3} as a consequence of our constraint that the dressed state has a real eigenvalue \eqref{eq:FK_ei}; equivalently, because we restrict to dressings with zero soft magnetic charge, see footnote \eqref{Footnote: reality of FK}.}
\begin{align}\label{goldstone operator}
\badat{2}
  \partial_{\bz} \hat{\phi}(z,\bz)&:=  i e \,\lim_{\epsilon\to 0} \int \frac{d\varpi}{4\pi^2}  \psi_{\epsilon}(\varpi)\left(\hat{a}^{\dagger}_+(\varpi,z,\bz) - \hat{a}_-(\varpi,z,\bz)\right).
  \eadat
\end{align}
 By construction, $\hat{\Phi}(z,\bz)$ satisfies
\begin{align}\label{QED: commutation relation for the glodstone}
    \left[ \hat{P}^{\mu} , \hat{\Phi}(z,\bz)  \right] &= 0, &   \left[ \partial_z\partial_{\bz} \hat{\mathcal N}^{ph} , \hat{\Phi}(w,\bw)  \right] &= -i\delta^{(2)}(z-w).
\end{align}

Note that the limit \eqref{goldstone operator} is not as innocuous as it might seem: it would certainly yield zero if the integrand were a function (since in that case the integrand would converge to an almost everywhere vanishing function). Thus, for this operator to be non-zero, the creation/annihilation operator must have a distributional contribution around $\varpi=0$. This is not very much of a surprise, but it highlights the sense in which the construction departs from the standard Fock space: for usual normalizable Fock states such as \eqref{Photons are different: normalizable state}, creation/annihilation operators always appear multiplied by $\varpi$ and the distributional contribution drops out.

The resulting dressed state 
\begin{equation}
\badat{2}
    | \mathcal P \rangle  :&= e^{ \hat R} \;\hat b^{\dagger}(p) |0\rangle\\
   & =e^{i\,\langle \partial \bar \partial \mathcal N,\hat \Phi \rangle}\;\hat b^{\dagger}(p) |0\rangle
  \eadat
\end{equation}
is then a supermomentum eigenstate, i.e. an eigenstate of both total momentum (\eqref{eq:Poin_gen1} plus \eqref{eq:Poin_gen1_s1}) and supercharge \eqref{eq:Qin2} operator:
\begin{equation}
\badat{2}
&\hat{P}^{\mu}|\P\rangle = p^{\mu}|\P\rangle \,,\\
&\hat{\Q}(z,\bz)|\P\rangle = \big( Q(z,\bz) + \partial_z\partial_{\bz}\mathcal N(z,\bz) \big)|\P\rangle\,,
    \eadat
\end{equation}
 where the soft contribution is given by \eqref{QED: FK eigenvalue}.
 Making use of the identity \eqref{QED: non linearity and distributional identity}, we find that the eigenvalue is equal to
\begin{equation}\label{QED dressed eigenstate: eigenvalue}
    \badat{2}
     Q(z,\bz) + \partial_z\partial_{\bz}\mathcal N(z,\bz)  &= q_e \,\delta^{(2)}(z-\infty,\bz -\infty)  + \frac{q_e}{2\pi}\;\partial_{\bz}\left(\hat{c}\cdot \partial_{z} q\right). 
    \eadat
\end{equation}

We therefore see that the net result of the first term in the dressing factor \eqref{QED: dressing of eigenstate}, is to \emph{linearize} the hard charge contribution:
if all $c$'s of a scattering process are identical, in particular zero, then \emph{conservation of supermomentum becomes equivalent to conservation of electric charge}, thus ensuring the IR-finiteness of these states. \\

It is also clear that the second term in \eqref{QED: dressing of eigenstate} provides an additional soft contribution to the charge, and that one can obtain a state of any soft eigenvalue $\partial_z\partial_{\bz}\mathcal N$ by a suitable choice of $\hat c$\footnote{It suffices to choose $\hat{c}^{\mu} = n^{\mu} + \frac{2\pi}{q_e} \partial_{\bz}\mathcal N \varepsilon_+^{\mu} + \frac{2\pi}{q_e}\partial_{z}\mathcal N \varepsilon_-^{\mu}+ (\dots)q^{\mu}$, where the last contribution, which drops out of \eqref{R_factorQED}, can be chosen so that $c$ is null.}. This extra contribution to the soft charge is part of the inherent ambiguity of the dressing. Now, if some scattering process preserves momentum $\sum_{\text{in}}p_i^{\mu}=\sum_{\text{out}}p_i^{\mu}$ and electric charge $\sum_{\text{in}} q_i = \sum_{\text{out}}q_i$, then conservation of supermomentum becomes equivalent to
 \begin{equation}\label{eq:condition for supermomentum cons_QED}
\partial_{\bz}\left( \sum_{i\in \text{in}} q_i\, \hat{c}_i\cdot \partial_{z} q \right) = \partial_{\bz}\left(  \sum_{i\in \text{out}} q_i\,\hat{c}_i\cdot \partial_{z} q\right),
\end{equation}
or, equivalently,\footnote{Making use of the fact that the $\bar{\eth}$ operator has no kernel on $s=-1$ spin-weighted coefficients.}
\begin{equation}\label{eq:condition for supermomentum cons_QED2}
    \Delta C := \Delta c^{\mu} \varepsilon_{\mu}^+ =0\,,
\end{equation}
where $\Delta c^{\mu}$ is given by \eqref{QED Dressing: q conservation1}. This condition is clearly equivalent to the Gabai--Sever condition for IR cancellation \eqref{QED Dressing: q conservation1} when the latter is understood up to terms proportional to $k^{\mu}$ and thus (see below), equivalent to the condition \eqref{QED Dressing: q conservation2} for IR finiteness of the scattering. \emph{As a result,\footnote{Provided that \eqref{QED Dressing: q conservation2} is indeed a necessary and sufficient condition for cancellation of IR divergences.} conservation of supermomentum is equivalent to the condition for dressed states to be infrared finite}.

Let us see in more detail how the equivalence between \eqref{QED Dressing: q conservation2} and \eqref{eq:condition for supermomentum cons_QED2} arises. First, taking into account the constraint $c^{\mu}k_{\mu}=1$, one can show that $c_{\mu}$ must be of the form
\begin{align}
    \hat{c}_{\mu}&=  -n_{\mu} +  C \varepsilon^-_{\mu} + \bar{C}\varepsilon^+_{\mu} + A q_{\mu}.
\end{align}
Here $C =  \hat{c}^{\mu} \varepsilon^+_{\mu}$ directly relates to the soft charge  \eqref{QED: FK eigenvalue} while $A$ is fixed uniquely as a result of \eqref{IR dressing QED: constraint on C}. The exact form of this last term will be irrelevant for us: it in fact vanishes when evaluated in \eqref{R_factorQED} and is in this sense pure gauge. It now follows that, as a result of charge conservation,
\begin{align}
    \Delta c_{\mu} &= \Delta C \varepsilon^-_{\mu} +\Delta {\bar{C}} \varepsilon^+_{\mu}
    + A' q_{\mu}
\end{align}
for some function $A'$. Finally, one sees that
\begin{equation}
    \Delta c_{\mu}\, \Delta c^{\mu} = 2|\Delta C|^2.
\end{equation}
As a result, the conditions \eqref{QED Dressing: q conservation2} and \eqref{eq:condition for supermomentum cons_QED2} are in fact equivalent. \\

In general, as we already emphasized, the extra contribution to the soft charge brought by $c^\mu$ is an inherent ambiguity of the dressing and cannot be discarded without breaking gauge or Lorentz invariance. For example, taking $c^\mu_i=0$ in \eqref{QED dressed eigenstate: eigenvalue} yields a charge $q_i \delta^{(2)}(z-\infty)$, which manifestly breaks Lorentz invariance by giving some special role to the south pole, $|z|=\infty$, of the celestial sphere. It can also be traced back, in the original Faddeev-Kulish derivation, to the fact that taking $c^{\mu}=0$ necessary means that $f\cdot k\neq0$, and thus break gauge invariance (which is tied up to Lorentz invariance by the choice $\varepsilon^+_{\mu} = \partial_z q_{\mu}$ for the polarization tensors). There is, however, a very specific phenomenon appearing when the total electric charge of an incoming state is zero, $\sum_{\text{in}} q_i =0$: then the corresponding incoming dressed state $|\text{in}\rangle_{\text{FK}}$ has total charge $\Q_{tot} =\partial_{\bz}\left(  \sum_{\text{in}} \frac{q_i}{2\pi}\,\hat{c}_i\cdot \partial_{z} q\right)$ and there exists a preferred, Lorentz-invariant and with zero total charge, dressing obtained by taking $c_i =0$. Making use of the results of section \ref{section: Hard/soft decomposition of supermomenta QED}, one can in fact do better and devise a Lorentz-invariant dressing for incoming states without the requirement of having zero incoming electric charge : it suffices to make sure that total incoming supermomentum of dressed states $(p_{\text{in}}^{\mu},\Q_{\text{in}})$ is effectively hard, i.e. $\Q_{\text{in}} = Q_{\text{in}}$ (this is obtained by choosing the c's such that $\sum_{\text{in}}\frac{q_i}{2\pi}\;\hat{c}_i\cdot \partial_{z} q =  \frac{q_{\text{in}}}{2\pi}\frac{p_{\text{in}} \cdot \partial_zq}{p_{\text{in}}\cdot q}$). These are such that conservation of electric charge and momentum will automatically imply supermomentum conservation.

\paragraph{On Goldstone operators} The introduction of a Goldstone operator is unnatural from the perspective of representation theory advocated in the present work; it arises only as a necessity to make contact with dressing constructions. First, the prescription  to  obtain genuine supermomentum eigenstates from dressed states (namely the dressing \eqref{QED: dressing of eigenstate} followed by the limit $\epsilon \to 0$) feels somewhat ad hoc and rather suggests that the dressing construction attempts to shoehorn the usual Poincaré states into a different Hilbert space -- one that we believe should be that of asymptotic particles, i.e. the space of UIRs of the QED asymptotic symmetry group. Second, and perhaps most importantly, the pair $(\partial_z\partial_{\bz} \hat{\mathcal N}, \hat{\Phi})$, consisting of the soft charge and Goldstone operator, satisfies the commutation relations \eqref{QED: commutation relation for the glodstone}, which make them a generalization of momentum and position operators $(\hat{P}, \hat{X})$ with canonical commutation relation $[\hat{P},\hat{X}] = -i$. In other terms, \emph{the Goldstone operator $\hat \Phi(z,\bz)$ plays the role of a position operator in the space of asymptotic QED vacua}\cite{Bekaert:2024uuy}. The same reason that forbids the use of a position operator $\hat{X}^{\mu}$ in quantum field theory makes the use of such a Goldstone operator problematic from the perspective of systematically constructing asymptotic states from UIRs of the asymptotic symmetry group (see the final discussion for further comments).

\newpage
\phantomsection
\part{Part II. Gravity}

The asymptotic symmetry group of asymptotically flat spacetimes, known as the BMS group~\cite{bondi_gravitational_1962,sachs_asymptotic_1962}, extends the Poincaré group to include an infinite amount of smooth conformal densities on the celestial sphere, $\T(z,\bz) \in \E[1]$, called ``supertranslations''.\footnote{See \cite{Donnay:2023mrd} for a recent review.} Supertranslations thus carry conformal weights\footnote{Recall that elements of $\E[w]$ correspond to fields of conformal weights $(h,\bar{h}) = (-\frac{w}{2},-\frac{w}{2})$; see Appendix \ref{app:conformal_density} for our conventions.} $(h,\bar{h})=(-\frac{1}{2},-\frac{1}{2})$  and form an abelian subgroup of the BMS group, which is given by
\begin{equation}\label{eq:BMS}
\tag{$2$}
   \text{BMS}_4  \simeq SO(3,1) \ltimes \mathcal E[1]\,.
\end{equation}
As in the QED case, the group is of the form $SO(3,1) \ltimes A$ (with $A$ an abelian group), for which the general representation theory is briefly reviewed in Appendix \ref{Appendix: section representations}.\footnote{See also \cite{Bekaert:2026cvx} for an introduction to BMS group theory from a geometric viewpoint.} Unitary irreducible representations (UIRs) of BMS were classified in the series of seminal papers \cite{Mccarthy:1972ry,McCarthy_72-I,McCarthy_73-II,McCarthy_73-III,McCarthy:1974aw,McCarthy_75,McCarthy_76-IV,McCarthy_78,McCarthy_78errata,PiardBMS}. More recently, new developments relating these UIRs to infrared physics have been worked out in \cite{Bekaert:2024uuy,Bekaert:2025kjb}.

\section{Hard representations of BMS}
\label{sec:hardBMS_reps}
For the representations that we will consider in this section, it will be enough to know that the abelian supertranslation group $\mathcal E[1]$ acts on the usual creation/annihilation operator of a particle of momentum $p_{\mu}$ as
\begin{equation}\label{eq:action of hard reps_bms}
    \hat b(p) \;\mapsto\; e^{i\,\langle P, \T \rangle}\, \hat b(p)\,.
\end{equation}
The phase involves the pairing
\be
\langle P, \T\rangle= \int d^2 z \,P(z,\bz) \T(z,\bz) \quad \in \mathbb R\,,
\ee
which is the contraction of the supertranslation parameter $\T(z,\bz)\in \mathcal E[1]$ with a dual element
\begin{equation}\label{Supermomentum_bms}
  P(z,\bz) \in \mathcal E[-3]\,,
\end{equation}
which we will refer to as the hard supermomentum.

The four Poincar\'e translations $T^{\mu}\in \mathbb{R}^{3,1}$ are canonically included in the space of supertranslations $\T(z,\bz) \in \E[1]$,
\begin{equation}
\label{eq: inclusion of translations}
       T^{\mu}\in   \mathbb{R}^{3,1} \quad \mapsto\quad  T^{\mu}q_{\mu}(z,\bz) \in  \E[1]\,,
\end{equation}
by means of the canonical object $q^{\mu}(z,\bz)$, whose explicit expression in given in Appendix \ref{app: q}.
This is directly related to the fact that the celestial sphere identifies with the projective null cone in Minkowski space, $(z,\bz) \mapsto q^{\mu}(z,\bz)$, with $q^2=0$.
Restricting the supertranslation in \eqref{eq:action of hard reps_bms} to a translation, i.e. taking $\T(z,\bz)= T^{\mu}q_{\mu}(z,\bz)$, one recovers the usual action of Poincaré translations\footnote{For the hard representations considered in this section, this follows from the identities discussed in Appendix \ref{app:hard_id}.}
\begin{equation}
   \hat b(p) \to \,e^{i\, p_{\mu}T^{\mu}}\,\hat b(p)\,.
\end{equation}

In this section, we show that the familiar Poincaré UIRs lift to representations of the BMS group \eqref{eq:BMS}. They give what we will refer to as the ``hard'' BMS UIRs.  In practice, this means that, for such representations, the corresponding hard supermomentum \eqref{Supermomentum_bms} is completely determined by the usual particle momentum $p_\mu$. The explicit expressions of hard BMS UIRs for a massless and massive scalar\footnote{In this work, we do not look at the tachyonic case $p^2>0$; see \cite{Bekaert:2025kjb}.}, which we will explain in the rest of this section, are summarized in Table \ref{tbl:Dictionary}. Among \emph{generic} representations (which will be reviewed in Section \ref{sec:generic_BMS}), hard representations play a very special role as they are the ones which are in one-to-one correspondence with usual Poincaré UIRs. 

BMS hard representations were originally studied by Sachs \cite{Sachs:1962zza} (for the massless case) and Longhi-Materassi \cite{Longhi:1997zt} (for the massive case). They can be geometrically realized in terms of the scattering data of fields at null and timelike infinities, respectively and this is the perspective taken in most of the recent literature, see in particular \cite{He:2014laa,Campiglia:2015kxa}. Our presentation -- which unifies both the massless and massive cases -- however, avoids relying on explicit asymptotic expansions. This has the advantage of making the presentation accessible to readers without prior knowledge of the technical aspects of asymptotic analysis.
Moreover, one of our main goals here is to emphasize the following overlooked fact: every standard field in QFT naturally carries a representation of the BMS group, and this independently of any gravitational considerations\footnote{However, as we shall see, interactions will not preserve the symmetry unless gravity comes into play.}. We start by showing that a scalar field carries a hard BMS representation in Section \ref{sec:scalar_BMS} and then extend the analysis to fields with spin in Section \ref{sec:spinning_BMS}. \\

\begin{table}[H]
	\centering
    \begin{minipage}{0.8\textwidth} 
    \centering
	\begin{tabular}{c|c}
		$\,\,$  Hard rep. $\,\,$ & $\,\,$ $\hat b(p) \to e^{i\langle P, \mathcal T\rangle}\, \hat b(p)$ $\,\,$ \\[0.3cm]
         \hline \\[-0.3cm]    
massless & $P(z,\bz)=\omega \delta^{(2)}(z-\zeta)$\\[0.3cm]
massive& \,\, \,\,\,$P(z,\bz)= \dfrac{- \,m^4}{4\pi\, (q(z,\bz)\cdot p)^3}$ \,\,\\		[0.3cm]
	\end{tabular}
	\caption{Hard UIRs of the BMS group \eqref{eq:BMS} for a scalar of massless
    ($p^\mu=\omega q^\mu(\zeta,\bar \zeta)$, $q^2=0$) or massive ($p^2=-m^2$) momentum. Group elements include supertranslations $\T(z,\bz) \in \E[1]$, which are paired with hard supermomenta, $P(z,\bz)\in \E[-3]$.}
	\label{tbl:Dictionary}
    \end{minipage}
\end{table}

\subsection{Scalar fields}
\label{sec:scalar_BMS}
Let $\phi(X)$ a real scalar field of momentum $p^\mu=(p^0,\vec p)$,
\begin{equation}
    \phi(X) =\int \frac{\D^3 p}{(2\pi)^3\, 2p^0} \Big[ \hat b(\vec p) \, e^{i p \cdot X} +  \hat b(\vec p)^\dagger \, e^{-i p \cdot X}\Big] \label{real_scalar}\,,
\end{equation}
where ladder operators obey the usual commutation relations
\begin{equation}\label{Canonical commutation relation GR}
   \left[ \hat b(\vec p),\hat b(\vec p\,')^\dagger \right] = (2\pi)^3\,2p^0\,\delta^{(3)}(\vec p-\vec p\,')\,.
\end{equation}
The momentum $p^\mu$ can be parametrized explicitly by \eqref{eq:p_massless}, \eqref{eq:massive_p} if it the scalar field is massless or massive, respectively.

\subsubsection{Poincaré UIRs}
Let us first briefly recall how the familiar UIRs of Poincaré are realized. In momentum space, the energy-momentum and Lorentz operators given by
\begin{equation}
\badat{2}
\label{eq:Poin_gen2}
&\hat P^\mu=\int \frac{\D^3 p}{(2\pi)^3\, 2p^0} \,p^\mu \,\hat b^\dagger(\vec p) \hat b(\vec p)\\
&\hat J^{\mu\nu}=i\int \frac{\D^3 p}{(2\pi)^3\, 2p^0} \,\hat b^\dagger(\vec p)\left(p^\mu \frac{\partial}{\p p_\nu}-p^\nu \frac{\partial}{\p p_\mu}\right)  \hat b(\vec p)\,,
\eadat
\end{equation}
generate the Poincaré transformations 
\begin{equation}
\badat{2}
&[\hat P^\mu,\phi(X)]=i\p^\mu \phi(X)\\
&[\hat J^{\mu \nu},\phi(X)]=i(x^\mu\p^\nu- x^\nu\p^\mu)\phi(X)\,.
\eadat
\end{equation}
The generators \eqref{eq:Poin_gen2} form an explicit representation of the Poincaré algebra\,,
\begin{equation}
\badat{2}
&\big[\hat{P}^{\mu} , \hat{P}^{\nu}  \big]=0\\
&\big[\hat{J}^{\mu\nu} , \hat{P}^{\rho}  \big]=i \eta^{\nu \rho}\hat P^\mu -i \eta^{\mu \rho}\hat P^\nu\\
&\big[\hat{J}^{\mu\nu} , \hat{J}^{\rho\sigma}  \big] = i\big(\eta^{\sigma\mu}\hat J^{\nu\rho}  -\eta^{\rho\mu} \hat J^{\nu \sigma} + \eta^{\rho\nu} \hat J^{\mu\sigma} -  \eta^{\sigma\nu} \hat  J^{\mu\rho}   \big)\,,
\eadat
\end{equation}
and a corresponding representation of the Poincaré group via $U(\omega,a)=\exp(\frac i 2 \omega_{\mu \nu}\hat J^{\mu \nu}+ia_\mu \hat P^\mu)$.
It defines an (infinite-dimensional) unitary irreducible representation (UIR) on the Hilbert space of one-particle states. 

\subsubsection{Hard BMS representations}
Let us now show that the Poincaré representations can be lifted, at no cost, to what we will refer to as ``hard'' BMS representations. To see this, we will follow as closely as possible the expressions for Poincaré UIRs recalled above. 

To each particle of momentum $p^\mu$, one can associate a supermomentum $P(z,\bz) \in \E[-3]$ via
{\setlength{\belowdisplayshortskip}{12pt}
\begin{equation}\label{eq:hardP}
P(z,\bar z) =
\left\{\begin{aligned}
  &\quad \omega\,\delta^{(2)}(z-\zeta) & \quad \text{if } & p^2=0,  \;p_{\mu} = \omega q_{\mu}(\zeta,\bar{\zeta}),\\
  &-\dfrac{m^4}{4\pi\,(q(z,\bar z)\cdot p)^3} & \quad\, \text{\,if } &p^2=-m^2\,.
\end{aligned}\right.
\end{equation}
\par\noindent
In the above, $q^\mu$ is the canonical null vector which realizes the inclusion of translations inside supertranslations; see \eqref{eq: inclusion of translations} and Appendix \ref{app: q}. Momentum can be recovered from supermomentum as
\begin{equation}\label{eq:hardP projection}
p_\mu=\int d^2z \,P(z,\bz) q_\mu(z,\bz)\,,
\end{equation}
which ensures that, when restricting \eqref{eq:action of hard reps_bms} to translations, one recovers the usual phase $e^{ip_{\mu}T^{\mu}}$. The proof of \eqref{eq:hardP projection} is trivial for the massless case and easily showed for the massive case in Appendix \ref{app:hard_id}. Importantly, let us emphasize that $P(z,\bar z)$ (in both cases) depends \emph{non-linearly} on the momentum $p^\mu$.

The key property of the \eqref{eq:hardP} is that both expressions satisfy the following identities, proven in Appendix \ref{app:hard_id},
\begin{equation}
    2p_{[\mu} \partial_{p^\nu]} \,P(z,\bz)  = \Big(  \mathcal{Y}_{\mu\nu}^{z} \partial_z +  \mathcal{Y}_{\mu\nu}^{\bz} \partial_{\bz}   + \tfrac{3}{2} \left(\partial_z \mathcal{Y}_{\mu\nu}^{z} + \partial_{\bz}\mathcal{Y}_{\mu\nu}^{\bz}\right)\Big)\,P(z,\bz)\,,
\end{equation}
where $\partial_{p^\mu} := \frac{\partial}{\partial p^{\mu}}$ and
\begin{equation}
    \mathcal{Y}_{\mu\nu}^{z} \partial_z=  2\partial_{\bz}q_{[\mu} q_{\nu]} \partial_z \virg  \mathcal{Y}_{\mu\nu}^{\bz} \partial_\bz=  2\partial_{z}q_{[\mu} q_{\nu]} \partial_\bz 
\end{equation}
are the infinitesimal generators of Lorentz transformations on the celestial sphere.

We can now use the above facts to construct an explicit representation of the BMS algebra. 
We define the hard supermomentum operator
\begin{equation}
\badat{2}
&\hat P(z,\bz)=\int \frac{\D^3 p}{(2\pi)^3\, 2p^0} \,P(z,\bz) \hat b^\dagger(\vec p) \hat b(\vec p)\,,
\eadat
\end{equation}
which extends the usual expression in \eqref{eq:Poin_gen2}. A direct computation shows that, together with the Lorentz generators $\hat J^{\mu \nu}$, it realizes the following algebra:
\begin{equation}\label{BMS algebra representation}
\badat{3}
&\big[\hat{P}(z,\bz) , \hat{P}(w,\bw) \big]=0\\
 &\big[\hat{J}^{\mu\nu} , \hat{P}(z,\bz)  \big] = i \bigg( \mathcal{Y}_{\mu\nu}^{z} \partial_z  + \mathcal{Y}_{\mu\nu}^{{\bz}} \partial_{\bz} +  \tfrac{3}{2} \left( \partial_z \mathcal{Y}_{\mu\nu}^{z} + \partial_{\bz}\mathcal{Y}_{\mu\nu}^{\bz} \right) \bigg)\, \hat{P}(z,\bz)\\
&\big[\hat{J}^{\mu\nu} , \hat{J}^{\rho\sigma}  \big] = i\big(\eta^{\sigma\mu}\hat J^{\nu\rho}  -\eta^{\rho\mu} \hat J^{\nu \sigma} + \eta^{\rho\nu} \hat J^{\mu\sigma} -  \eta^{\sigma\nu} \hat  J^{\mu\rho}   \big)\,,
\eadat
\end{equation}
which exactly coincides with the BMS algebra. In particular, we see that $\hat{P}(z,\bz)$ indeed transforms as a conformal primary of weights $\left (\frac{3}{2},\frac{3}{2}\right)$ under Lorentz transformations\footnote{Which implies that $\mathcal F[\T]=\int d^2z\, \hat P(z,\bz) \T(z,\bz)$ satisfies \begin{equation}
    \big[\hat{J}_{\mu\nu} , \mathcal F[\T] \big] =i \,\mathcal F\big[ \mathcal{Y}_{\mu\nu}^{z} \partial_z  \T+ \mathcal{Y}_{\mu\nu}^{{\bz}} \partial_{\bz}  \T\,-\frac{1}{2}( \partial_z \mathcal{Y}_{\mu\nu}^{z}+  \partial_{\bz}\mathcal{Y}_{\mu\nu}^{{\bz}})\T\big]\,.
\end{equation}
}. We thus see that, thanks to the existence of the canonical projection on \eqref{eq:Poin_gen2}
\begin{equation}
    \hat{P}(z,\bz)  \mapsto \hat{P}_\mu = \int d^2z \,\hat{P}(z,\bz)\,q_\mu(z,\bar z)\,,
\end{equation}
each familiar Poincaré UIR can be thought of as a UIR of the BMS group. In this representation, the usual creation/annihilation operators are eigenvectors of supermomentum
\begin{equation}
\badat{2}
[\hat{P}(z,\bz), \hat b^{\dagger}(p)] &= P(z,\bz)\,\hat b^{\dagger}(p)
\eadat
\end{equation}
the eigenvalue $P(z,\bz)$ being given by \eqref{eq:hardP}. Therefore, as announced in the beginning of this section, the finite action of the group, generated by
\begin{equation}
    U(\omega, \T) = \exp\Big(\frac i 2 \omega_{\mu \nu}\hat{J}^{\mu \nu}+i\int d^2 z \,\T(z,\bz) \hat{P}(z,\bz)\Big)\,,
\end{equation}
induces an action of the form \eqref{eq:action of hard reps_bms}.

\subsection{Spinning fields}
\label{sec:spinning_BMS}
For completeness, we now turn to the discussion of hard representations for a massless spin-$s$ field,
\begin{equation}\label{eq: spin s}
    \phi_{\mu_1 \dots \mu_s}(X) = k\sum_{\alpha=\pm } \int \frac{\D^3 p}{(2\pi)^3\, 2p^0} \Big[\varepsilon_{\mu_1 \dots \mu_s}^{*\alpha}(\vec p)\, \hat a_\alpha(\vec p) \, e^{i p \cdot X} + \varepsilon_{\mu_1 \dots \mu_s}^{\alpha}(\vec p)\, \hat a_\alpha(\vec p)^\dagger \, e^{-i p \cdot  X}\Big] \,,
\end{equation}
where $k$ is a constant and the rank-$s$ polarization tensors
\begin{equation}
    \varepsilon^\pm_{\mu_1\dots \mu_s}(\vec p) = \varepsilon^\pm_{\mu_1}(\vec p)\dots \varepsilon^\pm_{\mu_s}(\vec p) , \label{eq:pol_tensors_s}
\end{equation}
are fully symmetric and transverse, $p^{\mu_i}\varepsilon^\pm_{\mu_1\dots\mu_i\dots\mu_s}(\vec p)=0$ for any $i=1,\dots,s$. 

The Lorentz generator is given by
\be
\label{eq:Lorentz s}
\hat J^{\mu\nu}=\int \frac{\D^3 p}{(2\pi)^3\, 2p^0} \,\sum_{\alpha,\beta}\hat a^\dagger_\alpha(\vec p)\bigg[i\bigg(p^\mu \frac{\partial}{\p p_\nu}-p^\nu \frac{\partial}{\p p_\mu}\bigg)\delta_{\alpha \beta}+ S^{\mu\nu}_{\alpha \beta}(\vec p)\bigg] \hat a_\beta(\vec p)\,,
\ee
with $S^{\mu\nu}_{\alpha\beta}$ the spin generator. The lift from usual Poincaré UIRs of momentum $p^\mu$ and spin $s$ to UIRs of BMS group  proceeds analogously to the scalar case: we promote the momentum operator to the hard supermomentum operator
\begin{equation} \label{eq:spins_lifted}
\badat{2}
&\hat P(z,\bz)=\int \frac{\D^3 p}{(2\pi)^3\, 2p^0} \,P(z,\bz) \,\sum_{\alpha=\pm}\hat a^\dagger_\alpha(\vec p) \hat a_\alpha(\vec p)\,,
\eadat
\end{equation}
with $P(z,\bz)$ again given by \eqref{eq:hardP}. Since the spin part of the Lorentz generator commutes with $\hat P(z,\bz)$, we can automatically conclude that the generators \eqref{eq:Lorentz s}, \eqref{eq:spins_lifted} close the BMS algebra \eqref{BMS algebra representation}.

\subsection{Why hard representations are not enough}
\label{subsec:hard_not_enough_bms}
In the above, we saw that usual Poincaré particles are in 1:1 correspondence with the hard UIRs of the BMS group.
However, the issue with hard representations is that they \emph{cannot} preserve BMS supermomentum~\cite{Bekaert:2025kjb} (see also \cite{Chatterjee:2017zeb}).

To see this, let us consider a scattering involving a family of momenta $p^{\mu}_i$ satisfying momentum conservation
\begin{align}
    \sum_i \eta_i p^{\mu}_i&=0,
\end{align}
where $\eta=+1$ ($\eta=-1$) if the particle is incoming (outgoing). Denoting by $P_i(z,\bz) \in \mathcal E[-3]$ the corresponding hard supermomenta of each particle (as discussed in Table \ref{tbl:Dictionary}), then, as we show below,
\begin{equation}\label{Hard reps dont conserve supermomentum_bms}
    \sum_i \eta_i P_i(z,\bz) = \partial^2_z \partial^2_{\bz} \mathscr S(z,\bz)\,.
\end{equation}
The quantity $\mathscr S(z,\bz) \in \E[1]$ is always non-zero and therefore constitutes the \emph{obstruction to supermomentum conservation}. It is given explicitly by
\begin{equation}\label{BMS: soft factor}
    \mathscr S(z,\bz) = -\frac{1}{2\pi}\sum_i \eta_i\, (p_i\cdot q(z,\bz) ) \,\ln |p_i\cdot q(z,\bz) |\,.
\end{equation}

The proof of \eqref{Hard reps dont conserve supermomentum_bms} follows from momentum conservation together with the following distributional identity on the celestial sphere, proved in Appendix \ref{app:distrib_id} and valid for any hard supermomentum:
\be\label{BMS: non linearity and distributional identity}
P(z,\bz)=-\frac{1}{2\pi} \p_z^2\p_\bz^2 \big(p \cdot q(z,\bz) \ln|p\cdot q(z,\bz)| \big)+p^\mu \mathcal D_\mu(\delta^{(2)}(z-\infty,\bz-\infty))\,.
\ee
 The second term in \eqref{BMS: non linearity and distributional identity} is a distribution supported at $|z|=\infty$. The explicit expression of this last term is not needed here and can be found in the Appendix; what matters, in order to derive \eqref{Hard reps dont conserve supermomentum_bms}, is that this distribution is \emph{linear} in the momentum $p$. The above identity is meant as a \emph{global} identity, $P= \eth^2\bar{\eth}^2 B+ p^\mu\mathcal D_\mu\delta$, relating conformally weighted distributions on the celestial sphere $P\in \E[-3]$, $B\in \E[1]$. In other terms, a more precise -- though perhaps more cumbersome -- formulation of the identity \eqref{BMS: non linearity and distributional identity}  would be\footnote{Note that this expression is coherent with $\hat{P}(\hat{z},\hat{\bz})=-\frac{1}{2\pi}\hat{\p}^2_{\hat{\bz}}\hat{\p}^2_{\hat{z}}(p\cdot \hat{q}(\hat{z},\hat{\bz})\ln|p\cdot \hat{q}(\hat{z},\hat{\bz})|)$. Also note that  note that the absence of `hat' on $q$ comes a result of the definition of $B\in \E[1]$ (which is implicitly making use of the chart $(z,\bz)$).  Both, directly related, facts that 1) $z=\infty$ plays a special  role in \eqref{BMS: non linearity and distributional identity} and that  2) the definition of $B$ here depends on the chart $(z,\bz)$, point to the fact that this way of writing the hard supermomentum breaks Lorentz invariance. However this is \emph{not} the case of \eqref{BMS: soft factor} which is a perfectly well-defined, Lorentz-invariant, global weight-one density on the sphere; in particular $\hat{\mathscr{S}}(\hat{z},\hat{\bz}) = -\frac{1}{2\pi}\sum_i \eta_i\, (p_i \cdot \hat{q}(\hat{z},\hat{\bz}))\ln |p_i \cdot \hat{q}(\hat{z},\hat{\bz})|$.}
\begin{equation}
    \begin{cases}
        P(z,\bz)=\frac{-1}{2\pi}\p^2_\bz\p^2_z  \big(p\cdot q(z,\bz)\ln|p\cdot q(z,\bz)|\big) &\text{in the chart} \,\,(z,\bz), \\
        \hat{P}(\hat{z},\hat \bz)=\frac{-1}{2\pi}\hat{\p}^2_{\hat \bz}\hat{\p}^2_{\hat{z}}\big(p\cdot \hat{q}(\hat{z},\hat \bz) \ln
        |p\cdot q(\tfrac{1}{\hat{z}},\tfrac{1}{\hat \bz})|\big)+p^\mu \hat{\mathcal D}_\mu\delta^{(2)}(\hat{z},\hat \bz) &\text{in the chart} \,\, (\hat{z},\hat \bz)= (z^{-1},\bz^{-1}). 
    \end{cases}
\end{equation}

Therefore, what the above shows is that, for a scattering involving only hard UIRs, momentum conservation does \emph{not} imply supermomentum conservation $\sum_i \eta_i P_i(z,\bz)=0$. As we will see in Section \ref{section: soft graviton theorem}, this statement, when applied to conventional Fock states, directly relates to Weinberg's soft graviton theorem \cite{Weinberg:1965nx}, with the soft factor being given by the Lorentz-invariant quantity $\mathscr S\in \E[1]$.  This justifies the need to go beyond hard UIRs and instead consider more general representations that can accommodate supermomentum conservation.

\section{Generic BMS representations}
\label{sec:generic_BMS}
We now turn to generic representations of the BMS group, $SO(3,1) \ltimes \E[1]$. We highlight below those aspects of the representation theory that will be relevant for our subsequent analysis. We refer the reader to \cite{Mccarthy:1972ry,McCarthy_72-I,McCarthy_73-II,McCarthy_73-III,Girardello:1974sq,McCarthy:1974aw,McCarthy_75,McCarthy_76-IV,McCarthy_78,McCarthy_78errata,PiardBMS} for the classification of UIRs, \cite{Bekaert:2025kjb} for the study of generic representations, and to Appendix \ref{Appendix: section representations} for general considerations. 

\subsection{BMS supermomentum}
On general ground (see Appendix \ref{Appendix: section representations}), the starting point for constructing induced representations of $SO(3,1) \ltimes \E[1]$ is to consider elements $\P(z,\bz)$ of the dual space
\begin{equation}
   \P(z,\bz) \in \left(\E[1] \right)^*=\E[-3]\,,
\end{equation}
which are referred to as \emph{supermomenta}\footnote{Notice here the difference of notation with respect to hard supermomenta, denoted by $P(z,\bz)$.} and should be thought of as distributions. 
By definition, they are dual to the symmetry parameters, the supertranslations $\T(z,\bz) \in \E[1]$, with duality pairing given by
\begin{equation}
    \langle \P , \T \rangle := \int d^2z\, \P(z,\bz) \T(z,\bz)\,\in \mathbb R\,.
\end{equation}
An important property of supermomenta is that they can always be projected, in an invariant manner, on a usual momentum $p_{\mu}$ via
\begin{equation}\label{projection from supermomentum to momentum}
\pi\,: \quad\left|
    \begin{array}{ccc}
   \E[-3] &  \to &(\mathbb{R}^{3,1})^*\\[0.5em]
        \P(z,\bz) &  \mapsto & p_{\mu}= \int d^2z \,\P(z,\bz) q_\mu(z,\bz)\,,
    \end{array}\right.
\end{equation}
where $q_\mu(z,\bz)$ is the canonical object that realizes the inclusion of translations inside supertranslations; see \eqref{eq: inclusion of translations}.

\subsection{BMS little group}
The BMS little group $\ell_{\P} \subseteq SO(3,1)$ of a representation of supermomentum $\P$ is the subgroup of Lorentz whose elements $B\in \ell_{\P}$ stabilize the supermomentum
\begin{equation}
   B\cdot \P(z,\bz) = \P(z,\bz)\,.
\end{equation}
By construction, since $B$ above must in particular always stabilize the momentum $p^{\mu}$, the BMS little group is always a subgroup of the Poincaré little group $\ell_p$:
\begin{equation}
    \ell_{\P} \subset \ell_p= \begin{cases}
        SU(2) \\ ISO(2) \\
        SL(2,\mathbb{R})\\SL(2,\mathbb C)\,
    \end{cases}
\end{equation}
where $SU(2)$, $ISO(2)$, $SL(2,\mathbb{R})$ are the Poincaré little groups stabilizing, respectively, timelike (massive), null (massless), and spacelike (tachyonic) four-momenta, while $SL(2,\mathbb C)$ corresponds to the special case $p^\mu=0$. 

Let us for now set aside the case where $p=0$. Since the Lorentz orbit $\mathcal O_p \simeq \frac{SL(2,\mathbb C)}{\ell_p}$ for momentum is of dimension 3 (the mass shell; see the second column of Table \ref{table:classification bms little groups}), the orbits of supermomentum $\mathcal O_\P \simeq \frac{SL(2,\mathbb C)}{\ell_\P}$ are always of higher (but finite) dimensions, dim$(\mathcal O_\P) \in \{3,4,5,6\}$. In fact, due to the projection \eqref{projection from supermomentum to momentum}, the BMS shell $\mathcal O_\P \simeq \frac{SL(2,\mathbb C)}{\ell_\P}$ is a fiber bundle over the mass shell $\mathcal O_p \simeq \frac{SL(2,\mathbb C)}{\ell_p}$ where the typical fibre 
\begin{equation}\label{Supermomentum fibre}
    F = \frac{\ell_p}{\ell_{\P}}
\end{equation}
encodes the extra degrees of freedom of the BMS particle compared to the corresponding Poincaré particle. Soft representations, i.e. BMS particles for which momentum vanishes $p=0$, will be discussed in more details in Section \ref{sec: soft reps BMS}. Let us just mention that McCarthy proved that their BMS shell also must have dimension dim$(\mathcal O_\P) \in \{3,4,5,6\}$ (excluding the trivial representation $\P=0$).

In the series of papers \cite{Mccarthy:1972ry,McCarthy_72-I,McCarthy_73-II,McCarthy_73-III,McCarthy:1974aw,McCarthy_75,McCarthy_76-IV,McCarthy_78,McCarthy_78errata}, McCarthy provided a classification of allowed BMS little groups; they are summarized in Table \ref{table:classification bms little groups}. The physical meaning on the different choices of BMS little groups is still largely unclear\footnote{Some of them might be related to specific kinematic configurations.}. As argued in \cite{Bekaert:2025kjb}, one should not expect supermomenta to possess any particular symmetry; hence, despite the importance of the classification results of BMS little groups, the most relevant representations in a generic scattering process should be those associated with generic supermomenta (i.e. those for which the BMS little group is trivial).
\begin{table}[h!]
\centering
\begin{tabular}{>{\centering\arraybackslash}p{2.2cm}:  
                >{\centering\arraybackslash}p{2.2cm}   
                >{\centering\arraybackslash}p{2.2cm}:  
                >{\centering\arraybackslash}p{3.2cm}     
                >{\centering\arraybackslash}p{2.8cm}   
                >{\centering\arraybackslash}p{1.5cm}}  
                &&&&&\\
 & {\small \textbf{Poincaré little group $\ell_p$}} 
 & {\small \textbf{Orbit} \quad \quad (mass shell) $\frac{SL(2,\mathbb C)}{\ell_p}$} 
 & {\small \textbf{BMS little group \quad \quad \quad \quad $\ell_\P$}} 
 & {\small \textbf{Extra d.o.f.} (Fibre of  BMS shell) $\frac{\ell_p}{\ell_\P}$} 
 & {\small \textbf{BMS shell dim.}} \\[3pt]
\toprule

\multirow{6}{*}{\makecell{\small \textbf{Massless}\\\small \textbf{momentum}}} 
& \multirow{6}{*}{$ISO(2)$} 
& \multirow{6}{*}{\makecell{$\mathbb R^+ \times S^2$}} 
& $ISO(2)$ = $\ell_p$ & $\{e\}$ = trivial & 3 \\[1pt]
&  &  & $\mathbb R^2 $ & $S^1$ & 4 \\[1pt]
&  &  & $\mathbb R $  &  $S^1 \!\times \mathbb R$ & 5 \\[1pt]
&  &  & $U(1)$          & $\mathbb R^2$ & 5 \\[1pt]
&  &  & $\{e\}$ = trivial       & \hspace{-0.4cm}$ISO(2)$ \!=\! $S^1 \times \mathbb{R}^2$ & 6 \\[1pt]
\midrule

\multirow{3}{*}{\makecell{\small \textbf{Massive}\\\small \textbf{momentum}}} 
& \multirow{3}{*}{$SU(2)$} 
& \multirow{3}{*}{\makecell{$\mathbb H^3$}} 
& $SU(2)$ = $\ell_p$ & $\{e\}$ = trivial & 3 \\[1pt]
&  &  & $U(1)$  & $S^2$ & 5 \\[1pt]
&  &  & $\{e\}$ = trivial & \hspace{-0.4cm}$SU(2) \!=\! S^3$ & 6 \\[1pt]
\midrule

\multirow{8}{*}{\makecell{\small \textbf{Zero}\\\small \textbf{momentum}}} 
& \multirow{8}{*}{$SL(2,\mathbb C)$} 
& \multirow{8}{*}{\makecell{$p^{\mu}=0$}}
& $SL(2,\mathbb{C})$  = $\ell_p$& $\P=0$ (trivial) & 0 \\[1pt]
&  &  & $ISO(2)$ & $\mathbb R^+ \times S^2$  & 3 \\[1pt]
&  &  & $\mathbb R^+\times$ U(1) & \hspace*{-0.5cm}\small $SL(2,\mathbb{C})/ \mathbb R^+\!\times U(1)$ & 4 \\[1pt]
&  &  & $\mathbb R^2 $ & \small $SL(2,\mathbb{C})/ \mathbb R^2$ & 4 \\[1pt]
&  &  & $\mathbb R^+$ & \small $SL(2,\mathbb{C})/ \mathbb R^+$ &  5 \\[1pt]
&  &  & $\mathbb R^+$ &  \small  $SL(2,\mathbb{C})/ \mathbb R^+$ &  5 \\[1pt]
&  &  & U(1)  &  \small $SL(2,\mathbb{C})/ U(1)$ & 5 \\[1pt]
&  &  & $\mathbb R $  &   \small $SL(2,\mathbb{C})/ \mathbb R$ & 5 \\[1pt]
&  &  & $\{e\}$ = trivial& \small $SL(2,\mathbb C)$ & 6  \\[1pt]
\bottomrule
\end{tabular}
\caption{Classification of Poincaré and BMS little groups with their corresponding orbits and shell dimensions, mostly based on \cite{McCarthy_75,Bekaert:2025kjb}. Tachyonic orbits are not indicated here (and the corresponding little group $SL(2,\mathbb{R})$ must thus be added to list).  In total, there are 11 inequivalent admissible BMS little groups. Note that their are two different, inequivalent, realizations of the little group $\mathbb{R}^+$.}
\label{table:classification bms little groups}
\end{table}

\subsection{Hard representations}
We already encountered hard representations in Section \ref{sec:hardBMS_reps}, which are summarized in Table \ref{tbl:Dictionary}. From the perspective of the general theory, they are very special: their asymptotic little group $\ell_{\P}^{hard}$ has maximal dimension\footnote{Leaving aside the degenerate case $\ell_{\P}^{hard} = SL(2,\mathbb{C})$ corresponding to $\P=0$.},
\begin{equation}
    \ell_{\P}^{hard} = \ell_p= \begin{cases}
        SU(2) \\ ISO(2) \\
        SL(2,\mathbb{R})
    \end{cases}.
\end{equation}
This is a very special property which singles out hard representations from other (generic) representations and places them in one-to-one correspondence with Poincaré UIRs. In fact, the property \eqref{eq:samelittle} can almost be taken as a definition for hard representations; see \cite{Bekaert:2025kjb} for more details.

\subsection{Soft representations}
\label{sec: soft reps BMS}
Another type of representations which are quite special, but less so than the hard ones, are the ``soft'' representations\footnote{This terminology of ``hard'' and ``soft'' is chosen to align as closely as possible with the existing literature (e.g., \cite{Strominger:2017zoo} and references therein). In QFT, the designation ``soft'' is determined by the relevant energy scale; in the present context, it is meant in the sense specified in \eqref{eq:BMS_soft}.}. A representation is called soft if its momentum is zero, i.e. 
\begin{equation}\label{eq:BMS_soft}
    \pi(\P) = p_{\mu}  = 0_{\mu}\,.
\end{equation}
One can show (see \cite{Bekaert:2025kjb}) that BMS soft supermomenta must be of the form \begin{equation}
    \P(z,\bz) = \partial_z^2 \partial_{\bz}^2 \N (z,\bz)\,,
\end{equation}
for some function $\N(z,\bz) \in \E[1]$ on the celestial sphere. In other terms,
\begin{equation}\label{BMS: Equivalence between soft of image of eth}
    \P \;\;\text{is soft} \quad  \Leftrightarrow\qquad p_{\mu}=0 \quad \Leftrightarrow \quad\exists \N \in \E[1] \quad \text{s.t.}\quad \P =  \eth^2\bar{\eth}^2 \N.
\end{equation}
Here, the Paneitz operator $\eth^2 \bar{\eth}^2 $ is the primary operator which, in a chart and for the flat metric, simply reads
\begin{equation}
    \eth^2 \bar{\eth}^2 \left| \begin{array}{ccc}
    \E[1] &  \to & \E[-3]\\
        f &  \mapsto & \partial_z^2 \partial_{\bz}^2f
    \end{array}\right. .
\end{equation}

To prevent any later confusion (see also footnote \ref{footnote:distrib}), let us perhaps emphasize again that, in the equivalence \eqref{BMS: Equivalence between soft of image of eth}, we really mean $\P = \eth^2\bar{\eth}^2 \N$, with $\P \in \E[-3]$, $\N\in \E[1]$ as a \emph{global} statement on the sphere: i.e. not only $\P(z,\bz) = \partial^2_z\partial^2_{\bz}\N(z,\bz)$ in the north patch coordinates, but also $\hat{\P}(\hat{z},\hat \bz) = \partial^2_{\hat{z}}\partial^2_{\hat \bz}\hat{\N}(\hat{z},\bar{\hat{z}})$ in the south patch coordinates $\hat{z}=z^{-1}$, see Appendix \ref{app:global conformal densities} (and Appendix \ref{app: delta function} for a discussion on why the delta function is not soft). In particular, under these assumptions and as a result of the fact that the equality holds globally on the sphere, integration by parts always holds without any further concerns about boundary terms.

An important property of the space of soft supermomenta is that it forms a Hilbert space, with the norm $\|\N\|^2$ induced by the scalar product (see Section 6.4. in \cite{Gelfand2})
\begin{align}
\label{eq:scalar product}
    \langle \partial^2 \N_1, \bar \partial^2 \N_2  \rangle &:= \int d^2z \,\partial^2_z \N_1 \partial^2_{\bz} \N_2\,.
\end{align}
An alternative, useful, form for this Lorentz-invariant norm is in term of the following two-point function \cite{Gelfand2}
\begin{equation}
    \langle \partial^2 \N_1, \bar \partial^2 \N_2  \rangle = \int d^2z_1 \int d^2z_2 \, \partial^2_{z_1} \partial^2_{\bz_1}\N_1\, \partial^2_{z_2} \partial^2_{\bz_2} \N_2 \;q(z_1,\bz_1)\cdot q(z_2,\bz_2)\ln|q(z_1,\bz_1)\cdot q(z_2,\bz_2)|\,.
\end{equation}
Although this expression may appear singular and non-invariant at first sight, it is in fact well defined and invariant; see Appendix \ref{Ssection: apdx 2-point function E[1]} for a discussion.

\subsection{Hard/soft decomposition of supermomenta}\label{section: Hard/soft decomposition of supermomenta BMS}

\subsubsection{Definition}
Let $\P(z,\bz)$ be a generic supermomentum with associated momentum $p_{\mu}$. In general, it will be neither hard nor soft (i.e $p_{\mu} \neq 0 $). However, one can always uniquely decompose it in the form~\cite{Bekaert:2025kjb}
\begin{equation}\label{BMS: hard/soft decomposition}
    \P(z,\bz) = P(z,\bz)+ \partial^2_z\partial^2_{\bz}\N(z,\bz)\,,
\end{equation}
with $\N(z,\bz)\in \E[1]$ and  $P(z,\bz)$ the hard supermomentum of momentum $p_{\mu}$ (see Table \ref{tbl:Dictionary}).  This decomposition is unique and Lorentz-invariant. We will not repeat the proof here, as it is very similar to the one for QED and can moreover be found in~\cite{Bekaert:2025kjb}.

It is perhaps worth stressing that the decomposition \eqref{BMS: hard/soft decomposition} presented here has essentially nothing to do with a spherical harmonics decomposition, $\P = \P(z,\bz)\big|_{\ell =0,1}+\P(z,\bz)\big|_{\ell\geq 2}$. The decomposition \eqref{BMS: hard/soft decomposition} is Lorentz-invariant and nonlinear, whereas the spherical harmonics decomposition is linear but not Lorentz-invariant. In this sense, the two decompositions are as far apart as they can be.

\subsubsection{Non-linearity of the decomposition}
As we have just mentioned, an essential feature of the decomposition \eqref{BMS: hard/soft decomposition} is that it is nonlinear. If $\P_1$ and $\P_2$ are two supermomenta,
\begin{align}
        \P_1 =P_1(z,\bz)+\partial^2_z\partial^2_{\bz}\N_1(z,\bz) \virg   \P_2 =P_2(z,\bz)+\partial^2_z\partial^2_{\bz}\N_2(z,\bz)\,,
\end{align}
then
\begin{equation}
    \P_1 +\P_2 = P_3(z,\bz) + \partial^2_z\partial^2_{\bz} \big( \N_1(z,\bz)  + \N_{2}(z,\bz) + \mathscr S(z,\bz) \big)\,,
\end{equation}
with $P_3(z,\bz)$ the hard supermomentum of momentum $p^{\mu}_3 = p^{\mu}_1 + p^{\mu}_2$, and
\begin{equation}
 \mathscr S(z,\bz) =   -\frac{1}{2\pi}\Big(p_1\cdot q\ln|p_1\cdot q| + p_2\cdot q\ln|p_2\cdot q| - p_3\cdot q\ln|p_3\cdot q|\Big)\,.
\end{equation}
Note that this phenomenon directly arises as a consequence of \eqref{Hard reps dont conserve supermomentum_bms} and is directly responsible for the fact that hard UIRs cannot conserve BMS supermomentum.

\subsection{Wavefunctions and supermomentum eigenstates}
\label{sec:supermomentum eigenstates GR}
Let $|\P\rangle$ be a supermomentum eigenstate, $\hat{\P}(z,\bz)|\P\rangle = \P(z,\bz)|\P\rangle$. Since the decomposition \eqref{BMS: hard/soft decomposition} is unique, we are not loosing any information by rewriting this state as $|\P\rangle = |p , \partial_z^2\mathscr N\rangle$. By definition, it satisfies
\begin{align}
    \hat{\P}(z,\bz)|p , \partial_z^2\mathscr N\rangle = \left(P(z,\bz) +\partial_{\bz}^2\partial_z^2\mathscr N \right)|p , \partial_z^2\mathscr N\rangle\,.
\end{align}
Just like momentum eigenstates, supermomentum eigenstates are singular states which do not have a finite norm. The one-particle Hilbert space realizing the UIR is given be the space of normalizable states of the form\footnote{For simplicity we restrict ourselves to scalar states, see \cite{Bekaert:2025kjb} for more details.}
\begin{equation}\label{New state for GR}
    |\Psi\rangle = \int_{SL(2,\mathbb{C})/\ell_{\P}} \frac{d M}{\text{Vol}(\ell_{\P})} \Psi(M)|M\cdot p , M\cdot \partial_z^2\mathscr N\rangle\,,
\end{equation}
where $M\in SL(2,\mathbb{C})$, $dM$ is the Haar measure and the wavefunction $\Psi(M)$ must satisfy $\Psi(MB) = \Psi(M)$ for any $B\in \ell_{\P}$. The norm of the state is simply
\begin{equation}
    \langle \Psi|\Psi\rangle = \int_{SL(2,\mathbb{C})/\ell_{\P}} \frac{d M}{\text{Vol}(\ell_{\P})} |\Psi(M)|^2\,.
\end{equation}

For hard representations, this reproduces the usual scalar wavefunction $\Psi(M)=\Psi(k)$, where we introduced $k^{\mu} = M^{\mu}{}_{\nu}p^{\nu}$. These are functions on
\begin{align}
   \frac{SL(2,\mathbb{C})}{\ell_{p}}=\begin{cases}
        H_3 &= \frac{SL(2,\mathbb{C})}{SU(2)}, \qquad\text{if}\; p^2<0\\  \mathbb{R}^+\times S^2 &= \frac{SL(2,\mathbb{C})}{ISO(2)}, \qquad\text{if}\; p^2=0
   \end{cases}.
\end{align}
However, for generic representations, the BMS little group $\ell_{\P}$ is strictly smaller than the Poincaré little group $\ell_{p}$ and the wavefunctions $\Psi(M)$ are functions on the total space of a fibre bundle
\begin{equation}
    \frac{SL(2,\mathbb{C})}{\ell_{\P}} \to \frac{SL(2,\mathbb{C})}{\ell_{p}}\,,
\end{equation}
whose typical fibre $F = \ell_{\P}/\ell_{p}$ encodes the extra degrees of freedom of the representation (as compared to the corresponding hard one). Note that, while they have more degrees of freedom than the usual hard wavefunctions, BMS wavefunctions belong to a separable Hilbert space (since they are $L^2$ functions on a finite-dimensional manifold). In that sense, the Hilbert space to which these new ``particles'' belong is mathematically just as well behaved as the usual one.

\paragraph{BMS particles}
As we previously argued, if the BMS group \eqref{eq:BMS} is a symmetry of the $S$-matrix of gravity, then hard representations cannot be the end of the story since they cannot conserve supermomentum (see the discussion in Section  \ref{subsec:hard_not_enough_bms}). Since, as we shall review in the next section, Weinberg's soft theorem for gravity is in fact equivalent to the conservation of supermomentum, it strongly suggests\footnote{The alternative would be that the symmetry is broken is a rather non-standard way, likely implying along the way that the Lorentz group itself is broken; see the discussion at the end.} that the Hilbert space of asymptotic states must be extended to include new representations, including states of the form \eqref{New state for GR}. This is in complete parallel with the situation for QED and, here again, the idea of extending the Hilbert space of gravity is not new, see e.g. \cite{Ashtekar:1987tt,Ware:2013zja,Choi:2017bna,Prabhu:2022zcr,Prabhu:2024lmg,Prabhu:2024zwl}.  Rather, the point that we want to make here is that the representation theory of the BMS group provides an appropriate and natural framework to realize this extension as a Fock space built from a new notion of particles (which, as already emphasized many times, extends beyond hard UIRs). Since there are infinitely many representations of the  BMS group --- as many as $\E[-3]/SL(2,\mathbb{C})$, with the hard ones forming only a tiny special corner --- additional physical input is required to avoid getting lost in this vast landscape. We now turn to IR physics, which should provide precisely such input.

\section{Infrared divergences in hard scattering processes}
\label{sec: IR div gravity}
The upshot of Section \ref{sec:hardBMS_reps} was the following: while every usual QFT particle can be thought of as a hard BMS UIR, hard representations alone \emph{cannot} by themselves conserve supermomentum. We will in fact now see that the quantity $\mathscr S(z,\bar z)$ given in \eqref{BMS: soft factor} that characterizes the obstruction for supermomentum conservation, directly relates both to real and virtual infrared divergences for the scattering of conventional particles:  $\mathscr S(z,\bar z)$ gives the soft factor of Weinberg's theorem, while its Lorentz-invariant norm controls the exponentiated virtual divergences. 

\subsection{Gravitons beyond the hard Fock space}
Gravitons carry a non-zero soft BMS charge. Indeed, the Noether analysis for supertranslations, parametrized by $\mathcal T(z,\bar z)$, leads to a non-vanishing supertranslation charge~\cite{Barnich:2010eb,He:2014laa}. For a graviton field $h_{\mu \nu}$ expanded in modes as in \eqref{eq: spin s} with $s=2$ and $k=\sqrt{32\pi G}:= \kappa$, the charge is given by $\mathcal F_\T=\int d^2 z \,\T(z,\bz) \,\p_z^2 \p_\bz^2\hat{\N}^{\text{gr}}(z,\bz)$ with\footnote{Note that $\lim_{\omega \to 0} \omega\,\left( \hat a_-(\omega,z,\bz)^{\dagger} + \hat a_+(\omega,z,\bz)  \right)$ has conformal weight $w=-1$ and spin-weight $s=-2$ and that, since $\eth^2 : (w=1,s=0) \to (w=-1,s=-2)$ is surjective (see \cite{Penrose:1985bww} and our conventions in Appendix \ref{app:conformal_density}), this can always be written as $\eth^2 \hat{\N}$ with $\hat \N \in \mathcal E[1]$. Since the Noether charge is real by construction, $\hat{\N}$ is Hermitian and thus
\begin{equation}\label{eq:Psoft alt}
    \partial_z^2\partial_{\bz}^2 \hat{\N}^{\text{gr}}(z,\bz)=  \frac{1}{2\pi \kappa} \p^2_z \left( \,\lim_{\omega \to 0} \omega\,\left( \hat a_+(\omega,z,\bz)^{\dagger} + \hat a_-(\omega,z,\bz)  \right)\,\right).
\end{equation}
Imposing the corresponding reality condition is equivalent to assuming that the gravitational soft magnetic charge
\begin{equation}\label{eq:Psoft magnetic}
4\pi i \kappa\, \hat \P^M (z,\bz) := \p^2_\bz \left( \,\lim_{\omega \to 0} \omega\,\left( \hat a_-(\omega,z,\bz)^{\dagger} + \hat a_+(\omega,z,\bz)  \right)\,\right) - \p^2_z \left( \,\lim_{\omega \to 0} \omega\,\left( \hat a_+(\omega,z,\bz)^{\dagger} + \hat a_-(\omega,z,\bz)  \right)\right)
\end{equation} vanishes. In practice, it seems enough for us to suppose that $\hat \P^M (z,\bz)|\psi\rangle=0$ on all our states; see \cite{Kol:2019nkc} for possible alternative constraints that one might wish to impose.}
\begin{equation}\label{eq:Pgr}
    \p_z^2 \p_\bz^2\hat{\N}^{\text{gr}}(z,\bz) =   \frac{1}{2\pi \kappa}\p^2_\bz  \lim_{\omega \to 0} \Big( \omega\,\hat a_-(\omega,z,\bz)^{\dagger} + \omega\,\hat a_+(\omega,z,\bz)  \Big)\,.
\end{equation}
And indeed, together with the operators \eqref{eq:Poin_gen2} and \eqref{eq:hardP}, it forms a basis \begin{equation}
    \left( \hat{\P}(z,\bz):= \hat{P}(z,\bz) + \partial_z^2\partial_{\bz}^2 \hat{\N}^{\text{gr}}(z,\bz), \hat{J}^{\mu\nu}\right)
\end{equation}
of generators of the BMS algebra \eqref{BMS algebra representation}. 

However, in order for the asymptotic symmetry to be non-trivially represented on the Hilbert space, one must make a more radical step. To see this, let $|\psi\rangle := \hat{\psi} |0 \rangle$ be a graviton state, with $|0 \rangle$ the Poincaré invariant vacuum and
\begin{equation}\label{gravitons are different: normalizable state}
\badat{2}
  \hat{\psi} &=   \kappa  \int \frac{d^3k}{(2\pi)^3 2k^0} \,\left( \psi_-(k) \hat{a}^{\dagger}_{+}(k) + \psi_+(k) \hat{a}^{\dagger}_{-}(k)  \right)\\
   &=  \frac{\kappa}{16 \pi^3} \int \omega d\omega d^2z \,\left( \psi_-(\omega,z,\bz) \hat{a}^{\dagger}_{+}(\omega,z,\bz) + \psi_+(\omega,z,\bz) \hat{a}^{\dagger}_{-}(\omega,z,\bz)  \right).
  \eadat
\end{equation}
 Then, one has
 \begin{equation}\label{Gravitons are different: action of soft charge}
     \partial_z^2\partial_{\bz}^2 \hat{\N}^{\text{gr}}(z,\bz)|\psi\rangle = \left[ \partial_z^2\partial_{\bz}^2 \hat{\N}^{\text{gr}}(z,\bz) , \hat{\psi} \right] | 0\rangle + \hat{\psi} \, \partial_z^2\partial_{\bz}^2 \hat{\N}^{\text{gr}}(z,\bz)|0\rangle\,,
 \end{equation}
 where the first contribution only comes from the pole of the wavefunction\footnote{This follows from \eqref{eq:Pgr}, \eqref{gravitons are different: normalizable state} and \eqref{Canonical commutation relation in omega,z}. Note that the requirement that the state has no magnetic charge:  \begin{equation}
  0= \left[ \hat{\P}^{M}(z,\bz) , \hat{\psi} \right]  | 0 \rangle  = \frac{1}{4\pi i\kappa} \lim\limits_{\omega\to 0}\omega\Big( \partial^2_{\bz} \psi_-(\omega,z,\bz) - \partial^2_{z} \psi_+(\omega,z,\bz) \Big)  \;  | 0 \rangle\,,
\end{equation} means that $\lim\limits_{\omega\to 0}\omega \partial^2_{z} \psi_+(\omega,z,\bz) = \lim\limits_{\omega\to 0}\omega \partial^2_{\bz} \psi_-(\omega,z,\bz) $.}
\begin{equation}
   \left[ \partial_z^2\partial_{\bz}^2 \hat{\N}^{\text{ph}}(z,\bz) , \hat{\psi} \right]  | 0 \rangle  = \frac{1}{2\pi} \lim\limits_{\omega\to 0}\omega \,\partial^2_{\bz} \psi_-(\omega,z,\bz) \;  | 0 \rangle.
\end{equation}
However, for any state of finite norm
\begin{align}\label{Gravitons are different: finite norm of hard state}
\langle 0 | \hat{\psi}^{\dagger} \hat{\psi} | 0 \rangle &=\frac{\kappa^2}{16\pi^3}\int \omega d\omega d^2z\,\left(|\psi_{+}(\omega,z,\bz)|^2+ |\psi_{-}(\omega,z,\bz)|^2 \right) < \infty\,,
\end{align}
the first term in \eqref{Gravitons are different: action of soft charge} must in fact be zero, while the second term can be non-zero only if some sort of ``spontaneous symmetry breaking''\footnote{This wording should be taken with caution in this context, as the degeneracy of the vacuum does not imply that the $S$-matrix factorizes into ``superselection sectors''; see \cite{Strominger:2017zoo} and the discussion at the end.} occurs (since in that case the symmetry generator acts non-trivially on the vacuum). Therefore, the asymptotic symmetry algebra can be represented non-trivially only in one of two ways: i) by enlarging the Hilbert space (as compared to the Fock space of hard particle representations), or ii) by invoking spontaneous symmetry breaking. Both perspectives have appeared throughout the literature, often in alternation and not always clearly distinguished. The latter viewpoint underpins the discussion of asymptotic symmetries in the context of soft theorems and relates to infrared divergences of the $S$-matrix~\cite{He:2014laa,Strominger:2017zoo,Agrawal:2025bsy}, while the first scenario is essentially the core of the (gravity analogue of the) infrared-finite Faddeev-Kulish construction~\cite{Ashtekar:1987tt,Ware:2013zja,Choi:2017bna,Choi:2017ylo,Prabhu:2022zcr,Prabhu:2024lmg,Prabhu:2024zwl}.

\paragraph{Relationship with the gravitational memory effect}  

When a gravitational wave passes on a test particle at rest, it will in general induce a memory effect \cite{mem1,Christodoulou:1991cr}: there is a permanent displacement remaining after the wave has passed. This effect can be directly related to the pole in $\omega$ in $h_{\mu\nu}(\omega,z,\bz)$ \cite{Strominger:2014pwa}, i.e. precisely to the contribution that would be given by the first term in \eqref{Gravitons are different: action of soft charge}. However, while this is a perfectly reasonable classical effect, it cannot be accommodated by the usual hard states, due to \eqref{Gravitons are different: finite norm of hard state}. This problem, namely the impossibility for the quantum Hilbert space to realize a perfectly legitimate classical observable\footnote{Which was much later understood to be a memory effect.}, was pointed out very early on by Ashtekar \cite{Ashtekar:1987tt} as the source of IR divergences. In this sense, the possibility of implementing the asymptotic symmetries \eqref{eq:BMS} on the Hilbert space directly relates to the possibility of encoding memory at quantum level and to IR divergences.

\subsection{Soft graviton theorem as supermomentum conservation}
\label{section: soft graviton theorem}
Consider a scattering process  $\langle \text{out} | \hat{S} | \text{in} \rangle $ in momentum basis involving $n$ incoming and $m$ outgoing particles.
Supermomentum-invariance of the $S$-matrix reads
\begin{equation}\label{eqref:WI_gr}
    \langle \text{out} |\hat{\P}^{\text{out}}(z,\bz)\hat{S}-\hat{S} \hat{\P}^{\text{in}}(z,\bz) | \text{in} \rangle = 0\,,
\end{equation}
with $\hat{\P}(z,\bz)$ the total (incoming or outgoing) supermomentum operator.

Let us focus on the case of real scalar particles \eqref{real_scalar} of momentum $p_i$ (the case of spinning particles follows in a similar way).  Conventional asymptotic states belong to the Fock space associated with the free field operators $b(\vec p)$, $b^\dagger(\vec p)$.
As we have seen in Section~\ref{sec:scalar_BMS}, scalars have a non-zero supermomentum operator given by
\begin{equation}\label{eq:Pin}
    \hat{P}(z,\bar z) = \int  \frac{d^3p}{(2\pi)^32p^0}\, P(z,\bz)\,\hat b^\dagger(\vec p) \hat b(\vec p)\,,
\end{equation}
with $P(z,\bz)$ given by \eqref{eq:hardP}.
The incoming hard operator acts on incoming states $| \text{in} \rangle =|p_1^\text{in}; \dots; p_n^\text{in} \rangle$ as
\begin{equation}
    \hat{P}^\text{in}(z,\bar z) | \text{in} \rangle=\sum_{i=1}^n P^\text{in}_i(z,\bz)| \text{in} \rangle\,,
\end{equation}
where $P_i(z,\bar z)$ is the hard momentum of the $i$-th incoming particle, namely 
\begin{equation} 
P_i(z,\bar z) =
\left\{\begin{aligned}
  &\hspace{0.5cm} \quad \omega_i\,\delta^{(2)}(z-\zeta_i) & \quad \text{if } &p_i^2=0,  \;p^{\mu}_i = \omega_i q^{\mu}_i(\zeta_i,\bar{\zeta}_i)\\[0.6em]
  &\hspace{0.5cm} -\dfrac{m_i^4}{4\pi\,(q(z,\bar z)\cdot p_i)^3} & \quad\, \text{\,if } &p_i^2=-m_i^2\,.\\[0.4
  em]
\end{aligned}\right.
\end{equation}

Hence, for a scattering event involving gravitationally interacting real scalars, the total incoming supermomentum is the sum of \eqref{eq:Pgr} and \eqref{eq:Pin},
\begin{equation}\label{eq: sP in}
    \hat{\P}^\text{in}(z,\bz) = \hat{P}^\text{in}(z,\bar z)+ \frac{1}{2\pi \kappa}\p^2_\bz  \lim_{\omega \to 0^+}  \omega \big(\hat a^\text{in}_-(\omega,z,\bz)^{\dagger} + \hat a^\text{in}_+(\omega,z,\bz)  \big) \,,
\end{equation}
where only the gravitational field contributes to the second term. A similar expression holds for the total outgoing supermomentum.\\
We now want to use the fact that, for both massive and massless cases, $P_i(z,\bz)$ can be identically rewritten as
\begin{equation}\label{eq:Pfact}
P_i(z,\bz)=-\frac{1}{2\pi}\p^2_\bz \left( \frac{(\varepsilon^+(z,\bz)\cdot p_i)^2}{q(z,\bz)\cdot p_i}\right)+p^\mu_i \mathcal D_\mu\big(\delta^{(2)}(z-\infty,\bz-\infty)\big)\,.
\end{equation}
The above equality holds in the sense of distributions and is proven in Appendix \ref{app:distrib_id} (see also Proposition 3.5 in \cite{Bekaert:2025kjb}.).
Using this together with \eqref{eq:Pin}, \eqref{eq: sP in} and $\displaystyle \lim_{\omega \to 0} \omega\, \langle \text{out} | \hat a_+^{\text{out}}\hat{S}   | \text{in} \rangle = - \displaystyle\lim_{\omega \to 0} \omega\, \langle \text{out} | \hat{S} \hat a_-^{\text{in} \dagger}   | \text{in} \rangle$, the Ward identity \eqref{eqref:WI_gr} becomes
\begin{equation}\label{eq:eu2}
\p^2_\bz\left(\; \frac{2}{\kappa}\lim_{\omega \to 0} \omega\, \langle \text{out} | \hat a_+^{\text{out}}(\omega,z,
\bz)\hat{S}   | \text{in} \rangle+  \left(\sum_{i\in \text{in}} \, \frac{(\varepsilon^+\cdot p_i)^2}{q\cdot p_i} - \sum_{i\in \text{out}}\, \frac{(\varepsilon^+\cdot p_i)^2}{q\cdot p_i} \right)\langle \text{out} | \hat{S} | \text{in} \rangle  \right) =0\,.
\end{equation}
Notice that the distributional terms supported at $|z|=\infty$ have canceled out in the total sum by virtue of momentum conservation.
The equality \eqref{eq:eu2} is \emph{equivalent} to Weinberg's soft graviton theorem~\cite{Weinberg:1965nx}
\begin{equation}\label{eq:soft_graviton_theorem}
\lim_{\omega \to 0} \omega\, \langle \text{out} | \hat a_+^{\text{out}}(\omega,z,
\bz) \hat{S} | \text{in} \rangle= -\frac{\kappa}{2}\left(\sum_{i\in \text{in}} \frac{(\varepsilon^+\cdot p_i)^2}{q\cdot p_i} - \sum_{i\in \text{out}} \frac{(\varepsilon^+\cdot p_i)^2}{q\cdot p_i}\right)\langle \text{out} | \hat{S} | \text{in} \rangle  \,.
\end{equation}
The equivalence follows from the fact that the $\bar \eth^2$ operator is \emph{injective} when acting on a density $f$ of spin-weight $s=-2$, hence  $\bar \eth^2 f=0$ implies $f=0$ (see Proposition (4.15.59) in \cite{Penrose:1985bww} and our conventions \ref{app:conformal_density}).

The above offers a revisiting of the (by now classical) argument for the equivalence between the soft graviton theorem and the Ward identity associated with supertranslation symmetries, with the advantage of treating, at the same time, both the massless \cite{He:2014laa} and massive \cite{Campiglia:2015kxa} cases. The derivation is also substantially shorter, as it directly shows the equivalence (thanks to the property of the edth operator), as opposed to the original two-steps derivation that make use of a convenient choice of supertranslation parameter. 

The derivation presented here also makes very clear that, from the point of view of BMS representation theory, the soft graviton theorem is simply the manifestation of the fact that a scattering process involving $N$ Poincaré particles which satisfy momentum conservation fails to conserve supermomentum conservation (see Section \ref{subsec:hard_not_enough_bms}). The soft factor
\begin{equation}\label{eq:soft_factor_gr}
\sum_{i=1}^N  \eta_i  \, \frac{(\varepsilon^+\cdot p_i)^2}{q\cdot p_i} = -2\pi \,\p^2_z \mathscr S(z,\bz) \virg \mathscr S(z,\bz) =-\frac{1}{2\pi}\sum_{i=1}^N \eta_i\, (q \cdot p_i) \ln |q\cdot p_i|\,,
\end{equation}
exactly provides the missing contribution required for the supermomentum conservation law to hold; see equations \eqref{Hard reps dont conserve supermomentum_bms} and \eqref{BMS: soft factor}.

\subsection{Exponentiation formula for virtual divergences}
\label{sec:virtual}
On top of the divergences arising when attaching an external soft graviton, scattering amplitudes are plagued with IR divergences arising from the exchange of virtual gravitons between external legs. In the seminal paper \cite{Weinberg:1965nx} (see also \cite{Weinberg:1995mt}), Weinberg showed that these divergences factorize in the amplitude $\mathcal A$ according to the formula\footnote{These divergences are one-loop exact \cite{Weinberg:1965nx}; see also \cite{Naculich:2011ry}.}
\begin{equation}
    \mathcal A= e^{\mathcal W} \mathcal A^{I.F.}_\Lambda\,,
\end{equation}
where $\mathcal A^{I.F.}_\Lambda$ is infrared finite and the exponent is
\begin{equation}
\label{weinfact_gr}
\mathcal W= -\frac{G}{2\pi}\log\Big(\frac{\Lambda}{\lambda}\Big) \sum_{i,j} \eta_i \eta_j\, m_i m_j\,  \frac{1+\beta_{ij}^2}{\beta_{ij}\sqrt{1-\beta_{ij}^2}}  \left(\frac{1}{2} \ln \frac{1+\beta_{ij}}{1-\beta_{ij}} -i\pi \delta_{\eta_i,\eta_j} \right) \,,
\end{equation}
where $\Lambda$ sets the soft scale, $\lambda$ is a regulator cutoff and $\beta_{ij}$ is the relative velocity of particle $i$ and $j$ defined in \eqref{betaij}.
The imaginary part of $\mathcal W$ is often referred to as the Coulombic phase divergence (it does not affect the decay rate), while the real part is the term that cancels the divergences associated with the emission of soft gravitons in inclusive cross-sections.
It was shown in \cite{Himwich:2020rro} that the above soft factorization theorem for gravity is the same as a current algebra factorization on the celestial sphere. There, the exponentiated soft divergence was reproduced from the expectation values of Wilson-line inspired operators, and the supertranslation Goldstone current algebra level was identified with the gravitational cusp anomalous dimension. 

As we will now prove, the real part of the infrared factor is, in fact, for both massive and massless particles, 
\begin{equation}
\label{eq:norm}
    \Re( W)  = -2G\log\Big(\frac{\Lambda}{\lambda}\Big) \,\| \mathscr S \|^2\,,
\end{equation}
where $\mathscr S(z,\bz)$ is the obstruction to supermomentum conservation, defined in \eqref{BMS: soft factor},
and $\|\mathscr S\|^2 = \int d^2z \,\partial^2_z \mathscr S \partial^2_{\bz} \mathscr S $ its Lorentz-invariant norm squared \eqref{eq:scalar product}\footnote{For massless particles, it was previously noted in \cite{Agrawal:2025bsy} that the Goldstone two-point function of \cite{Himwich:2020rro} relates to the inner product of unitary discrete series representations of the Lorentz group.}.

To see this, we first recall that the real part of \eqref{weinfact_gr} can be written as~\cite{Weinberg:1965nx} 
\begin{equation}
\label{weinfact2}
\Re(W)=  -\frac{G}{4\pi^2} \log\Big(\frac{\Lambda}{\lambda}\Big) \sum_{i,j} \eta_i\eta_j \,\int d^2 z\, \frac{(p_i\cdot p_j)^2 - \frac{1}{2}m_i^2m_j^2}{(q\cdot p_i)\; (q \cdot p_{j})}\,.
\end{equation}
This formula holds for any particle, independently of whether it is massive or massless. To show that the norm of $\mathscr S$ is proportional to \eqref{weinfact2}, we write 
\begin{align}    \label{eq:norm compu GR}
\| \mathscr S \|^2\, &=\frac{1}{4\pi^2}
     \sum_{i,j} \eta_i\eta_j\; \int d^2 z \,\partial_z^2\Big( p_i\cdot  q (z,\bz) \ln|p_i \cdot q(z,\bz)|\Big)\, \partial_{\bz}^2\Big( p_j\cdot  q (z,\bz) \ln|p_j \cdot q(z,\bz)|\Big) \nonumber\\
    &=  \frac{1}{4\pi^2} \sum_{i,j} \eta_i\eta_j\;\int d^2 z \, \frac{(\varepsilon^+ \cdot p_i)^2}{q\cdot p_i}\, \frac{(\varepsilon^- \cdot p_j)^2}{q\cdot p_j} \\
    &= \frac{1}{8\pi^2} \sum_{i,j} \eta_i\eta_j\;\int d^2 z \, \frac{\left( 2p_i^{\mu} p_j^{\nu}\; \varepsilon^+_{(\mu}\varepsilon^-_{\nu)}\right)^2-2|\varepsilon^+ \cdot p_i|^2|\varepsilon^+ \cdot p_j|^2}{(q\cdot p_i) \, (q\cdot p_j)}\,. \nonumber
\end{align}
Making use of $p_i\cdot p_j= 2 p_{i}^{\mu}p_{j}^{\nu} \left(\varepsilon^-_{(\mu}\varepsilon^+_{\nu)} - q_{(\mu} \,\partial_z\partial_{\bz}q_{\nu)}\right)$ and several applications of momentum conservation $\sum_i \eta_i p^{\mu}_i=0$ yield
\begin{equation}
\| \mathscr S \|^2\,=  \frac{1}{8\pi^2}\sum_{i,j} \eta_i\eta_j \,\int d^2 z\, \frac{(p_i\cdot p_j)^2 - \frac{1}{2}m_i^2m_j^2}{(q\cdot p_i)\; (q \cdot p_{j})}\,,
\end{equation}
and, comparing with \eqref{weinfact2}, the result \eqref{eq:norm}.

As opposed to the situation in QED, the formula is well-behaved even for massless particles. To see this, let us suppose that the $k$-th particle is massless, $p_k^{\mu} = \omega_k q^{\mu}(z_k,\bz_k)$. Integrating by part the first line of \eqref{eq:norm compu GR}, which is allowed since $\mathscr S \in \E[1]$ is a global function on the celestial sphere (see the discussion in Section \ref{subsec:hard_not_enough_bms}), and making use of \eqref{BMS: non linearity and distributional identity}, we find\footnote{Notice that the distributional terms supported at $|z|=\infty$ drop in the second line by virtue of momentum conservation.}
\begin{equation}
    \badat{2}
   \|\mathscr S\|^2 & = \frac{1}{4\pi^2}
     \sum_{i,j} \eta_i\eta_j\; \int d^2 z \,\partial_{\bz}^2\partial_z^2\Big( p_i\cdot  q (z,\bz) \ln|p_i \cdot q(z,\bz)|\Big)\, \Big( p_j\cdot  q (z,\bz) \ln|p_j \cdot q(z,\bz)|\Big)\\
    &=-\frac{1}{2\pi}
     \sum_{i,j} \eta_i\eta_j\; \int d^2 z \,P_i\, \Big( p_j\cdot  q (z,\bz) \ln|p_j \cdot q(z,\bz)|\Big)\,.
\eadat
\end{equation}
Now if, the $k$-th particle is massless, $P_k(z,\bz) = \omega_k \delta^{(2)}(z-z_k)$, then the corresponding contribution is
\begin{align*}
    -\, \frac{\eta_k}{2\pi} \;\sum_{j} \eta_j \int d^2 z \, P_k(z,\bz)\,p_j \cdot q(z,\bz)\,\ln|p_j \cdot q(z,\bz)| &= -\frac{ \eta_k \omega_k}{2\pi}\;\sum_{j} \eta_j\,  p_j\cdot  q (z_k,\bz_k)\,\ln|p_j \cdot q(z_k,\bz_k)|\\
    &= -\frac{ \eta_k}{2\pi}\;\sum_{j} \eta_j\,  p_j\cdot  p_k \,\ln|p_j \cdot p_k |\,.
\end{align*}
This expression is well-behaved\footnote{The argument of the $\log$ should contain a mass scale $\mu$ to be dimensionless, $\ln(\frac{p_i \cdot p_j}{\mu^2})$, but the latter can be dropped from by virtue of momentum conservation.}, even for the potentially dangerous $j=k$ term, and one thus recovers the fact that no further divergences arise for gravity; see~\cite{Weinberg:1965nx}. One also deduces from this discussion that if all particles are massless, then
\begin{align}\label{eq:massless}
\Re(W) &=\frac{G}{\pi} \log\Big(\frac{\Lambda}{\lambda}\Big)\; \sum_{i,j} \eta_i\eta_j \, (p_i\cdot  p_j) \ln|p_i \cdot p_j|\,.
\end{align}

\section{Dressed to kill IR divergences}
\label{sec: dressed to kill gravity}
The equivalence between Weinberg's soft graviton theorem and supermomentum conservation lies at the core of the argument supporting the view that the BMS group \eqref{eq:BMS} is a genuine symmetry of the gravitational $S$-matrix~\cite{Strominger:2013jfa,He:2014laa,Campiglia:2015kxa,Strominger:2017zoo}. In which case, as we argued in Section \ref{subsec:hard_not_enough_bms}, this must imply that the Hilbert space must be extended to include representations of the BMS group that are not hard. In fact, the analogue of the Faddeev–Kulish dressed states for gravity -- which allow for IR-finite $S$-matrix elements -- precisely provide examples of states which lie outside of the conventional free Fock space~\cite{Ware:2013zja}. 

The main purpose of this section is to compare and contrast the dressed state construction for gravity, where charged states are dressed with clouds of gravitons, with supermomentum eigenstates, which we introduced in Section \ref{sec:supermomentum eigenstates GR}.
In Section \ref{sec: dressed state approach BMS}, we recall that the dressed-state construction leads to asymptotic states which are eigenstates of the soft BMS charge \cite{Choi:2017bna,Choi:2017ylo}. However, general dressed states are not BMS supermomentum eigenstates, for the simple reason that they are not momentum eigenstates. Nevertheless, we show in Section \ref{sec: dressed states vs BMS} that dressed states can be turned into supermomentum eigenstates through a specific limiting procedure. As we will demonstrate, the net effect of the dressing is effectively to \emph{linearize} the BMS supermomentum. Since the nonlinearity of supermomentum constitutes the obstruction to supermomentum conservation, this automatically ensures that, for scattering amplitudes involving such dressed states, momentum conservation become equivalent to supermomentum conservation, and hence to IR-finiteness.

\subsection{The dressed-state approach}
\label{sec: dressed state approach BMS}

In this section, we review the dressed-state approach to obtain IR-finite scattering amplitudes in perturbative quantum gravity; first the direct analogue of FK dressing \cite{Ware:2013zja} and then its generalizations \cite{Choi:2017bna,Choi:2017ylo}.

\subsubsection{Faddeev-Kulish (FK) dressing for gravity}
The FK method to construct IR-finite $S$-matrix elements was generalized for perturbative quantum gravity by Akhoury, Saotome and Ware~\cite{Ware:2013zja}. 
The dressing for a single particle state $|\vec p\,\rangle=\hat b^{\dagger}(\vec 
p)  |0\rangle$ of momentum $p^\mu(\omega, \zeta,\bar \zeta) $ is 
\begin{align}\label{eq:dressed_gr}
     |\vec p\,\rangle_{\text{FK}}=\, e^{\hat{R}}\,|\vec p\,\rangle,
\end{align}
with the dressing operator
\begin{equation}\label{R_factor_gr}
\badat{2}
  \hat{R}&=\frac{\kappa}{2} \int \frac{d^3k}{16\pi^3 k^0} \,f^{\mu\nu}(k,p)\left[ \hat a^{\dagger}_{\mu\nu}(k) - \hat a_{\mu\nu}(k)\right]\,,
  \eadat
\end{equation}
where $k^{\mu}= \varpi q^{\mu}(z,\bz)$ denotes the graviton momentum.
The dressing factor $f^{\mu\nu}(k,p)$ takes the form
\begin{equation}
\label{f term gravity}
    f^{\mu\nu}(k,p)=\bigg(\frac{p^\mu p^\nu}{p\cdot k}+c^{\mu\nu}\bigg)\psi(k,p)\,,
\end{equation}
where $\psi(k,p)$ is a smooth function such that $\psi(k,p)=1$ in a neighborhood of $\varpi=0$. The tensor $c^{\mu\nu}(k,p)$ must be real and satisfy $c^{\mu\nu}k_{\nu}+p^{\mu}=0$, ensuring the transverse condition $f_{\mu\nu}k^{\nu}=0$. 
On top of this, it must satisfy \begin{equation}\label{IR dressing GR: constraint on C}
    2c^{\mu\nu}(p)c_{\mu\nu}(p') - \left(c(p)^{\mu}{}_{\mu}\right)\left( c(p')^{\mu}{}_{\mu}\right) =0 \qquad \forall p, p';
\end{equation} see \cite{Ware:2013zja,Choi:2017bna}.
The anti-Hermitian operator $\hat R$ is said to dress each particle state with a ``cloud'' of gravitons.

Let us now comment on the different choices made for the $c$-tensor.
In \cite{Ware:2013zja}, it was shown that, taking $c^{\mu\nu} \varepsilon_{\mu\nu}^+=0$ in strict analogy with Chung~\cite{Chung:1965zza}, the corresponding $S$-matrix elements are finite. This amounts to the dressing \eqref{R_factor_gr} where $f^{\mu\nu}$ is replaced by
\begin{equation}
    F^{\mu\nu}=\frac{p^\mu p^\nu}{p\cdot k}\,\psi(k,p)\,.
\end{equation}
This dressing can be generalized by dressing each particle separately.

\subsubsection{Generalized dressing}

As we saw, introducing a tensor $c^{\mu \nu}$ in the dressing \eqref{f term gravity} is necessary to ensure that the dressing is gauge-invariant, i.e. $f_{\mu\nu}k^{\nu}=0$. There is no canonical, gauge-invariant, way to fix this $c^{\mu \nu}$ and there is thus an inherent ambiguity when dressing each particle separately.  In this respect, it is more natural to allow each particle to have its own dressing. In that case, a necessary and sufficient condition for the cancellation of IR divergences is~\cite{Choi:2017bna}
\begin{equation}\label{BMS Dressing: IR finite condition}
    2\Delta c_{\mu\nu}\, \Delta c^{\mu\nu} - \left( \Delta c{}^{\mu}{}_{\mu} \right)^2 =0\,,
\end{equation}
where
\begin{equation}\label{BMS Dressing: c conservation}
\Delta c^{\mu\nu}  := \sum_{i \in \text{in}} \hat{c}^{\mu\nu}_i(k,p_i) - \sum_{i \in \text{out}} \hat{c}^{\mu\nu}_i(k,p_i)\,,
\end{equation}
 $c_i$ denoting the dressing of the $i$'th particle and $\hat{c}_i = \lim_{\varpi\to 0} \varpi c_i$ its pole. This condition is in fact automatically satisfied when all $c_i$'s are equal as a result of \eqref{IR dressing GR: constraint on C}.

One might take the viewpoint that it is not necessary for each dressed state $|\vec p_i\,\rangle_{\text{FK}}=\, e^{\hat{R}_i}\,|\vec p_i\,\rangle$ to be separately gauge-invariant, and only require the invariance of the dressed `in' state $|\vec{\text{in}}\,\rangle_{\text{FK}}=\, e^{\hat{R}_{\text{in}}}\,| \vec{\text{in}}\,\rangle$, where now
\begin{equation}\label{FK: f term gravity in}
    f^{\mu \nu}_{\text{in}}(k)= \sum_{i\in \text{in}} \Big(\frac{p^\mu_i p^\nu_i}{p_i\cdot k}+c^{\mu\nu}_i\Big)\psi_i(k,p_i)\,,
\end{equation}
and the constraint following from gauge invariance is $f^{\mu \nu}_{\text{in}}k_\mu=0$.  For QED, the dressing with $c_i =0$, $\psi_i(k,p_i) = \psi(k)$, turned out to have very nice properties (however implying $\sum_{i\in\text{in}}q_i=0$; see Section \ref{sec: dressed state approach QED}). This is not the case for gravity: there is no room for this dressing, since the transverse condition would then read $\sum_{i\in \text{in}} p_i^\mu=0$, setting to zero the total incoming energy. Thus gauge invariance prevents to fix the ambiguity in the dressing by taking $c_i=0$. A different line of reasoning led the authors of \cite{Prabhu:2022zcr} to a very similar conclusion, from which they argued that the gravitational analogue of the Faddeev-Kulish construction fails.\footnote{The essence of the problem being that the Hilbert space constructed in \cite{Prabhu:2022zcr} can only be simultaneously Lorentz invariant and separable if the total incoming charge is zero.} 

\subsubsection{Dressing and asymptotic symmetries}
Writing
$a^{\dagger}_{\mu\nu}(k) = \sum_{\alpha} a^{\dagger}_{\alpha} \varepsilon^{\alpha}_{\mu\nu}$ with the sum running over polarizations ($\alpha=\pm$) and $f_{\alpha} = f^{\mu\nu} \varepsilon^{(-\alpha)}_{\mu\nu}$, the dressing \eqref{R_factor_gr} can be written as
\begin{equation}\label{R_factor_gr2}
\badat{2}
  \hat{R}
  &= \frac{\kappa}{2} \int \frac{\varpi d\varpi d^2z}{16\pi^3} \left( f_+(\hat a^{\dagger}_- -\hat a_+) + f_-(\hat a^{\dagger}_+ - \hat a_-)  \right)\,.
  \eadat
\end{equation}

The dressed state \eqref{eq:dressed_gr} turns out to be an eigenstate of the \emph{soft part} of the supermomentum operator, $\hat{\P}^{soft}(z,\bz) = \partial_z^2 \partial_{\bz}^2\hat{\N}(z,\bz)$~\cite{Choi:2017ylo,Choi:2017bna}. Indeed, making use of the definition of the soft charge \eqref{eq:Pgr}, one finds\footnote{This uses the canonical commutation relations \eqref{Canonical commutation relation in omega,z} and the fact that $[A,B]= c\mathds{1}$ implies $[A,e^B] = c e^B$.}
\begin{equation}
  \hat{\P}^{soft}(z,\bz) \,|\vec p\,\rangle_{\text{FK}} =  \partial_{z}^2 \partial_{\bz}^2\N(z,\bz)\,|\vec p\,\rangle_{\text{FK}}\,,
\end{equation}
where the eigenvalue is given by\footnote{\label{Footnote: reality of FK BMS} We would like to comment on a subtlety in how the reality conditions are imposed here: the eigenvalue appearing in this expression has no reason a priori to be real. In fact, had we instead used the alternative definition \eqref{eq:Psoft alt} for the soft charge, we would have found the complex conjugate eigenvalue
\begin{equation}
   \partial_z^2\partial_{\bz}^2 \N = \frac{1}{2\pi} \lim\limits_{\varpi\to 0}\,\varpi\partial_{z}^2 f_+=  \frac{1}{2\pi}\big( \lim\limits_{\varpi\to 0}\,\varpi\partial_{\bz}^2 f_-\big)^* \,.
\end{equation}
What this means is that a generic choice of dressed state \eqref{eq:dressed_gr} has non-zero soft magnetic charge 
\begin{equation}
\P^M(z,\bz) |\vec p\,\rangle_{\text{FK}} =  \frac{1}{2\pi} \left(\lim\limits_{\varpi\to 0}\,\varpi \,\frac{\partial_{z}^2 f_+ - \partial_{\bz}^2 f_-}{2i}\right)  |\vec p\,\rangle_{\text{FK}}\,.
\end{equation}
Considering dressed states with zero magnetic charge thus yields an unambiguous, real, eigenvalue for the soft charge. This seems to have gone unnoticed in the previous literature.}
\begin{equation}\label{eq:FK_ei BMS}
   \partial_z^2\partial_{\bz}^2 \N = \frac{1}{2\pi} \lim\limits_{\varpi\to 0}\varpi\partial_{\bz}^2 f_- \,.
\end{equation}
As a result, the dressed state with \eqref{f term gravity} has eigenvalue
\begin{align}\label{GR dressing: soft charge from C}
   \partial^2_z\partial^2_{\bz} \N 
   &= \frac{1}{2\pi}   \partial^2_z\partial^2_{\bz}\left( p\cdot q \ln |p\cdot q|\right)  + \frac{1}{2\pi}\;\partial^2_{\bz}\left(  \hat{c}^{\mu \nu}\partial_{z} q_\mu \partial_{z} q_\nu\right) \,.
\end{align}

However, because the operator \eqref{R_factor_gr} inserts infinitely many hard particles, \eqref{eq:dressed_gr}
is however \emph{not} eigenstate of the full supermomentum operator 
$\hat{\mathcal P}(z,\bz)  =\hat{P}(z,\bar z)+\p_\bz^2\p_z^2 \hat \N(z,\bz)$,
simply because it is not even a momentum eigenstate.

\subsection{Dressed states vs supermomentum eigenstates}
\label{sec: dressed states vs BMS}
The point of view developed in this paper is that the asymptotic one-particle states for an IR-finite unitary gravitational $S$-matrix should be given by UIRs of the BMS group. Supermomentum eigenstates $|\P\rangle = |p , \partial_z^2\N\rangle$ were presented in Section \ref{sec:supermomentum eigenstates GR}; they satisfy
\begin{equation}
\hat{\P}(z,\bz)\,|\P\rangle = \big( P(z,\bz) + \partial_z^2\partial_{\bz}^2\N(z,\bz) \big)\,|\P\rangle\,,
\end{equation}
where $P(z,\bz)$ is the hard contribution coming from $|p\rangle$ (see Table \ref{tbl:Dictionary}) and $\partial_z^2\partial_{\bz}^2\N$ the eigenvalue of the soft part of the supermomentum.
In this section, we discuss the relationship between supermomentum eigenstates and dressed states.
As we saw, generalized dressed states are rather intricate objects, with infinitely many hard particles superimposed on the initial one, and as such are not momentum nor supermomentum eigenstates. As we shall now discuss, however, it is possible via a subtle limiting procedure to produce genuine supermomentum eigenstates from dressing. We show that these dressing à la Faddeev-Kulish have the effect of making the supermomentum of the particle \emph{linear} in its momentum, thereby ensuring supermomentum conservation.\\

In order to obtain supermomentum eigenstates, we consider the dressing $\hat{R}$ obtained from \eqref{R_factor_gr} by taking
\begin{equation}\label{BMS: dressing of eigenstate}
    f^{\mu\nu}= \Bigg(\frac{p^\mu p^{\nu}}{p\cdot k}+\frac{\hat{c}^{\mu\nu}}{\varpi}\Bigg)\,\psi_{\epsilon}(\varpi)\,,
\end{equation}
where $\hat{c}^{\mu\nu}$ does not depend on $\varpi$ and $\psi_{\epsilon}(\varpi)$ is defined as in \eqref{definition: psi function}. Taking the limit $\epsilon\to 0$, one can see that the dressing becomes
\begin{equation}\label{R_factorBMS3}
\badat{2}
  \hat{R}  &= i \int d^2z \; \partial_{\bz}^2\partial_z^2 \N \; \hat{\C},
  \eadat
\end{equation}
where
\begin{equation}\label{BMS: FK eigenvalue}
   \partial^2_z\partial^2_{\bz} \N 
   = \frac{1}{2\pi}   \partial^2_z\partial^2_{\bz}\left( p\cdot q \ln |p\cdot q|\right)  + \frac{1}{2\pi}\;\partial^2_{\bz}\left(  \hat{c}^{\mu \nu}\partial_{z} q_\mu \partial_{z} q_\nu\right) 
\end{equation}
and $\hat{\C}(z,\bz) := \frac{1}{2}\left(\hat{C}(z,\bz) +\hat{C}^{\dagger}(z,\bz) \right)$ is the real part of a ``Goldstone operator'' defined
as\footnote{Note that writing the left hand side as $\partial_{\bz} \hat{\phi}(z,\bz)$ is not an assumption since the $\bar{\eth}$ operator is surjective on spin coefficient of spin-weight $s=1$, see \cite{Penrose:1985bww} taking into account our conventions in Appendix \ref{app: eth operator}. Also note that the imaginary part of $\hat{C}$ decouples in \eqref{R_factorBMS3} as a consequence of our constraint that the dressed state has a real eigenvalue \eqref{eq:FK_ei BMS}; equivalently, because we restrict to dressings with zero soft magnetic charge, see footnote \eqref{Footnote: reality of FK BMS}.}
\begin{align}\label{goldstone operator gravity}
\badat{2}
  \partial_{\bz}^2 \hat{C}(z,\bz)&:=  - i \kappa \,\lim_{\epsilon\to 0} \int \frac{d\varpi}{8\pi^2}  \psi_{\epsilon}(\varpi)\,\Big(\hat{a}^{\dagger}_+(\varpi,z,\bz) - \hat{a}_-(\varpi,z,\bz)\Big)\,.
  \eadat
\end{align}
By construction, $\hat{\C}(z,\bz)$ satisfies
\begin{align}\label{Gravity: commutation relation for the goldstone}
    \left[ \hat{p}^{\mu} , \hat{\C}(z,\bz)  \right] &= 0, &   \left[ \partial_z^2\partial_{\bz}^2 \hat{\N} , \hat{\C}(w,\bw)  \right] &= -i\delta^{(2)}(z-w).
\end{align}

Again, we note that the limit \eqref{goldstone operator} is not as innocuous as it might seem: it would certainly yield zero if the integrand were a function (since in that case the integrand would converge to an almost everywhere vanishing function). Thus, for this operator to be non-zero, the creation/annihilation operator must have a distributional contribution around $\varpi=0$. This is not very much of a surprise, but it highlights the sense in which the construction departs from the standard Fock space: for usual normalizable Fock states such as \eqref{Photons are different: normalizable state}, creation/annihilation operators always appear multiplied by $\varpi$ and the distributional contribution thus drops out.

The resulting dressed state
\begin{equation}
\badat{2}
    | \mathcal P \rangle  :&= e^{ \hat R} \;\hat b^{\dagger}(p) |0\rangle\\
   & =e^{i\,\langle \partial^2 \bar{\partial}^2 \N,\hat \C \rangle}\;\hat b^{\dagger}(p) |0\rangle
  \eadat
\end{equation}
is then a supermomentum eigenstate
\begin{equation}
\badat{2}
    \hat{\P}(z,\bz)  | \P \rangle = \big( P(z,\bz) + \partial_z^2\partial_{\bz}^2\N(z,\bz) \big)| \P\rangle,
    \eadat
\end{equation}
where $P(z,\bz)$ is the hard contribution coming from $|p\rangle$ (see Table \ref{tbl:Dictionary}) and the soft contribution is given by \eqref{BMS: FK eigenvalue}. Making use of the identity \eqref{BMS: non linearity and distributional identity}, we find
\begin{equation}\label{BMS dressed eigenstate: eigenvalue}
    \badat{2}
     P(z,\bz) + \partial_z^2\partial_{\bz}^2\N(z,\bz)  &= p^{\mu} \mathcal{D}_{\mu}\left(\delta^{(2)}(z-\infty)\right)  + \frac{1}{2\pi}\;\partial^2_{\bz}\left(  \hat{c}^{\mu \nu}\partial_{z} q_\mu \partial_{z} q_\nu\right).
    \eadat
\end{equation}

We therefore see that the net result of the first term in \eqref{BMS: dressing of eigenstate} is to linearize the hard charge contribution:
if all $c$'s of a scattering process are identical, in particular if they are all zero, then \emph{conservation of supermomentum becomes equivalent to conservation of momentum}.\\

It is also clear that the second term in \eqref{BMS: dressing of eigenstate} brings an extra soft contribution to the charge and that any value $\partial_z^2\partial_{\bz}^2\N$ of the soft charge can be obtained in this way.\footnote{This should be clear from the parametrization below.} This extra contribution to the soft charge is part of the inherent ambiguity of the dressing. Now, if some scattering process preserves momentum $\sum_{\text{in}}p_i^{\mu}=\sum_{\text{out}}p_i^{\mu}$, then conservation of supermomentum becomes equivalent to
 \begin{equation}
\partial_{\bz}^2\left( \sum_{i\in \text{in}} \, \hat{c}_i^{\mu\nu} \partial_{z} q_{\mu} \partial_{z} q_{\nu} \right) = \partial_{\bz}^2\left(  \sum_{i\in \text{out}} \hat{c}_i^{\mu\nu} \partial_{z} q_{\mu} \partial_{z} q_{\nu}\right),
\end{equation}
or, equivalently,\footnote{Making use of the fact that the $\bar{\eth}^2$ operator has no kernel on $s=-2$ spin-weighted coefficients.} \begin{equation}\label{BMS dressing: charge conservation condition}
    \Delta C :=  \Delta c^{\mu\nu} \varepsilon^+_{\mu} \varepsilon^+_{\nu} =0\,,
\end{equation}
where $\Delta c^{\mu\nu}$ is given by \eqref{BMS Dressing: c conservation}. As it turns out (see below), this last condition is equivalent to the condition \eqref{BMS Dressing: IR finite condition} for IR finiteness of the scattering. \emph{As a result, conservation of supermomentum is equivalent to the condition for dressed states to be infrared finite}.

Let us see how this equivalence arise in more details. First, taking into account the constraint $c^{\mu\nu}k_{\nu}+p^{\mu}=0$, one can show that $c_{\mu\nu}$ must be of the form
\begin{align}
    \hat{c}_{\mu\nu}&=  -q\cdot p \, n_{\mu}n_{\nu} +  2n_{(\mu}p_{\nu)} +C \varepsilon^-_{\mu}\varepsilon^-_{\nu} + \bar{C}\varepsilon^+_{\mu}\varepsilon^+_{\nu} + 2A \varepsilon^+_{(\mu}\varepsilon^-_{\nu)} + B_{(\mu}q_{\nu)}\,,
\end{align}
where $B_{\mu}q^{\mu}=0$. Here $C =  \hat{c}^{\mu\nu} \varepsilon^+_{\mu} \varepsilon^+_{\nu} $ directly relates to the soft charge  \eqref{BMS: FK eigenvalue} while $A$ and $B$ are fixed uniquely as a result of \eqref{IR dressing GR: constraint on C}. The exact form of these last terms will be irrelevant for us: they in fact vanish when evaluated in \eqref{R_factor_gr} (as a result of $0=q^{\nu}a_{\mu\nu}$ and $0=g^{\mu\nu}a_{\mu\nu} = (-2q^{(\mu}n^{\nu)} + 2\varepsilon_+^{(\mu}\varepsilon_-^{\nu)}) a_{\mu\nu}  = 2\varepsilon_ +^{(\mu}\varepsilon_-^{\nu)} a_{\mu\nu}$) and are in this sense pure gauge. It now follows that, as a result of momentum conservation,
\begin{align}
    \Delta c_{\mu\nu} &= \Delta C \varepsilon^-_{\mu}\varepsilon^-_{\nu} +\Delta {\bar{C}} \varepsilon^+_{\mu}\varepsilon^+_{\nu}
    + 2 A' \varepsilon^+_{(\mu}\varepsilon^-_{\nu)} + B'_{(\mu}q_{\nu)}
\end{align}
for some functions $A'$ and $B'$ such that $B'_{\mu}q^{\mu}=0$. Finally, making use of this expression, one readily derives that
\begin{equation}
    2\Delta c_{\mu\nu}\, \Delta c^{\mu\nu} - \left( \Delta c{}^{\mu}{}_{\mu} \right)^2 = 4|\Delta C|^2.
\end{equation}
As a result the conditions \eqref{BMS Dressing: IR finite condition} and \eqref{BMS dressing: charge conservation condition} are in fact equivalent.\\

In general, as we already emphasized, the extra contribution to the soft charge brought by $c$ is an inherent ambiguity of the dressing and cannot be discarded without breaking gauge or Lorentz invariance. For example, taking $c_i=0$ in \eqref{BMS dressed eigenstate: eigenvalue} yields a soft charge $p^{\mu} \mathcal{D}_{\mu}\left(\delta^{(2)}(z-\infty)\right)$ which manifestly breaks Lorentz invariance by giving some special role to the south pole, $|z|=\infty$, of the celestial sphere. It can also be traced back to the fact that, taking $c^{\mu\nu}=0$ necessary means that $f^{\mu\nu}k_{\nu}\neq0$ and thus break gauge invariance (which is tied up to Lorentz invariance by the choice $\varepsilon^+_{\mu\nu} = \partial_z q_{\mu} \partial_z q_{\nu}$ for the polarization tensors).  However, making use of the results of section \ref{section: Hard/soft decomposition of supermomenta BMS}, one can in fact devise a Lorentz-invariant dressing for incoming states : it suffices to make sure that total incoming supermomentum $\P_{\text{in}}$ is effectively hard, i.e. $\P_{\text{in}} = P_{\text{in}}$ (this is obtained by choosing the c's such that $\sum_{\text{in}}\frac{1}{2\pi}\;\hat{c}_i^{\mu\nu} \partial_{z} q_{\mu}\partial_{z} q_{\nu} =  -\frac{1}{2\pi}\frac{(p_{\text{in}} \cdot \partial_zq)^2}{p_{\text{in}}\cdot q}$). These are such that conservations of momentum will automatically imply supermomentum conservation.

\paragraph{On Goldstone operators} As in the case of QED, the introduction of a Goldstone operator is unnatural from the perspective of representation theory advocated in the present work; it arises only as a necessity to make contact with dressing constructions. First, the prescription  to  obtain genuine supermomentum eigenstates from dressed states (namely the dressing \eqref{BMS: dressing of eigenstate} followed by the limit $\epsilon \to 0$) feels somewhat ad hoc and rather suggests that the dressing construction attempts to shoehorn the usual Poincaré states into a different Hilbert space -- one that we believe should be that of asymptotic particles, i.e. the space of UIRs of the BMS group. Second, and perhaps most importantly, the pair $(\partial_z^2\partial_{\bz}^2 \hat{\mathscr{N}}, \hat{\C})$, consisting of the soft charge and Goldstone operator, satisfies the commutation relations \eqref{Gravity: commutation relation for the goldstone}, which make them a generalization of momentum and position operators $(\hat{P}, \hat{X})$ with canonical commutation relation $[\hat{P},\hat{X}] = -i$. In other terms, \emph{the Goldstone operator $\hat \C(z,\bz)$ plays the role of a position operator in the space of gravity vacua}\cite{Bekaert:2024uuy}. The same reason that forbids the use of a position operator $\hat{X}^{\mu}$ in quantum field theory makes the use of such a Goldstone operator problematic from the perspective of systematically constructing asymptotic states from UIRs of the BMS group (see the discussion below for further comments).

\phantomsection
\part{Discussion and outlook}

In this article, we developed a systematic and unified representation-theoretic formulation of the relation between asymptotic symmetries in QED and gravity and the infrared divergences that arise in scattering amplitudes built from conventional states. We believe that asymptotic symmetry representation theory offers a natural framework for the study of the $S$-matrix in gravity and gauge theory, a perspective that has, surprisingly, remained almost entirely unexplored so far. It is also expected to provide a rigid framework for the development of flat space holography: at the core of the AdS/CFT correspondence lies the equivalence between unitary irreducible representations of the symmetry algebras realized on both sides of the duality.

We deliberately adopted a pedagogical presentation, formulating the key features in a way that makes almost no reference to the asymptotics of spacetime.
This choice streamlines the exposition,  unifies both massless and massive cases, and spares the reader from having to work out the asymptotic behavior of (nonlinear) fields in different coordinate systems in order to reach the core results. A second aim was to highlight both the coherence of the picture that has emerged over the past ten years on this topic, but also the points of tension that remain within it. 
We now summarize what appear to us to be these main points, before turning to future perspectives.

\paragraph{Spontaneous symmetry breaking?} 
A first point of tension concerns the different roles played by the vacuum with respect to the symmetry generators in the soft theorem analysis and in dressing constructions. This issue is closely related to the question of the existence of superselection sectors. In the context of soft theorems, the central point is that the symmetry generator is thought of as a creation operator for soft particles and, therefore, has a non-trivial action on the vacuum. In other terms, there is a degeneracy of vacua, reminiscent of a spontaneous symmetry-breaking pattern. In the dressing construction, the perspective is rather different: the central idea is to work with eigenstates of the soft supermomentum. Such states are as far as possible from being localized in a QED/gravity vacuum -- much as momentum eigenstates are very much delocalized in position space. 
If one wishes to adopt the point of view of symmetry breaking, the resulting pattern is therefore highly unusual since the best states currently available (namely, dressed infrared-finite states) are superpositions over all possible QED/gravity vacua.

This tension might, in fact, not be as important as it first seems: as highlighted in \cite{Strominger:2017zoo} (section 2.11), the symmetry at stake is unlike any previously discussed symmetry and, despite the degeneracy of vacua, there are no superselection sectors associated with this symmetry breaking. The reason is that the vacuum state will be changed by the infinitely many soft particles produced in any physical process (at the classical level, this relates to the fact that the memory effect is induced by radiation of finite energy). This stands in contrast to usual superselection sectors, which cannot be related by any finite-energy process.  

\paragraph{Supermomentum conservation and unitarity} In the literature on dressed states, the soft charge is typically thought of as parameterizing different superselection sectors (see, e.g., \cite{Gabai:2016kuf,Choi:2017ylo,Prabhu:2022zcr}). This might appear natural because the charge $\mathcal{Q}(z,\bz)$ associated with the asymptotic symmetry group QED is in a sense an extension of the usual electric charge $q_e = \int d^2z \mathcal{Q}$, and the latter is known to parametrize superselection sectors. Our presentation, however, should make clear that the charges associated with the asymptotic symmetries of QED and gravity should rather be understood as parts of a \emph{supermomentum} $\P$, extending the momentum in ordinary quantum field theory. In QED, the hard supermomentum $\P= (p_{\mu}, Q(z,\bz))$ blends together momentum and electric charge (see Table \ref{tbl:Dic}),  while in gravity the BMS supermomentum extends momentum through the relation $p_{\mu} = \int d^2z \,q_{\mu} \P$. In fact, the specific nature of the asymptotic symmetry groups \eqref{eq:AGS} and \eqref{eq:BMS} is precisely that they extend the usual notion of translations in a nontrivial way.\footnote{The present set-up evades the assumption of Coleman-Mandula theorem of finitely many particle species with mass below any fixed value \cite{Coleman:1967ad}.} 

Just as in ordinary QFT it would not be correct to conclude from conservation of momentum that states with different fixed energies are in different superselection sectors, we do not believe that supermomentum should be regarded as parameterizing superselection sectors in the present context. It might be worth pursuing the analogy further: Fixed-momentum states are not superselection sectors in QFT because there exists unitary operators, the Lorentz transformations, that relate states with different momenta. This realizes the physical requirement that a given state must be a physical state for any inertial observer. What is more, one can -- and indeed should -- make linear combinations of momentum eigenstates: this is precisely what wavepackets are. This point is directly related to the way unitarity is implemented in QFT: a superselection sector of fixed momentum would have a hard time producing any normalizable state. Returning to supermomentum, one sees that the very same logic means that, unless one is willing to accept that two boosted observers might disagree on whether a given state is physical, physical states must be UIRs of the asymptotic symmetry groups. This is the perspective advocated in \cite{Bekaert:2024uuy,Bekaert:2025kjb} and in the present work.

\paragraph{Towards a new notion of particles} As advocated above, we believe that the relationships between asymptotic symmetries and IR physics that have been uncovered in the last ten years point to the fact that the usual definition of a particle -- as a UIR of the Poincaré group--, should give way to a new notion of ``asymptotic particle'', defined as a UIR of the asymptotic symmetry group. This might be the key to a systematic construction of IR-finite scattering amplitudes in QED and gravity. Advocating the necessity of a new notion of particles to cope with IR divergences is of course not new (see, e.g., the conclusion of \cite{Kulish:1970ut}). However, the main new insight stimulated by recent developments is that the representation theory for asymptotic symmetries might have a leading role to play, a possibility that has not yet been thoroughly explored. The point of view taken here is thus to take the lessons of the past ten years seriously and to take them to their next logical conceptual step: if an IR-finite unitary $S$-matrix for massless particles can be defined, then the asymptotic one-particle states must belong to UIRs of the asymptotic symmetry group and the multiparticle states must be suitable tensor products thereof. In order for the asymptotic symmetries to be unitarily implemented on the Hilbert space, this calls for an extension of quantum field theory.

In fact, the comparison between dressed/coherent states and supermomentum eigenstates drawn in Sections \ref{sec: dressed to kill QED} and \ref{sec: dressed to kill gravity}
seems to point to the fact that the usual dressing construction is trying to shoehorn this new notion of asymptotic particles into the standard framework of quantum field theory, which is not suited for this purpose. 
A first hint of this comes from that fact that dressed states appear to systematically rely on the crucial use of a Goldstone operator. However, if one pursues the analogy between the asymptotic symmetry group and the Poincaré group, this would amount to making use of a position operator $\hat{X}$ to generate momentum eigenstates. Such an operator certainly does not exist in standard QFT since, in momentum space, this would amount to being able to define an operator $\frac{\partial}{\partial p^{\mu}}$ on wave packets (which is not admissible since physical wave packets must be on shell). For this reason, we believe that asymptotic particles should be taken as fundamental objects in their own right.

\paragraph{Future directions} Clearly, many questions remain open and the physics behind the representation theory still needs to be further studied. 
 A pressing question is the following: is there a sense in which asymptotic particles form a good alternative basis for Kibble's Hilbert space \cite{Kibble:1968sfb,KibbleII,KibbleIII,KibbleIV}? Can we construct multi-particle states built out of asymptotic symmetry UIRs and show how the latter reproduce, in a certain limit, well-known IR-safe observables, such as inclusive cross sections? While this work focused on the study of the infrared structure of QED and gravity in four spacetime dimensions, it would also be very interesting to extend the analysis to higher dimensions and in particular derive a classification of the associated BMS UIRs (see \cite{Bekaert:2026cvx} for preliminary results). In $d>4$ spacetime dimensions, the status of infrared divergences is qualitatively different from the $d=4$ case and therefore deserves a separate and systematic study. Another natural extension of the present work concerns the enlargement of the BMS group to include superrotations \cite{Barnich:2010eb,Kapec:2014opa}, together with the associated soft physics at subleading order. Very little is currently understood about the corresponding representations (see \cite{Freidel:2024jyf,Chen:2025fcc,AliAhmad:2025hdl,Batlle:2025jwn,Ruzziconi:2026isv}) and, except in $d=3$ \cite{Barnich:2014kra,Barnich:2015uva,Oblak:2016eij}}, no classification results are available. 
 Finally, an important future direction concerns the case of non-abelian gauge theories. In that setting, the representation theory is expected to raise new conceptual difficulties, but is also likely to entail particularly rich physical implications.
 We will return to these questions in future works.

\newpage
\section*{Acknowledgments}
The authors gratefully acknowledge numerous discussions with Xavier Bekaert related to the subject of this work. 
The authors also thank Shreyansh Agrawal, Sangmin Choi, Hofie Hannesdottir, Luke Lippstreu, Massimo Porrati, Evgeny Skvortsov, Andy Strominger, Massimo Taronna, Beniamino Valsesia, Robert Wald and Sasha Zhiboedov for useful discussions. YH thanks Xavier Bekaert, Bilal Benazout and Aadharsh Raj for collaborations on representations of the asymptotic symmetry group of QED.
L.D. is supported by the European Research Council (ERC) Project 101076737 -- CeleBH. Views and opinions expressed are however those of the author only and do not necessarily reflect those of the European Union or the European Research Council. Neither the European Union nor the granting authority can be held responsible for them.
L.D. is also partially supported by INFN Iniziativa Specifica ST\&FI.  The authors also thank IDP and SISSA for the hospitality during their respective visits.

\appendix

\phantomsection
\part{Appendices}

\section{Conventions}
\label{app:conventions}
\subsection{Conformal densities and spin weights}
\label{app:conformal_density}

The M\"obius group $PSL(2,\mathbb{C})\simeq \frac{SL(2,\mathbb{C})}{\mathbb{Z}_2}\simeq SO(3,1)$ acts on the celestial sphere $S^2\simeq\mathbb{CP}^1$ as
\begin{equation}\label{eq:Mobius}
     z  \to z'(z)= \frac{az+b}{cz+d}\,,\quad\text{with}\,\begin{pmatrix}
        a & b \\ c & d
    \end{pmatrix}\in SL(2,\mathbb{C})\,.
\end{equation} 
A spin-weighted conformal density $\phi(z,\bar z)$ of spin weight $s\in \mathbb R$ and conformal weight $w \in \mathbb R$ transforms under the action of \eqref{eq:Mobius} as\footnote{\label{footnote on convention for s} Equivalently, seen as sections of (anti)-holomorphic line bundle over  $\mathbb{CP}^1$ and, for comparison with \cite{Eastwood_Tod_1982,Penrose:1985bww}, $\phi(z,\bar z) \in \mathcal{O}\left(w+ s,w - s\right)$. Notice that $s^{here} = -s^{\!\!\text{\cite{Penrose:1985bww,Eastwood_Tod_1982}}}$, the present convention has the advantage that the shear $\bar{\sigma}$, of weight $w=-1$ and $s=+2$, have positive helicity and describes a linearized self-dual space-time deformation.}
 \begin{equation}\label{eq:density}
   \phi(z,\bar z) \quad \mapsto (cz+ d)^{w+s} (\bar c \bar z+\bar d)^{w-s}  \phi(z', \bar z').
\end{equation}   
The vector space of conformal densities (namely with $s=0$) of weight $w$ will be denoted $\mathcal{E}[w]$. In the remaining of this appendix, but not in main text, we will sometimes use the notation $\phi\in\mathcal{E}_s[w]$ for spin-weighted conformal densities on the celestial sphere.

For sake of comparison with the usual CFT notation (with active convention), notice that this definition is equivalent to
 \begin{equation}
   \phi(z,\bar z) \quad \mapsto \left(\frac{\partial z'}{\partial z}\right)^{\!h}  \left(\frac{\partial \bar z'}{\partial \bar z}\right)^{\!\bar h}  \phi(z', \bar z')\,,
\end{equation}   
with the holomorphic and anti-holomorphic weights $(h,\bar h)$ related to $s$ and $w$ via
\begin{equation}
    w=-(h+\bar h) \virg s=-(h-\bar h)\,.
\end{equation}

\subsection{Global conformal densities on the celestial sphere}
\label{app:global conformal densities}

In order to make globally well-defined statements, we will use the following two charts to cover the whole celestial sphere:
\begin{equation}
\badat{2}
&\text{North patch coordinates:}\quad (z,\bar z)  \in \mathbb{C} \quad \text{\,cover\,} S^2\,\backslash\text{\{south pole\}}\\
&\text{South patch coordinates:}\quad (\hat{z} ,\hat{\bar{z}}) \in \mathbb{C} \quad   \text{\,cover\,} S^2\,\backslash\text{\{north pole\}}
\eadat
\end{equation}
with transition function $\hat{z}= z^{-1}$. A global conformal density on the celestial sphere $\phi \in \E[w]$ is then defined by $\phi(z,\bz)$ and $\hat{\phi}(\hat{z},\hat{\bz})$  in each chart, together with the transition rule
\begin{equation}
    \begin{cases}
        \phi(z,\bz) &\text{in the chart} \quad (z,\bar z), \\
        \hat{\phi}(\hat{z},\hat{\bz}) = |z|^{-2w}\phi(z,\bz) &\text{in the chart} \quad (\hat{z} ,\hat{\bar{z}}) =(z^{-1},\bz^{-1}). 
    \end{cases}
\end{equation}
Everywhere in the present article, when writing $\phi \in \E[w]$ we will always mean that $\phi$ is a \emph{global} conformal density on the sphere. Similarly, when writing $\phi \in \E_s[w]$ we will always mean that $\phi$ is a \emph{global} spin-weighted conformal density on the sphere.

\subsection{\texorpdfstring{$\eth$}{Eth} operator }
\label{app: eth operator}

Let $\gamma^{-2} 4dz d\bz$ be the round sphere metric with $\gamma = (1+ |z|^2)$. The edth operators $\eth$ and $\bar \eth$ are defined as\footnote{We here take the definition of \cite[p.323]{Eastwood_Tod_1982} which, as suggested in this reference, is more natural when fields are conformally weighted ($w\neq0)$.  It differs from the one in \cite{Newman:1966ub} by a factor $\gamma = P_{there}$ and is such that $\eth^k f = \gamma^{-k}(\eth_{there})^k f$. Keep also in mind that our convention for $s$ is different, $s^{here} = -s^{\!\!\text{\cite{Penrose:1985bww,Eastwood_Tod_1982}}}$.}
\begin{align}
   \eth &\left| \begin{array}{ccc}
         \E_s[w] & \to & \E_{s-1}[w-1]  \\[0.3em]
         \phi(z,\bar z) & \mapsto 
         & \gamma^{w+s} \,\partial_{z} \left( \gamma^{-w-s}\phi(z,\bar z) \right)
    \end{array}\right.,&    \bar{\eth} &\left| \begin{array}{ccc}
         \E_s[w] & \to & \E_{s+1}[w-1]  \\[0.3em]
         \phi(z,\bar z) & \mapsto 
         & \gamma^{w-s} \,\partial_{\bz} \left( \gamma^{-w+s}\phi(z,\bar z) \right)
    \end{array}\right.,
\end{align} 
where $(w,s)\in\mathbb R^2$ is any couple of real numbers. The edth operator  $\eth$ (resp $\bar{\eth}$ ) is $SU(2)$-covariant but, due to the explicit appearance of $\gamma$, it is not $SL(2,\mathbb{C})$-covariant unless $w+s=0$ (resp $w-s=0$). Nevertheless, for any non-negative integer $k\in\mathbb N$, and any $s\in\mathbb{N}$, the following powers
\begin{align}
   \eth^{k} &\left| \begin{array}{ccc}
         \E_s[k-1-s] & \to & \E_{s-k}[-1-s]  \\[0.3em]
         \phi(z,\bar z) & \mapsto 
         & \p_{z}^{k} \phi(z,\bar z)
    \end{array}\right., &    \bar{\eth}^{k} &\left| \begin{array}{ccc}
         \E_s[k-1+s] & \to & \E_{s+k}[-1+s]  \\[0.3em]
         \phi(z,\bar z) & \mapsto 
         & \p_{\bz}^{k} \phi(z,\bar z)
    \end{array}\right.,
\end{align} 
are $SL(2,\mathbb{C})$-covariant, as manifested here by the fact that $\gamma$ drops out from the expressions (which can be checked explicitly using that $\partial_z^2\gamma=0$).

\subsection{\texorpdfstring{$\delta$}{Delta}-function}
\label{app: delta function}
The $\delta$-function is normalized as\footnote{In contrast, notice the normalization $\int \frac{i}{2}dz \wedge d\bz \,\delta^{(2)}(z) = 1$ in \cite{Bekaert:2025kjb}.}
\be
\int d^2 z \,\delta^{(2)}(z)= \int i dz \wedge d\bz \,\delta^{(2)}(z) = 1
\ee
where $d^2 z \equiv  i dz \wedge d\bz$. The delta function is naturally a global conformal density of weight $w=-2$ on the sphere:
\begin{equation}\label{App: delta function def 1}
    \begin{cases}
        \delta^{(2)}(z) &\text{in the chart} \quad (z,\bar z), \\
        0 &\text{in the chart} \quad (\hat{z} ,\hat{\bar{z}}) =(z^{-1},\bz^{-1})\,. 
    \end{cases}
\end{equation}
Note in particular that $\frac{1}{2\pi}\partial_z\partial_{\bz} \ln |z|^2= \delta^{(2)}(z)$ which is a valid distributional identity on the complex plane, is not a global identity on the celestial sphere. Rather,
\begin{equation}\label{App: delta function def 2}
    \begin{cases}
        \frac{1}{2\pi}\partial_z\partial_{\bz} \ln|z|^2 = \delta^{(2)}(z)  &\text{in the chart} \quad (z,\bar z), \\
        \frac{1}{2\pi}\partial_{\hat{z}}\partial_{\hat{\bz}} \ln|\hat{z}|^{-2} = -\delta^{(2)}(\hat{z}) &\text{in the chart} \quad (\hat{z} ,\hat{\bar{z}}) =(z^{-1},\bz^{-1})\,. 
    \end{cases}
\end{equation}
So that, taking the difference between \eqref{App: delta function def 1} and \eqref{App: delta function def 2}, one finds that, as an identity between global densities on the celestial sphere, $\delta^{(2)}(z) - \frac{1}{2\pi}\partial_z\partial_{\bz} \ln|z|^2 =  \delta^{(2)}(z-\infty)$. Similarly the identity $\frac{1}{2\pi}\partial_z^2\partial_{\bz}^2  \left(|z|^2\ln |z|^2\right)= \delta^{(2)}(z)$, valid on the complex plane, cannot be read as a global identity between global distributions of weight $w=-3$ on the celestial sphere. Rather,
\begin{equation}\label{App: delta function def 3}
    \begin{cases}
        \frac{1}{2\pi}\partial_z^2\partial_{\bz}^2\left( |z|^2\ln|z|^2\right) = \delta^{(2)}(z)  &\text{in the chart} \quad (z,\bar z), \\
        \frac{1}{2\pi}\partial_{\hat{z}}^2\partial_{\hat{\bz}}^2 \ln|\hat{z}|^{-2} = -\partial_{\hat{z}}\partial_{\hat{\bz}} \delta^{(2)}(\hat{z}) &\text{in the chart} \quad (\hat{z} ,\hat{\bar{z}}) =(z^{-1},\bz^{-1})\,,
    \end{cases}
\end{equation}
and thus, $ \delta^{(2)}(z)- \frac{1}{2\pi}\partial_z^2\partial_{\bz}^2 \left( |z|^2\ln|z|^2\right) = \partial_{\hat{z}}\partial_{\hat{\bz}}\delta^{(2)}(z-\infty)$.

\subsection{Identities for \texorpdfstring{$q^{\mu}(z,\bz)$}{q(z)}}
\label{app: q}
The canonical object $q^{\mu}(z,\bz)$ realizes the Lorentz-invariant inclusion of $\mathbb{R}^{3,1}$ inside conformal densities of weight one: 
\begin{equation}
\label{eq: canonical inclusion}
  T^{\mu}\in \mathbb{R}^{3,1} \quad \mapsto\quad T^{\mu} q_{\mu}(z,\bz) \in \E[1].
\end{equation}
In the gravitational context, this means that translations are canonically included inside supertranslations. In the chart $(z,\bz)$ on the complex plane, this is explicitly given by the null vector\footnote{Note the $\sqrt{2}$ factor which differs from \cite{Bekaert:2025kjb}: $q_{\mu}^{\text{here}} = \frac{1}{\sqrt{2}}q_{\mu}^{\text{\cite{Bekaert:2025kjb}}}$. At times, it may also be helpful to have the explicit expression for $
    \hat{q}^\mu(\hat{z},\hat{\bz}) = \frac{1}{\sqrt{2}} \Big(1+\hat{z}\hat{\bz},\hat{z}+\hat{\bz},i(\hat{z}-\hat{\bz}),-1+\hat{z}\hat{\bz}\Big)
$.}
\begin{equation}
    q^\mu(z,\bz) = \frac{1}{\sqrt{2}} \Big(1+z\bz,z+\bz,-i(z-\bz),1-z\bz\Big).
\end{equation}

It is quite useful to keep in mind the identity
\begin{equation}\label{App: qq identity}
    q(z,\bz)\cdot q(\zeta,\bar \zeta) = -|z-\zeta|^2\,.
\end{equation}

\subsubsection{Lorentz invariance}
\label{app:Lorentz invariance of q}

The inclusion \eqref{eq: canonical inclusion} is Lorentz-invariant in the sense that $q^{\mu}(z,\bz)$ satisfies the following essential property: if $p^{\mu}$ is any momentum then\footnote{The convention for antisymmetrization over indices we use is $A_{[\mu\nu]}=\frac 1 2 \left(A_{\mu \nu}-A_{\nu \mu}\right)$.}
\begin{equation}\label{Equivariance of p.q}
    2p_{[\mu} \frac{\partial}{\partial p^{\nu]}} \,\Big( p\cdot q(z,\bz) \Big)^k  = \bigg(  \mathcal{Y}_{\mu\nu}^{z} \partial_z +  \mathcal{Y}_{\mu\nu}^{\bz} \partial_{\bz}   - \frac{k}{2} \left(\partial_z \mathcal{Y}_{\mu\nu}^{z} + \partial_{\bz}\mathcal{Y}_{\mu\nu}^{\bz}
    \right)\bigg)\,\Big(   p\cdot  q(z,\bz) \Big)^k
\end{equation}
where
\begin{equation}
\mathcal{Y}_{\mu\nu}^{z}(z,\bar{z}) \partial_z:=  2\partial_{\bz}q_{[\mu} q_{\nu]} \partial_z \virg  \mathcal{Y}_{\mu\nu}^{\bz}(z,\bar{z}) \partial_\bz:=  2\partial_{z}q_{[\mu} q_{\nu]} \partial_\bz  
\end{equation}
are the infinitesimal generators of Lorentz transformations on the celestial sphere. 

Let us derive this for $k=1$; the general case following straightforwardly. We have
 \begin{align*}
    2p_{[\mu} \frac{\partial}{\partial p^{\nu]}} \left( p\cdot q(z,\bz) \right) & = 2p^{[\mu} q^{\nu]}\\
     &= 2\Big( -(p\cdot \partial_z\partial_{\bz}q) q^{[\mu}  + (p\cdot \partial_z q) \partial_{\bz}q^{[\mu} + (p\cdot \partial_{\bz} q) \partial_{z}q^{[\mu} - (p\cdot q) \partial_{\bz}\partial_z q^{[\mu}   \Big) q^{\nu]} \\
     &=  2\Big(\partial_{\bz}q^{[\mu} q^{\nu]} \partial_z (p\cdot  q) +   \partial_{z}q^{[\mu} q^{\nu]} \partial_{\bz} (p\cdot  q) -  \partial_{\bz}\partial_{z}q^{[\mu} q^{\nu]}\, (p\cdot  q) \Big)
 \end{align*}
 where in the second line we used the completeness relation \eqref{eq:completeness}. Then, since $\mathcal{Y}_{\mu\nu}^{z} =  2\partial_{\bz}q_{[\mu}\, q_{\nu]}$ satisfies
\begin{equation}
    \partial_z\mathcal{Y}_{\mu\nu}^{z} = 2\partial_{z}\partial_{\bz} q_{[\mu}\, q_{\nu]} + 2\partial_{\bz} q_{[\mu} \partial_{z} q_{\nu]} \quad\Rightarrow\quad \frac{1}{2}(\partial_z\mathcal{Y}_{\mu\nu}^{z}+\partial_{\bz}\mathcal{Y}_{\mu\nu}^{\bz})  = 2\partial_{z}\partial_{\bz} q_{[\mu} \,q_{\nu]},
\end{equation}
one finds
 \begin{equation}
     2p_{[\mu}\partial_{p_\nu]} \left( p\cdot q\right)  =  \mathcal{Y}_{\mu\nu}^{z} \partial_z (p\cdot  q) +   \mathcal{Y}_{\mu\nu}^{\bz} \partial_{\bz} (p\cdot  q) -\frac{1}{2}\left( \partial_z \mathcal{Y}_{\mu\nu}^{z} + \partial_{\bz}\mathcal{Y}_{\mu\nu}^{\bz} \right)(p\cdot  q).
 \end{equation}

\subsection{Basic identities for hard supermomenta}
\label{app:hard_id}

The hard QED and BMS supermomenta of Table \ref{tbl:Dic} and \ref{tbl:Dictionary} satisfy the important identity (see \cite{Bekaert:2025kjb})
\begin{align}
    q_e&=\int d^2 z \,Q(z,\bz)\,, &  p_{\mu}&=\int d^2 z \, q_\mu(z,\bz) P(z,\bz) \,.
\end{align}
This is straightforward for the massless case and in the massive case these amounts to showing the following 
\begin{align}
    1&=\int d^2 z \,\frac{1}{4\pi} \frac{m^2}{(p\cdot q(z,\bz))^2}\,, &  p_{\mu}&=\int d^2 z \,\frac{1}{4\pi} \frac{-m^4}{\left( p\cdot q(z,\bz)  \right)^3}\; q_{\mu}\,.
\end{align}
We first observe that, since the expression is Lorentz-invariant, one can always go in the frame where
$p_\mu=\pm m(1,0,0,0)$. Hence, we have $p\cdot q=\pm \frac{m}{\sqrt 2}(1+z\bz)$ and the identity follows from
$
2\pi =\int d^2 z \,(1+z\bz)^{-2}\,.
$

Other important identities satisfied by the hard supermomenta are
\begin{equation}\label{Equivariance of hard supermomenta}
\badat{2}
    &2p_{[\mu} \frac{\partial}{\partial p^{\nu]}} \,Q(z,\bz)  = \Big(  \mathcal{Y}_{\mu\nu}^{z} \partial_z +  \mathcal{Y}_{\mu\nu}^{\bz} \partial_{\bz}   + \left(\partial_z \mathcal{Y}_{\mu\nu}^{z} + \partial_{\bz}\mathcal{Y}_{\mu\nu}^{\bz}
    \right)\Big)\,Q(z,\bz)  \,,\\
      &  2p_{[\mu} \frac{\partial}{\partial p^{\nu]}} \,P(z,\bz)  = \Big(  \mathcal{Y}_{\mu\nu}^{z} \partial_z +  \mathcal{Y}_{\mu\nu}^{\bz} \partial_{\bz}   + \frac{3}{2} \left(\partial_z \mathcal{Y}_{\mu\nu}^{z} + \partial_{\bz}\mathcal{Y}_{\mu\nu}^{\bz}
    \right)\Big)\,P(z,\bz)\,.
\eadat
\end{equation}
For massive hard supermomenta, it follows from \eqref{Equivariance of p.q}. For the massless case, $p^{\mu}=\omega q^{\mu}(\zeta, \bar{\zeta})$, we prove the first equation in \eqref{Equivariance of hard supermomenta} (the proof of the second equation is similar and left to the reader). Making use of \eqref{App: qq identity}, one has $z-\zeta = \frac{p\cdot \partial_{\bz}q}{p\cdot \partial_z\partial_{\bz}q} $ and $\omega= -p\cdot \partial_z\partial_{\bz}q$, from which we derive
\begin{equation}
\badat{5}
2p_{[\mu} \frac{\partial}{\partial p^{\nu]}} \delta^{(2)}(z-\zeta)
&=\left(2p_{[\mu} \frac{\partial}{\partial p^{\nu]}} (z-\zeta)\right) \partial_z\delta^{(2)}(z-\zeta) + c.c.\\
    &= \left( -\frac{2}{\omega}p_{[\mu} \partial_{\bz}q_{\nu]}  - \frac{2 p \cdot \partial_{\bz}q }{\omega^2} p_{[\mu} \partial_{\bz}\partial_{z}q_{\nu]}   \right)\partial_z\delta^{(2)}(z-\zeta) + c.c.\\
    &=\partial_z \left(\Big( -\frac{2}{\omega}p_{[\mu} \partial_{\bz}q_{\nu]}  - \frac{2 p \cdot \partial_{\bz}q }{\omega^2} p_{[\mu} \partial_{\bz}\partial_{z}q_{\nu]}   \Big)\delta^{(2)}(z-\zeta)\right) + c.c.\\
    &= \partial_z \Big( -2 q_{[\mu} \partial_{\bz}q_{\nu]} \delta^{(2)}(z-\zeta)\Big) + c.c.\\
    &= \Big( \mathcal{Y}_{\mu\nu}^{z} \partial_z  + \partial_z\mathcal{Y}_{\mu\nu}^{z}   \Big)\delta^{(2)}(z-\zeta)   +c.c.
    \eadat
\end{equation}
where, in going from from the third to the fourth line, we used that $ \frac{1}{\omega}p^{\mu}\;\delta^{(2)}(z-\zeta) = q^{\mu}\delta^{(2)}(z-\zeta)$ and several times \eqref{App: qq identity}. 

\subsection{Massless momenta, polarization tensors and null coordinates}
\label{app:momenta_massless}

Massless momenta ($p^2=0$) can be parametrized by the light-cone energy $\omega>0$ and coordinates $(\zeta,\bar \zeta)$ on the complex plane as 
\begin{equation}
    p^\mu(\omega,\zeta,\bar \zeta) = \omega\, q^\mu(\zeta,\bar \zeta) ,\quad   q^\mu(\zeta,\bar \zeta) = \frac{1}{\sqrt{2}} \Big(1+\zeta\bar \zeta,\zeta+\bar \zeta,-i(\zeta-\bar \zeta),1-\zeta\bar \zeta\Big)\,.\label{eq:p_massless}
\end{equation}
Polarization tensors for spinning massless particles are constructed from the polarization vectors $\varepsilon^{\pm}_{\mu}(\vec q)$, defined as
 \begin{equation}
    \begin{split}
        \varepsilon^{+}_{\mu}(\vec q) &= \partial_\zeta q_\mu = \frac{1}{\sqrt{2}}\big(-\bar \zeta,1,-i,-\bar \zeta\big), \\
        \varepsilon^{-}_{\mu}(\vec q) &= [\varepsilon^{+}_{\mu}(\vec q)]^* = \partial_{\bar \zeta} q_\mu = \frac{1}{\sqrt{2}}\big(-\zeta,1,i,-\zeta\big).
    \end{split}
    \label{epsilon pola}
\end{equation}
We complete these into a null co-tetrad $\mathcal N = \{q_\mu,  n_\mu, \varepsilon^+_\mu, \varepsilon^-_\mu\}$ by setting $n_{\mu} \equiv \partial_{\zeta}\partial_{\bar \zeta}q_{\mu}$:
\begin{equation}
    q^\mu n_\mu = -1,\quad \varepsilon_+^\mu \varepsilon^-_\mu = 1, \quad q^\mu \varepsilon^\pm_\mu = 0 = n^\mu \varepsilon^\pm_\mu,\label{properties of polarization vectors}
\end{equation}
which follow from the identity \eqref{App: qq identity}.
The completeness relation reads\footnote{The convention for symmetrization over indices we use is $A_{(\mu\nu)}=\frac 1 2 \left(A_{\mu \nu}+A_{\nu \mu}\right)$.}
\be
\label{eq:completeness}
2\left(\varepsilon_+^{(\mu} \varepsilon_-^{\nu)}-q^{(\mu}n^{\nu)}\right)=\eta^{\mu \nu}\,.
\ee

\noindent We use ``flat Bondi'' coordinates $(u,r,z,\bz)$, which relate to Cartesian coordinates $X^\mu$ as
\begin{equation}
    X^\mu = u\,n^{\mu} + r\, q^\mu(z,\bar z)\,, \label{flat_Bondi}
\end{equation}
with $q^\mu(z,\bz)$ the null vector given in \eqref{eq:p_massless}, and for which Minkowski line element reads
\begin{equation}
    \D s^2 = -2\D u\D r+2r^2\D z\D \bar z \, .
\end{equation}
In these coordinates, $(z,\bz)$ cover the celestial sphere where a point at infinity has been removed, and future ($\mathscr{I}^+$) and past ($\mathscr{I}^-$) null infinities are respectively obtained by taking the limits $r \to +\infty$ and $r\to -\infty$. Notice that these are not global coordinates on $\mathbb{R}^{3,1}$: the null vector $n^{\mu} = \partial_z\partial_{\bz}q^{\mu}$ plays a special role and, as a result, the null hypersurface $X^{\mu} n_{\mu}=0$ is not properly covered by $(u,r,z,\bz)$ (since, along $r=0$, $X^\mu = u\,n^{\mu}$ is independent of $z$ and $\bz$).

For the creation/annihilation operators $a_{\pm}^{\dagger}(p)/a_{\pm}(p)$ of massless fields, we make use of \eqref{eq:p_massless} and systematically write $a_{\pm}(\omega,\zeta,\bar \zeta) := a_{\pm}(p)$. The canonical commutation relations are
\begin{equation}\label{Canonical commutation relation in omega,z}
    [a_{\alpha'}(\omega,\zeta,\bar{\zeta}),a_{\alpha}^{\dagger}(\omega',\zeta',\bar{\zeta}') ] = 16 \pi^3 \omega^{-1}\delta(\omega-\omega')\delta^{(2)}(\zeta-\zeta')\delta_{\alpha\alpha'}.
\end{equation}

\subsection{Massive momenta and hyperbolic coordinates}
\label{app:momenta_massive}
Massive momenta ($p^2=-m^2$) of mass $m>0$ are parametrized as 
\begin{equation}
    p^\mu = m \left(\sqrt{1+|y|^2 } \,, \,y ^\alpha\right) 
:=m\,\hat{p}^\mu(y^{\alpha}) \,,\label{eq:massive_p}
\end{equation}
with $\hat p^2\left(y^{\alpha} \right)=-1$.
We use Beig-Schmidt type coordinates $(\tau, y^{\alpha}) \in \mathbb R^+\times \mathbb R^3$ on the inside of the future light-cone (see e.g.~\cite{Borthwick:2024skd}),
 \begin{equation}
     X^\mu = \tau \,\hat{p}^\mu(y^{\alpha}).
 \end{equation}
They provide a hyperbolic slicing of Minkowski space,
\begin{equation}\label{Massive field: flat metric expression}
    \D s^2  = - d \tau^2 + \tau^{2} \left(\delta_{\alpha\beta} - \frac{y_{\alpha}y_{\beta}}{ 1+ |y|^2}\right) dy^{\alpha}dy^{\beta}.
\end{equation}

\section{A lightning review of induced representation methods for UIR of groups of the form \texorpdfstring{$SO(3,1) \ltimes A$}{SO(3,1) x A}.}
\label{Appendix: section representations}

In this appendix, we give a lightning overview of the classification of UIRs for groups of the form
\begin{equation}
    SO(3,1) \ltimes A\,,
\end{equation}
where $A$ is some, possibly infinite-dimensional, abelian group.  See \cite{Mccarthy:1972ry,McCarthy_72-I,McCarthy_73-II,McCarthy_73-III,McCarthy:1974aw,McCarthy_75,McCarthy_76-IV,McCarthy_78,McCarthy_78errata,Girardello:1974sq,PiardBMS,Bekaert:2025kjb} for explicit applications of this general theory to the BMS group. We refer the reader to \cite{Bekaert2026} for the classification of UIRs for the asymptotic symmetry group of QED and \cite{MacKey, Piard} for general mathematical results in this context. 

To fix notation and terminology, we first recall the classification of Poincar\'e group UIRs by Wigner \cite{Wigner:1939cj}. The present exposition follows \cite{Bekaert:2025kjb}.

\begin{theorem}[Induced representations of Poincar\'e group; Wigner 1939]\label{Wignertheo}\mbox{}\\
All UIRs of the group $SO(3,1) \ltimes \mathbb{R}^4$ can be constructed in the following way:

\begin{enumerate}
    \item Choose the orbit $\mathcal{O}_{{p}}$ of a momentum ${p_{\mu}} \in (\mathbb{R}^{3,1})^*$ with mass square $m^2\in\mathbb{R}$ under the group $SL(2,\mathbb{C})$. It defines the mass shell, a finite-dimensional submanifold inside the vector space $(\mathbb{R}^{3,1})^*$ of momenta, 
    \begin{equation*}
        \mathcal{O}_{{p}} \simeq \frac{SL(2,\mathbb{C})}{\ell_{{p}}},
    \end{equation*}
where $\ell_{{p}}$ is the Poincar\'e little group of the momentum $p_{\mu}$. It is the base manifold of an $\ell_{{p}}$-principal bundle over $\mathcal{O}_{{p}}$ with total space $SL(2,\mathbb{C})$.
    \item Choose a UIR of the Poincar\'e little group  $\ell_{{p}}$ on the vector space $V$, i.e. a group morphism $\rho:\ell_{{p}} \to U(V)$ from the little group  $\ell_p$ to the group $U(V)$ of unitary operators on the vector space $V$. This representation is called the (Poincar\'e) spin of the corresponding induced representation of $ISO(3,1)$.
\end{enumerate}
The corresponding UIR of $ISO(3,1)$ with mass square $m^2$ and Poincar\'e spin $\rho$ is then given by the vector space of square-integrable sections of the homogeneous vector bundle
$E\,:=\,SL(2,\mathbb{C}) \times_{\rho} V$ over the coset space $\frac{SL(2,\mathbb{C})}{\ell_{{p}}}$
or, equivalently, by $\rho$-equivariant functions $f:SL(2,\mathbb{C}) \to V$. 
\end{theorem}

The classification of the UIRs for the asymptotic symmetry group of QED and gravity follows a similar pattern, using the following theorem \cite{MacKey}.

\begin{theorem}[Induced representations of group of the form $SO(3,1) \ltimes A$; Mackey 1968]\label{MacKeytheo}\mbox{}\\
All induced UIRs of the group $SO(3,1) \ltimes A$ can be constructed by the following algorithm:

\begin{enumerate}
    \item Choose the orbit $\mathcal{O}_{\P}$ of a supermomentum $\P \in A^*$ under the group $SL(2,\mathbb{C})$. This orbit is a finite-dimensional submanifold inside the vector space $A^*$ of supermomenta,
    \begin{equation*}
        \mathcal{O}_{\P} \simeq \frac{SL(2,\mathbb{C})}{\ell_{\mathcal P}},
    \end{equation*}
where $\ell_{\mathcal P}$ is the little group of the supermomentum $\P$.
It is the base manifold of an $\ell_{\mathcal P}$-principal bundle over $\mathcal{O}_{\P}$ with total space $SL(2,\mathbb{C})$. 
    \item Choose a UIR of the little group  $\ell_{\mathcal P}$ on the vector space $V$, i.e. a group morphism $\rho:\ell_{\mathcal P} \to U(V)$ from the little group  $\ell_{\mathcal P}$ to the group $U(V)$ of unitary operators on the vector space $V$. This representation is called the spin of the corresponding induced representation.
\end{enumerate}
The corresponding UIR of $SO_0(3,1) \ltimes A$ is then given by the vector space of square-integrable sections of the homogeneous vector bundle
$E\,:=\,SL(2,\mathbb{C}) \times_{\rho} V$ over the coset space $\frac{SL(2,\mathbb{C})}{\ell_{\mathcal P}}$
or, equivalently, by $\rho$-equivariant functions $f:SL(2,\mathbb{C}) \to V$. 
\end{theorem}

The exposition above furnishes a suitable starting point for a physical treatment of the associated notion of a “particle.”  In contrast to Wigner’s theorem, the preceding result however does not address whether all possible UIRs can be obtained through induced representations. There is in fact an inherent imprecision with this question: if $A$ is an infinite-dimensional group, the nature of its dual $A^*$ will depend on the topology chosen for $A$. For example, if $A$ is a space of functions on $S^2$, its dual will depend on whether we are thinking of these functions as being $L^2$ (in which case the dual elements are also $L^2$ functions) or $C^{\infty}$ (in which case the natural dual is a space of distributions); for BMS representations, this point has generated some confusion in the literature, and its importance was first highlighted in \cite{Girardello:1974sq}. Thus, for any theorem claiming that induced UIRs exhaust all UIRs to hold, it is necessary to provide assumptions on the topology of $A$. A general sufficient condition was given by Piard in \cite{Piard}, from which he could prove the following result for BMS \cite{PiardBMS}.

\begin{theorem}[Piard 1977]\label{Piardtheo}\mbox{}\\
All UIRs of the BMS group $SO(3,1) \ltimes \E[1]$, with $\E[1]$ square integrable density on $S^2$, are induced. 
\end{theorem}
\noindent As far as we are aware, no results that would cover smooth supertranslations are available at the moment. Considering the strong similarity between asymptotic symmetry group of QED  \eqref{eq:AGS} and  BMS \eqref{eq:BMS}, one expects that a result similar to Theorem \eqref{Piardtheo} should hold for QED.

\section{Distributional identities on the celestial sphere}
\label{App:distrib_id}
\subsection{Distributional identities for hard supermomenta}
\label{app:distrib_id}
In this appendix, we give succinct proofs for the following global identities:
\begin{align}
    Q(z,\bz)&=\frac{q_e}{2\pi}\p_z\p_\bz  \ln|p\cdot q(z,\bz)|+q_e\,\delta^{(2)}(z-\infty,\bz-\infty) \label{Apendix: hard supermomenta distributional identity1}\\
   P(z,\bz)&=-\frac{1}{2\pi} \p_z^2\p_\bz^2 \big( p\cdot q(z,\bz) \ln|p\cdot q(z,\bz)| \big)+p^\mu \mathcal D_\mu(\delta^{(2)}(z-\infty,\bz-\infty))\,,\label{Apendix: hard supermomenta distributional identity2}
\end{align}
satisfied by the QED and BMS hard (massless and massive) supermomenta, given in \eqref{eq:Qhard} and \eqref{eq:hardP}.

To avoid any possible confusion, we stress that the above notation is a convenient way of stating identities between global conformally weighted distributions on the sphere:
\begin{align}
    Q&= \eth \bar{\eth} A + \delta&
   P&= \eth^2 \bar{\eth}^2 B + p^\mu \mathcal D_{\mu}\delta,
\end{align}
 with $A\in \E[0]$, $B\in \E[1]$ and $\delta\in \E[-2]$, $\mathcal D_{\mu}\delta\in \E[-3]$ distributions supported on the south pole ($z=\infty \leftrightarrow \hat{z}=0$). More precise expressions for these identities are given below.

\subsubsection{Distributional identities for QED hard supermomenta}
We here want to derive the identity \eqref{Apendix: hard supermomenta distributional identity1} for the hard supermomenta of QED \eqref{eq:Qhard}. As previously emphasized, it is meant as an identity between global distributions on the sphere: 
\begin{equation}
    \begin{cases}
        Q(z,\bz)=\frac{q_e}{2\pi}\p_\bz\p_z  \ln|p\cdot q(z,\bz)| &\text{in the chart} \quad (z,\bz), \\[0.3em]
        \hat{Q}(\hat{z},\hat \bz)=\frac{q_e}{2\pi}\hat{\p}_{\hat \bz}\hat{\p}_{\hat{z}} \ln
        |p\cdot q(\tfrac{1}{\hat{z}},\tfrac{1}{\hat{\bz}})|+q_e\,\delta^{(2)}(\hat{z},\hat \bz) &\text{in the chart} \quad (\hat{z},\hat \bz)= (z^{-1},\bz^{-1})\,. 
    \end{cases}
\end{equation}

The first part of the identity, in the chart $(z,\bz)$, amounts to proving that for any function $\varepsilon \in\E[0]$ on the celestial sphere which is vanishing in a neighborhood of the south pole $z=\infty$, we have
\begin{equation}\label{appendix: proof of distributional id, chart z}
    \int d^2z \, Q(z,\bz) \varepsilon(z,\bz) =  \frac{q_e}{2\pi} \int d^2z  \ln|p\cdot q(z,\bz)|\,  \p_\bz\p_z\varepsilon(z,\bz)\,.
\end{equation}
For a massive hard supermomentum, $p^2 =-m^2$, this follows from the equality 
\begin{equation}
    \dfrac{q_e m^2}{4\pi\,(q(z,\bar z)\cdot p)^2} = \frac{q_e}{2\pi} \p_\bz \left(\frac{p\cdot \partial_z q(z,\bz)}{p\cdot q(z,\bz)}\right) = \frac{q_e}{2\pi} \p_\bz\p_z\ln|p\cdot q(z,\bz)|\,,
\end{equation}
and noting that we can safely integrate by parts thanks to the fact that $\varepsilon$ vanishes in a neighborhood of $z=\infty$. For the massless hard supermomentum, $p^{\mu} = \omega q^{\mu}(\zeta,\bar{\zeta})$, this follows from the Cauchy-Pompeiu formula
\begin{equation}
\varepsilon(\zeta,\bar \zeta)=\frac{1}{2\pi i}\oint_{\partial D} \frac{\varepsilon(z,\bz)}{z-\zeta}dz +\frac{1}{2\pi i}\int_{D} dz \wedge d\bz \,\frac{\p_\bz \varepsilon(z,\bz)}{z-\zeta}\,,
\end{equation}
where $D \subset \mathbb C$ is an open domain (topologically a disk) inside the complex plane. Neglecting the boundary terms thanks to our assumption on $\varepsilon$, the formula yields the usual distributional identity for the delta function on the complex plane, $\delta^{(2)}(z-\zeta) = \frac{1}{2\pi}\partial_{\bz}\left( \frac{1}{z-\zeta}\right) = \frac{1}{2\pi}\partial_z \partial_{\bz} \ln|z-\zeta|^2$, from which one then readily recovers \eqref{appendix: proof of distributional id, chart z}.

In order to prove the second part of the identity, in the chart $(\hat{z},\hat{\bar{z}})$, there are two possible strategies: \\ A first strategy is to follow step by step the previous proof but now applied to a generic function $\varepsilon \in \E[0]$ on the celestial sphere. Since $\varepsilon$ is not anymore supposed to vanish at infinity, one needs to take great care of boundary terms. Going carefully through this procedure yields:
\begin{align}
\int d^2z \,Q(z,\bz)\varepsilon(z,\bar z)  &=\frac{q_e}{2\pi} \int d^2 z \; \ln|p\cdot q(z,\bz)|\, \p_\bz\p_z \varepsilon(z,\bar z)\; +q_e\,\varepsilon(z=\infty,\bar{z}=\infty)\,.
\end{align}
A more straightforward road is to realize that the above discussion also means that, if $\hat{\varepsilon}(\hat{z},\hat{\bar{z}})$ is a function on the second chart vanishing around $\hat{z}=\infty$ (i.e. $z=0$), then\footnote{It might here help to highlight that $\hat{q}^{\mu}(\hat{z},\hat{\bz}) := |z|^{-2} q^{\mu}(z,\bz)$ has, with respect to the derivatives $\hat{\partial}_{\hat z}$,$\hat{\partial}_{\hat{\bz}}$, the exact same properties that $q^{\mu}(z,\bz)$ had with respect to $\partial_z$, $\partial_{\bz}$. In particular $\hat{q}^{\mu}(\hat{z},\hat{\bz}) \cdot \hat{q}^{\mu}(\hat{w},\hat{\bar{w}}) = |\hat{z} - \hat{w}|^2$. } 
\begin{equation}\label{appendix: proof of distributional id, chart hat z}
    \int d^2\hat{z}  \,\hat{Q}(\hat{z},\hat{\bz}) \hat{\varepsilon}(\hat{z},\hat{\bz}) =  \frac{q_e}{2\pi} \int d^2\hat{z}  \ln|p\cdot \hat{q}(\hat{z},\hat{\bz})|\,  \hat{\p}_{\bar{\hat{z}}} \hat{\p}_{\hat{z}} \hat{\varepsilon}(\hat{z},\hat{\bz})\,
.
\end{equation}
Now, since $p\cdot q$ is a weight-one conformal density, we have $p\cdot\hat{q}(\hat{z},\hat{\bz}) = |z|^{-2} p\cdot q(z,\bz)$, and thus
\begin{equation}
    \int d^2\hat{z}\,  \hat{Q}(\hat{z},\hat{\bz}) \hat{\varepsilon}(\hat{z},\hat{\bz}) =  \frac{q_e}{2\pi} \int d^2\hat{z}  \ln|p\cdot q(\tfrac{1}{\hat{z}},\tfrac{1}{\hat{\bz}})|\,  \hat{\p}_{\bar{\hat{z}}} \hat{\p}_{\hat{z}} \hat{\varepsilon}(\hat{z},\hat{\bz}) + \frac{q_e}{2\pi} \int d^2\hat{z}  \ln|\hat{z}|^2\,  \hat{\p}_{\bar{\hat{z}}} \hat{\p}_{\hat{z}} \hat{\varepsilon}(\hat{z},\hat{\bz}) \,.
\end{equation}
Since $\hat{\varepsilon}$ vanishes around $\hat{z}=\infty$ one can, without further concerns, integrate by part and use the identity $\delta^{(2)}(\hat{z}) = \frac{1}{2\pi}\hat{\partial}_{\hat{z}}\hat{\partial}_{\bar{\hat{z}}} \ln|\hat{z}|^2$. As a result, we find that
\begin{align}
    \int d^2\hat{z}  \,\hat{Q}(\hat{z},\hat{\bz}) \hat{\varepsilon}(\hat{z},\hat{\bz}) =  \frac{q_e}{2\pi} \int d^2\hat{z}  \ln|p\cdot q(\tfrac{1}{\hat{z}},\tfrac{1}{\hat{\bz}})|\,  \hat{\p}_{\bar{\hat{z}}} \hat{\p}_{\hat{z}} \hat{\varepsilon}(\hat{z},\hat{\bz})  +q_e\,\hat{\varepsilon}(\hat{z}=0,\hat{\bar{z}}=0)\,.
\end{align}
This concludes the proof.

\subsubsection{Distributional identities for BMS hard supermomenta}
We now want to derive the identity \eqref{Apendix: hard supermomenta distributional identity2} for the hard supermomenta of gravity \eqref{eq:hardP}. As previously emphasized, it is meant as an identity between global distributions on the sphere:
\begin{equation*}
    \begin{cases}
        P(z,\bz)=-\frac{1}{2\pi}\p_\bz^2\p_z^2\Big( p\cdot q(z,\bz) \ln|p\cdot q(z,\bz)|\Big) &\text{in the chart} \, (z,\bz), \\[0.6em]
        \hat{P}(\hat{z},\hat \bz)=-\frac{1}{2\pi}\hat{\p}_{\hat \bz}^2\hat{\p}_{\hat{z}}^2\Big( p\cdot \hat{q}(\hat{z},\hat{\bz}) \ln|p\cdot q(\frac{1}{\hat{z}},\frac{1}{\hat{\bz}})|\Big)+ p^{\mu}\,\hat{\mathcal{D}}_{\mu}\delta^{(2)}(\hat{z},\hat{\bz}) &\text{in the chart} \, (\hat{z},\hat \bz)= (z^{-1},\bz^{-1}), 
    \end{cases}
\end{equation*}
with $\hat{\mathcal{D}}_{\mu}\delta^{(2)}(\hat{z},\hat \bz)$ the distribution supported on the south pole defined as\footnote{Equivalently as
\begin{equation}
    \int i d\hat{z}\wedge d\bar \hat{z}\; \hat{\T}(\hat{z},\bar{\hat{z}}) \, \hat{\mathcal{D}}_{\mu}\delta^{(2)}(\hat{z}\big) = \frac{1}{\sqrt{2}} \begin{pmatrix}
            \partial_{\hat z}\partial_{\bar{\hat{z}}}\hat\T + \hat\T\\ \partial_{\hat z}\hat\T + \partial_{\bar{\hat{z}}}\hat \T \\ i(\partial_{\bar{ \hat{z}}}\hat \T - \partial_{\hat z}\hat\T )\\  \partial_{\hat z}\partial_{\bar{\hat{z}}}\hat\T -\hat\T 
        \end{pmatrix}\big(\hat z=0\big) = -\hat{\mathcal{D}}_{\mu}\hat{\T}(0,0)\,.
\end{equation}}
\begin{equation}\label{Appendix: Thomas operator distribution}
       \hat{\mathcal{D}}_{\mu}\delta^{(2)}(\hat{z},\hat{\bz})= -\Big(4 \p_{\hat{z}} \p_{\hat{\bz}}\hat{q}_{\mu}(\hat{z},\hat{\bz}) + 2 \p_{\hat{\bz}}\hat{q}_{\mu}(\hat{z},\hat{\bz}) \p_{\hat{z}}+2 \p_{\hat{z}}\hat{q}_{\mu}(\hat{z},\hat{\bz}) \p_{\hat{\bz}} +\hat{q}_{\mu}(\hat{z},\hat{\bz}) \p_{\hat{\bz}}\p_{\hat{z}} \Big) \delta^{(2)}(\hat{z},\hat{\bz})\,.
\end{equation}
As the notation suggests, this distribution can be understood as the evaluation at $\hat{z}=0$ of a conformally covariant quantity, $\mathcal{D}_{\mu}\T$, proportional to the image of $\T \in \E[1]$ under the ``Thomas operator'' $\mathcal{D}_{\mu}$ of tractor calculus, see \cite{Curry_Gover_2018}.

The proof is very similar to that of \eqref{Apendix: hard supermomenta distributional identity1}, a detailed account can already be found in \cite{Bekaert2026}. We will thus only sketch the main steps and leave it as an exercise to the reader to fill the gaps. 

The first part of the identity, in the chart $(z,\bz)$, amounts to proving that, for a density $\T(z,\bz) \in\E[1]$ on the celestial sphere vanishing in a neighborhood of the south pole $z=\infty$, we have
\begin{equation}\label{appendix: proof of distributional id2, chart z}
    \int d^2z \, P(z,\bz) \T(z,\bz) =  -\frac{1}{2\pi} \int d^2z \, p\cdot q(z,\bz)\ln|p\cdot q(z,\bz)|\,  \p_\bz^2\p_z^2\T(z,\bz).
\end{equation}
For the massive hard supermomentum, $p^2 =-m^2$, this follows from the equality 
\begin{equation}
    \dfrac{-m^4}{4\pi\,(q(z,\bar z)\cdot p)^3} = -\frac{1}{2\pi} \p_\bz^2 \left(\frac{\left(p\cdot \partial_z q(z,\bz)\right)^2}{p\cdot q(z,\bz)}\right) = -\frac{1}{2\pi} \p_\bz^2\p_z^2\Big(p\cdot q(z,\bz) \ln|p\cdot q(z,\bz)|\Big),
\end{equation}
and the possibility to integrate by parts without producing any boundary terms thanks to our hypothesis on $\T$. For the massless hard supermomentum, $p^{\mu} = \omega q^{\mu}(z,\bz)$, this follows from using the Cauch-Pompeiu formula recursively (without worrying about boundary terms) to prove the distributional identity on the complex plane: $\delta^{(2)}(z-\zeta) =  \frac{1}{2\pi}\partial_{\bar z}^2 \left( \frac{\bar z - \bar \zeta}{z-\zeta}\right) = \frac{1}{2\pi}\partial_{\bar z}^2 \partial_z^2\Big(|z-\zeta|^2 \ln|z-\zeta|^2 \Big)$. From this last identity, one then obtains \eqref{appendix: proof of distributional id2, chart z} applied to massless momenta.

The second part of the identity, in the chart $(\hat{z},\bar{\hat{z}})$, can, here again, be proved in two ways: A first possibility is to consider a generic density $\T \in\E[1]$ (i.e. not vanishing in a neighborhood of $z=\infty$) and work out carefully all possible boundary terms that will now appear if one reconsiders the previous discussion. This was done in \cite{Bekaert2026} and yields 
\begin{equation}\label{appendix: proof of distributional id2, chart hat z}
    \int d^2z \, P(z,\bz) \T(z,\bz) =  -\frac{1}{2\pi} \int d^2z \, p\cdot q(z,\bz)\ln|p\cdot q(z,\bz)|\,  \p_\bz^2\p_z^2\T(z,\bz) - p^{\mu}\hat{\mathcal{D}}_{\mu}\hat{\T}(\hat{z}=0,\bar{\hat{z}}=0)\,,
\end{equation}
where the boundary term is expressed in terms of \eqref{Appendix: Thomas operator distribution}.

An alternative proof amounts to realizing that the discussion in the chart $(z,\bz)$ also means that, if $\hat{\T}(\hat{z},\hat{\bar{z}})$ is a function on the second chart vanishing around $\hat{z}=\infty$ (i.e. $z=0$), then 
\begin{equation}
    \int d^2\hat{z} \, \hat{P}(\hat{z},\hat{\bz}) \hat{\T}(\hat{z},\hat{\bz}) =  -\frac{1}{2\pi} \int d^2\hat{z}\,  p\cdot \hat{q}(\hat{z},\hat{\bz})\ln|p\cdot \hat{q}(\hat{z},\hat{\bz})|\,  \hat{\p}_{\bar{\hat{z}}}^2 \hat{\p}_{\hat{z}}^2 \hat{\T}(\hat{z},\hat{\bz})\,.
\end{equation}
Making use the fact that $p\cdot q$ is a weight-one conformal density, $p\cdot\hat{q}(\hat{z},\hat{\bz}) = |z|^{-2} p\cdot q(z,\bz)$, and taking into account that, since $\hat{\T}$ vanishes at $\hat{z}=\infty$, one can integrate by part, one obtains
\begin{equation}
    \int d^2\hat{z}  \,\hat{P}(\hat{z},\hat{\bz}) \hat{\T}(\hat{z},\hat{\bz}) =  -\frac{1}{2\pi} \int d^2\hat{z}\,  p\cdot \hat{q}(\hat{z},\hat{\bz})\ln|p\cdot q(\tfrac{1}{\hat{z}},\tfrac{1}{\hat{\bz}})|\,  \hat{\p}_{\bar{\hat{z}}}^2 \hat{\p}_{\hat{z}}^2 \hat{\T}(\hat{z},\hat{\bz}) - p^{\mu}D_{\mu}\,,
\end{equation}
where
\begin{align}
    D_{\mu} & = \frac{-1}{2\pi}\int d^2\hat{z} \; \hat{\p}_{\bar{\hat{z}}}^2 \hat{\p}_{\hat{z}}^2\Big( \hat{q}_{\mu}(\hat{z},\hat{\bz}) \ln |\hat{z}|^2\Big) \; \hat{\T}(\hat{z},\hat{\bz}).
\end{align}
Making use of the identity $\delta^{(2)}(\hat{z}) = \frac{1}{2\pi}\hat{\partial}_{\hat{z}}\hat{\partial}_{\bar{\hat{z}}} \ln|\hat{z}|^2$ (without concerns about boundary terms thanks to our hypothesis on $\hat{\T}$), one can however check that this distribution is entirely supported at $\hat{z}=0$ and coincides with \eqref{Appendix: Thomas operator distribution}, i.e.
$\frac{-1}{2\pi} \hat{\p}_{\bar{\hat{z}}}^2 \hat{\p}_{\hat{z}}^2\Big( \hat{q}_{\mu}(\hat{z},\hat{\bz}) \ln |\hat{z}|^2\Big) = \hat{\mathcal{D}}_{\mu}\delta^{(2)}(\hat{z},\hat{\bz})$.
This concludes the proof.

\subsection{Inner products on soft supermomenta}

\subsubsection{Two-point function on \texorpdfstring{$\E[0]/1$}{E[0]/1}}\label{Ssection: apdx 2-point function E[0]}

The quotient space $\E[0]/1$, of functions on $S^2$ quotiented by constant functions, identifies via $\eth\bar{\eth}$ with soft supermomenta of QED. It is naturally equipped with the scalar product given by (see e.g. \cite{Gelfand2})
\begin{align}
    \langle \partial \mathcal{N}_1, \bar \partial \mathcal{N}_2  \rangle &=  \int d^2z  \, \partial_z\mathcal{N}_1\, \partial_{\bz} \mathcal{N}_2\nonumber\\
    &= \frac{1}{2\pi}\int d^2z_1 \int d^2z_2 \, \partial_{z_1} \partial_{\bz_1}\mathcal{N}_1\, \partial_{z_2} \partial_{\bz_2} \mathcal{N}_2 \;\ln |q(z_1,\bz_1)\cdot q(z_2,\bz_2)|\\
    &= \frac{1}{2\pi}\int d^2z_1 \int d^2z_2 \, \partial_{z_1} \partial_{\bz_1}\mathcal{N}_1\, \partial_{z_2} \partial_{\bz_2} \mathcal{N}_2 \;\ln |z_1-z_2|^2\nonumber\,.
\end{align}
The last expression, although equivalent to the first, appears singular at first glance. In fact, it is not even clear that it is invariantly defined. This is, however, rescued by the fact that the following invariant distributional identity holds for any $\left[\mathcal{N}\right] \in \E[0]/1$ and any $w\in \mathbb{CP}^1$:
\begin{equation}\label{eq:C2}
    \left[\frac{1}{2\pi}\int d^2z \,\partial_z \partial_{\bz} \mathcal{N}(z,\bz) \;\ln|q(z,\bz)\cdot q(w,\bw)| \right]= \left[\mathcal{N}(w,\bw)\right]\,,
\end{equation}
where the bracket stands for the equivalence class $f \sim g \Leftrightarrow f-g =c$ where $c$ is any constant. In this sense, the above distribution plays the role of a delta function $\delta^{(2)}(z-w)$ on the quotient space\footnote{Note that if instead $\mathcal{N}\in \E[0]$, then $\frac{1}{2\pi}\int d^2 z \;\partial_z \partial_{\bz} \mathcal{N}(z,\bz) \;\ln |q(z,\bz)\cdot q(w,\bw)| =  \mathcal{N}(w,\bar w)$ is obviously wrong as can be seen from taking $\mathcal{N} =c$ for some constant $c$.} $\E[0]/1$, and the formal singularity appearing at $z=w$ is as much of a problem as the singularity of the usual delta function. 

The identity \eqref{eq:C2} follows from applying \eqref{Apendix: hard supermomenta distributional identity1} to a massless momenta $p^ {\mu} = q^{\mu}(w,\bw)$,
\begin{equation}\label{Apdx: identity on E[0]}
\frac{1}{2\pi}\int d^2z \,\partial_z \partial_{\bz} \mathcal{N}(z,\bz) \;\ln|q(z,\bz)\cdot q(w,\bw)|  = \mathcal{N}(w,\bw) - \mathcal{N}(\infty,\infty)\,,
\end{equation}
and taking the suitable quotient. It is also instructive to check that the expression does not depend on the choice of chart considered: for example, if $\hat{z}= z^{-1}$, $\hat{w}= w^{-1}$, then one can check that
\begin{equation}\label{Apdx: two point function id QED}
 \left[\int d^2z \,\partial \bar{\partial} \mathcal{N}(z,\bz) \;\ln|q(z,\bz)\cdot q(w,\bw)| \right] =  \left[\int d^2\hat{z} \,\hat{\partial} \bar{\hat{\partial}} \hat{\mathcal{N}}(\hat{z},\bar{\hat{z}}) \;\ln|\hat{q}(\hat{z},\bar{\hat{z}})\cdot \hat{q}(\hat{w},\bar{\hat{w}})| \right]
\end{equation}
where, since they are respectively conformal weight zero and one densities, $\hat{\mathcal{N}}(\hat{z},\bar{\hat{z}})=\mathcal{N}(z,\bar{z})$, $\hat{q}(\hat{z},\bar{\hat{z}}) = |z|^{-1}q(z,\bar{z})$. Equation \eqref{Apdx: two point function id QED} follows from a change of variable, making use of the identity \eqref{Apdx: identity on E[0]} and making use of the definition of the equivalence class to get rid of the unwanted term.

\subsubsection{Two-point function on \texorpdfstring{$\E[1]/\mathbb{R}^{3,1}$}{E[1]/R^{3,1}}}\label{Ssection: apdx 2-point function E[1]}

The quotient space $\E[1]/\mathbb{R}^{3,1}$, of weight-one conformal densities quotiented by the image of \eqref{eq: canonical inclusion}, identifies via $\eth^2\bar{\eth}^2$ with soft supermomenta of gravity. It is naturally equipped with the scalar product given by (see e.g. \cite{Gelfand2})
\begin{align}
    \langle \partial^2 \N_1, \bar \partial^2 \N_2  \rangle &=  \int d^2z  \, \partial_z^2\N_1\, \partial_{\bz}^2 \N_2\\
    &= \frac{-1}{2\pi}\int d^2z_1 \int d^2z_2 \, \partial_{z_1}^2 \partial_{\bz_1}^2\N_1\, \partial_{z_2}^2 \partial_{\bz_2}^2 \N_2 \;q(z_1,\bz_1)\cdot q(z_2,\bz_2)\ln|q(z_1,\bz_1)\cdot q(z_2,\bz_2)|\nonumber\\
    &= \frac{1}{2\pi}\int d^2z_1 \int d^2z_2 \, \partial_{z_1}^2 \partial_{\bz_1}^2\N_1\, \partial_{z_2}^2 \partial_{\bz_2}^2  \N_2 \;|z_1-z_2|^2\ln|z_1-z_2|^2\,.\nonumber
\end{align}
Here again, it might not be obvious at first glance how to invariantly make sense of the last expression. The following invariant distributional identity however holds for any $\left[\N\right] \in \E[1]/\mathbb{R}^{3,1}$ and any $w\in \mathbb{CP}^1$:
\begin{equation}
    \left[\frac{-1}{2\pi}\int d^2z \,\partial_z^2 \partial_{\bz}^2 \N(z,\bz) \;q(z,\bz)\cdot q(w,\bw)\ln|q(z,\bz)\cdot q(w,\bw)| \right]= \left[\N(w,\bw)\right]\,,
\end{equation}
where the bracket stands for the equivalence class $f \sim g \Leftrightarrow f-g =T$, with $T = T^{\mu}q_{\mu}$ a translation. The distribution thus plays the role of a delta function $\delta^{(2)}(z-w)$ on the quotient space\footnote{Here again, if $\N\in \E[1]$ then $-\frac{1}{2\pi}\int d^2 z \;\partial_z^2 \partial_{\bz}^2 \N(z,\bz) \;q(z,\bz)\cdot q(w,\bw)\ln|q(z,\bz)\cdot q(w,\bw)| =  \N(w,\bar w)$ is obviously wrong as can be seen from taking $\N =T$ for some translation $T$.} $\E[1]/\mathbb{R}^{3,1}$ and the formal singularity appearing at $z=w$ is in that sense innocuous.

The identity above follows from applying \eqref{Apendix: hard supermomenta distributional identity2} to a massless supermomenta $p^{\mu} = q^{\mu}(w,\bw)$:
\begin{equation}\label{Apdx: identity on E[1]}
\frac{-1}{2\pi}\!\int d^2z \,\partial_z^2 \partial_{\bz}^2 \N(z,\bz) \;q(z,\bz)\cdot q(w,\bw)\ln|q(z,\bz)\cdot q(w,\bw)|  = \N(w,\bw) + q^{\mu}(w,\bw)\mathcal{D}_{\mu}\N(\infty,\infty)\,,
\end{equation}
 and taking the suitable quotient. One can also check that the expression does not depend on the choice of chart considered.

\bibliographystyle{utphys}
\bibliography{references}

\end{document}